\documentclass{aa} 
\usepackage{txfonts}
\def\sec{\thinspace\hbox{s}}
\def\sqcm{\thinspace\hbox{$\hbox{cm}^{2}$}}
\def\ergsqcmsec{\thinspace\hbox{erg}\sqcm\sec$^{-1}$}
\long\def\symbolfootnote[#1]#2{\begingroup%
\def\thefootnote{\fsymbol{footnote}}\footnote[#1]{#2}\endgroup}
\def\kmsec{\thinspace\hbox{$\hbox{km}\thinspace\hbox{s}^{-1}$}}

\usepackage{graphicx}

\usepackage{natbib,twoopt}
 \bibpunct{(}{)}{;}{a}{}{,}    
 \newcommandtwoopt{\citeads}[3][][]{\href{http://adsabs.harvard.edu/abs/#3}%
                                        {\citealp[#1][#2]{#3}}}
 \newcommandtwoopt{\citepads}[3][][]{\href{http://adsabs.harvard.edu/abs/#3}%
                                        {\citep[#1][#2]{#3}}}
 \newcommandtwoopt{\citetads}[3][][]{\href{http://adsabs.harvard.edu/abs/#3}%
                                        {\citet[#1][#2]{#3}}}
 \newcommandtwoopt{\citeyearads}[3][][]%
   {\href{http://adsabs.harvard.edu/abs/#3}{\citeyear[#1][#2]{#3}}}

\begin{document}
   \title{A Gemini/GMOS study of the physical conditions and kinematics of the 
blue compact dwarf galaxy Mrk 996\thanks{Based on observations obtained at the
Gemini Observatory, which is operated by the Association of Universities for 
Research in Astronomy, Inc., under a cooperative agreement
with the NSF on behalf of the Gemini partnership: the National Science
Foundation (United States), the Science and Technology Facilities Council 
(United Kingdom), the National Research Council (Canada), CONICYT (Chile), 
the Australian Research Council (Australia), Minist\'erio da Ci\^encia e 
Tecnologia (Brazil), and SECYT (Argentina)}}

   \subtitle{}

   \author{Eduardo Telles
          \inst{1},
          Trinh X. Thuan\inst{2,3},
          Yuri I. Izotov
          \inst{4,5},
          \and
          Eleazar Rodrigo Carrasco
          \inst{6}
          }

   \institute{Observat\'orio Nacional, Rua Jos\'e Cristino, 77, Rio de Janeiro,
RJ, 20921-400, Brazil, 
              \email{etelles@on.br}
         \and
      Astronomy Department, University of Virginia, P.O. Box 400325,
Charlottesville, VA 22904, USA            
         \and
Institut d'Astrophysique, Paris, 98 bis Boulevard Arago, 75014, Paris, France,  \email{txt@virginia.edu}                 
         \and
Max-Planck-Institut f\"ur Radioastronomie, Auf dem H\"ugel 69, 53121, Bonn, Germany 
        \and Main Astronomical Observatory, Ukrainian National Academy of Sciences, Zabolotnoho 27, Kyiv, 03680, Ukraine, 
            \email{izotov@mao.kiev.ua}
         \and
Gemini Observatory/AURA, Southern Operations Center, Casilla 603, La Serena, Chile,
      \email rcarrasco@gemini.edu
             }

   \date{Received ...; accepted ...}

 
  \abstract
   {}
   {We present an integral field spectroscopic study with the Gemini 
Multi-Object Spectrograph (GMOS) of the unusual blue compact dwarf (BCD) 
galaxy Mrk 996.}
   {We show through velocity and dispersion maps, emission-line intensity and 
ratio maps, and by a new technique of electron density limit imaging that 
the ionization properties of different regions in Mrk 996 
are correlated with their kinematic properties. }
{From the maps, we can spatially
   distinguish a very dense high-ionization zone with broad lines in
   the nuclear region, and a less dense low-ionization zone with
   narrow lines in the circumnuclear region. Four kinematically
   distinct systems of lines are identified in the integrated spectrum
   of Mrk 996, suggesting stellar wind outflows from a population of
   Wolf-Rayet (WR) stars in the nuclear region, superposed on an
   underlying rotation pattern. From the intensities of the blue and
   red bumps, we derive a population of $\sim$473 late nitrogen (WNL)
   stars and $\sim$98 early carbon (WCE) stars in the nucleus of Mrk
   996, resulting in a high $N$(WR)/$N$(O+WR) of 0.19.

We derive, for the outer narrow-line region, an oxygen abundance
12+log(O/H)=7.94$\pm0.30$ ($\sim$ 0.2 $Z_\odot$) by using the direct
T$_e$ method derived from the detected narrow [O {\sc
    iii}]$\lambda$4363 line. The nucleus of Mrk 996 is, however,
nitrogen-enhanced by a factor of $\sim$20, in agreement with previous
CLOUDY modeling. This nitrogen enhancement is probably due to
nitrogen-enriched WR ejecta, but also to enhanced nitrogen line
emission in a high-density environment. Although we have made use here
of two new methods - Principal Component Analysis (PCA) tomography and a
method for mapping low- and high-density clouds - to analyze our data,
new methodology is needed to further exploit the wealth of information
provided by integral field spectroscopy.  }
   {}


\authorrunning{Telles, Thuan, Izotov \& Carrasco}

\titlerunning{GMOS/IFU spectroscopy of Mrk 996}

   \maketitle
%

\def\ergsqcmsec{\thinspace\hbox{erg}\sqcm\sec$^{-1}$}

\section{Introduction}\label{introduction}

The blue compact dwarf (BCD) galaxy Mrk 996 ($M_{B}$ = $-$16.9) is a
very unusual galaxy. It stands out from its counterparts because of
its extremely large nuclear electron density, of the order of 10$^6$
cm$^{-3}$ instead of the usual several 100 cm$^{-3}$ for H~{\sc ii}
regions.  Much work has been done to study the unusual physical
properties of Mrk 996.  {\sl Hubble Space Telescope} ({\sl HST}) $V$
and $I$ images by \citet{T96} show that the bulk of the star formation
occurs in a compact, roughly circular, high surface brightness nuclear
region of radius ~340 pc, with evident dust patches to the north of
it. The nucleus (n) is located within an elliptical (E) low surface
brightness (LSB) component, so that Mrk 996 belongs to the relatively
rare class of nE BCDs \citep{LT85}. It may also be classified as a
Type I H~{\sc ii} galaxy according to \cite{tel97}. \citet{T96} found
the extended envelope to show a distinct asymmetry. The envelope is more
extended to the northeast side than to the southwest side, perhaps the
sign of a past merger. This asymmetry is also seen in the spatial
distribution of the globular clusters around Mrk 996,  seen
mainly to the south of the galaxy.  The extended LSB component
possesses an exponential disc structure with a small scale length of
0.42 kpc. While Mrk 996 does not show an obvious spiral structure in
the disc, there is a spiral-like pattern in the nuclear star-forming
region, which is no larger than 160 pc in radius.  This galaxy has a
heliocentric radial velocity of 1622 km s$^{-1}$ \citep{T99}, which
gives it a distance of 21.7 Mpc, adopting a Hubble constant of 75 km
s$^{-1}$ Mpc$^{-1}$ and including a very small correction for the
Virgocentric flow. Table~\ref{galinfo} summarizes the basic
information on Mrk 996. At the adopted distance, 1\arcsec\ corresponds
to a linear size of 105 pc.

The UV and optical spectra of the nuclear star-forming region of Mrk
996 \citep{T96} show remarkable features, suggesting very unusual
physical conditions. The He {\sc i} line intensities are 2-4 times
larger than those in normal BCDs. In the UV range, the N {\sc iii}]
  $\lambda$ 1750 and C {\sc iii}] $\lambda$ 1909 are particularly
    intense. Moreover, the line width depends on the degree of
    ionization of the ion. Thus, low-ionization forbidden emission
    lines such as [O {\sc ii}] $\lambda$3726,3729, [S {\sc ii}]
    $\lambda$6717, 6731, and [N {\sc ii}] $\lambda$6548, 6584 have
    narrow widths, similar to those in other H~{\sc ii} regions, while
    high-ionization emission lines such as the helium lines, the [O
      {\sc iii}] $\lambda$4959, 5007, and [Ne {\sc iii}] $\lambda$3868
    nebular lines consist of narrow and broad components, and all
    auroral lines such as [O {\sc iii}] $\lambda$4363, [N {\sc ii}]
    $\lambda$5755, and [S {\sc iii}] $\lambda$6312 are broad with line
    widths of $\geq$ 500 km s$^{-1}$. These correlations of line
    widths with the degree of excitation suggest different ionization
    zones with very distinct kinematic properties. \citet{T96} found
    that the usual one-zone, low-density, ionization-bounded H~{\sc
      ii} region model cannot be applied to the nuclear star-forming
    region of Mrk 996 without leading to unrealistic helium and
    heavy-element abundances. Instead, they showed that a two-zone,
    density-bounded H~{\sc ii} region model that includes an inner
    compact region with a central density of ~10$^6$ cm$^{-3}$ (about
    4 orders of magnitude greater than the densities of normal H~{\sc
      ii} regions) together with an outer region with a lower density
    of $\sim$ 450 cm$^{-3}$ (comparable to those of other H~{\sc ii}
    regions), is needed to account for the observed line
    intensities. The large density gradient is probably caused by a
    mass outflow driven by the large population of Wolf-Rayet stars
    present in the galaxy. The gas outflow motions may account for the
    line widths of the high-ionization lines originating in the dense
    inner region being much broader than the low-ionization lines
    originating in the less dense outer region. The high intensities
    of [N {\sc iii}] $\lambda$1750, [C {\sc iii}] $\lambda$1909, and
    He {\sc i} can be understood by collisional excitation of these
    lines in the high-density region. In the context of this model,
    the oxygen abundance of Mrk 996 is 12+ log O/H =8.0. If we adopt
    12+ log O/H = 8.70 for the Sun \citep{A09}, then Mrk 996 has a
    heavy element mass fraction of 0.2 solar.  The 2-zone CLOUDY models
    with element abundance ratios typical of low-metallicity BCDs
    reproduce well the observed line intensities, except for
    nitrogen. With an enhancement factor of $\sim$5 or greater, the
    nitrogen line intensities can be reproduced. \citet{T96} attribute
    this nitrogen enhancement to local pollution from Wolf-Rayet
    stars.

\citet{T08} have used the {\it Spitzer} satellite to study Mrk 996 in
the mid-infrared (MIR). They also found that a CLOUDY model that
accounts for both the optical and MIR lines requires that they
originate in two distinct H~{\sc ii} regions: a very dense H~{\sc ii}
region where most of the optical lines arise, with densities declining
from ~10$^6$ cm$^{-3}$ at the center to a few hundred cm$^{-3}$ at the
outer radius of $\sim$ 580 pc, and a H~{\sc ii} region with a density
of $\sim$ 300 cm$^{-3}$ that is hidden in the optical, but seen in the
MIR. The infrared lines arise mainly in the optically obscured H~{\sc
  ii} region, while they are strongly suppressed by collisional
deexcitation in the optically visible one. The presence of the [O {\sc
  iv}] 25.89 $\mu$m emission line implies the presence of ionizing
radiation as hard as 54.9 eV. This hard ionizing radiation is most
likely due to fast radiative shocks propagating in a dense
interstellar medium.

Because of the presence in it of distinct ionization zones with
different electron densities and kinematic properties, a very dense
nuclear high-ionization zone with broad emission lines and a less
dense low-ionization zone with narrow emission lines in the
circumnuclear region, Mrk 996 is a prime target for observation with
the Gemini Multi-object Spectrograph. This allows us to carry out a
two-dimensional (2D) study of the kinematics and ionization structure of
Mrk 996 with exquisite spatial and spectral resolution. In the same
spirit, \citet{J09} have also recently carried a 2D study of Mrk 996
with the VLT VIMOS integral field unit, although with less spatial and
spectral resolution. Those authors found that most of the
  emission lines of Mrk 996 show two components: a narrow central
  Gaussian with a full width at half-maximum FWHM$\sim$110 km s$^{-1}$
  superposed on a broad component with FWHM$\sim$400 km s$^{-1}$. The
  [OIII] $\lambda$4363 and [NII] $\lambda$5755 lines show only a broad
  component and are detected only in the inner region. The broad line
  region shows N/H and N/O enhanced by a factor of $\sim$20, while the
  abundances of the other elements are normal. An oxygen abundance of
  12+log(O/H)=8.37, greater than 0.5 that of the Sun, and a very large
  Wolf-Rayet ($\sim$ 3000) and O star ($\sim$ 150 000) population were
  derived. A follow-up Chandra study to explore the presence of an
  intermediate-mass black hole in the heart of Mrk 996 which may
  account for the presence of the [O {\sc iv}] 25.89 $\mu$m line was
  undertaken by \citet{G11}. No Active Galactic Nuclei (AGN) were  found.
     
We discuss the observations and the data reduction in
Sect.~\ref{data}. The integrated spectrum is considered in
Sect.~\ref{sec:integrated}. We discuss here the systems of emission
lines with different kinematics, the collisional excitation of
hydrogen and helium lines, and the Wolf-Rayet stellar population. The
2D kinematics data are presented in Sect.~\ref{sec:kin} in the form of
velocity and velocity dispersion maps. In Sect.~\ref{sec:den} we
present a technique to delimit the spatial extent of the high electron
density region. In Sect.~\ref{PCA results} we apply a recently
developed method for exploiting data cubes and extracting uncorrelated
physical information, called Principal Component Analysis (PCA)
tomography. The 2D description of the physical conditions is presented
in Sect.~\ref{results} through extinction, electron temperature and
density, excitation, and Wolf-Rayet feature maps. We summarize our
conclusions in Sect.~\ref{conclusions}.

\begin{table}
\caption{Basic data on Mrk 996}             
\label{galinfo}      
\begin{tabular}{l c}        
\hline\hline                 
Parameter & Value \\    
\hline                        
$\alpha$(J2000)& 01$^{\rm h}$27$^{\rm m}$35\fs5\\
$\delta$(J2000)&$-$06\degr19\arcmin36\arcsec \\
Heliocentric velocity [\kmsec]&1622\\
$z$&0.0054\\
Distance [Mpc]&21.7\\
$C$(H$\beta$)&0.53\tablefootmark{a}\\
$E(B-V)$$_{Gal}$ &0.044\\
\hline                                   
\end{tabular}\\
\tablefoottext{a}{logarithmic reddening parameter from the 
0.86\arcsec\ aperture {\sl HST} spectrum of \citet{T96}.}
\end{table}


\section{Observations and data reduction}\label{data}


\begin{figure*}[ht]
   \centering
 
   \includegraphics[width=6cm,angle=0.]{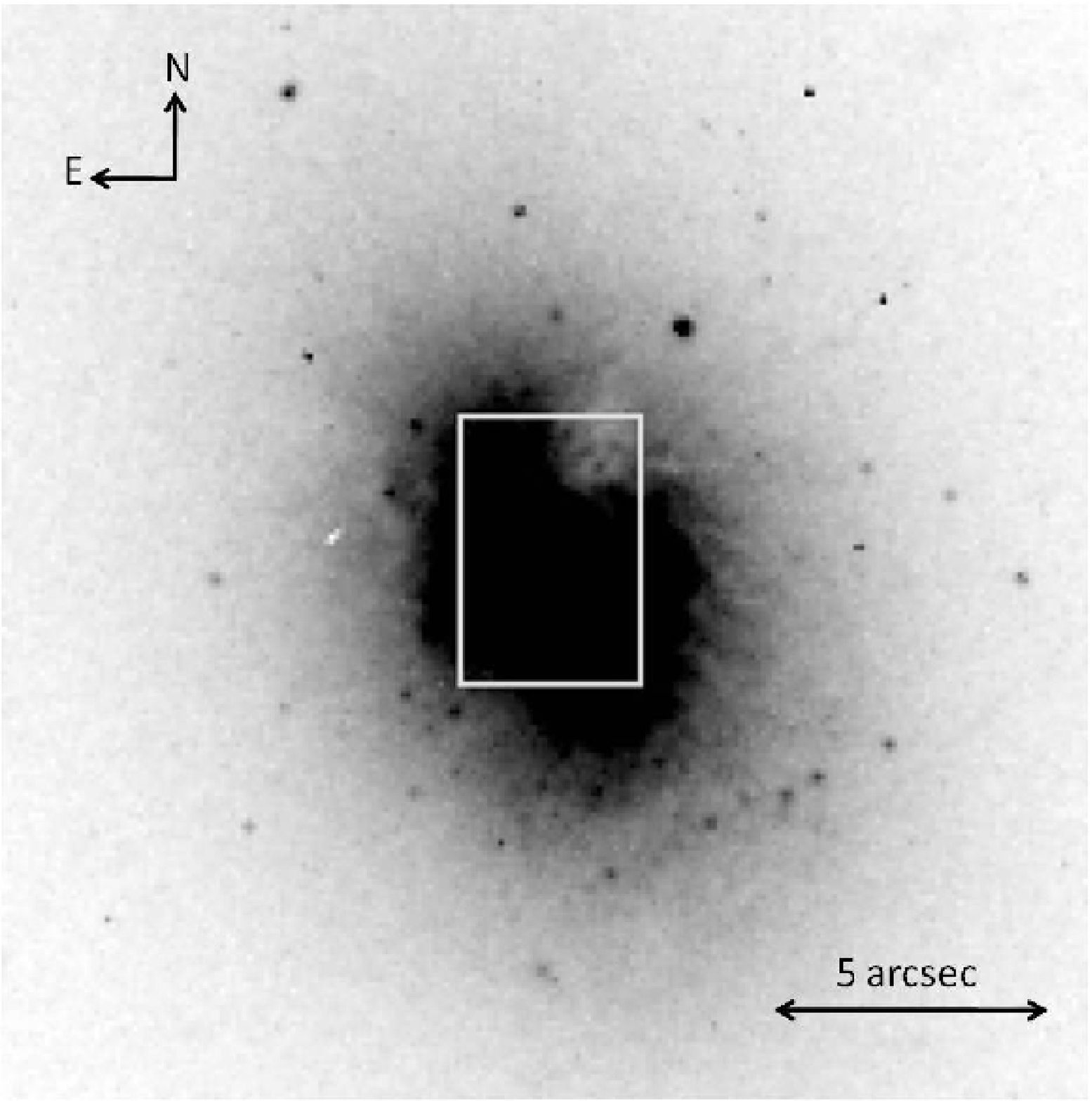}
   \includegraphics[width=6cm,angle=0.]{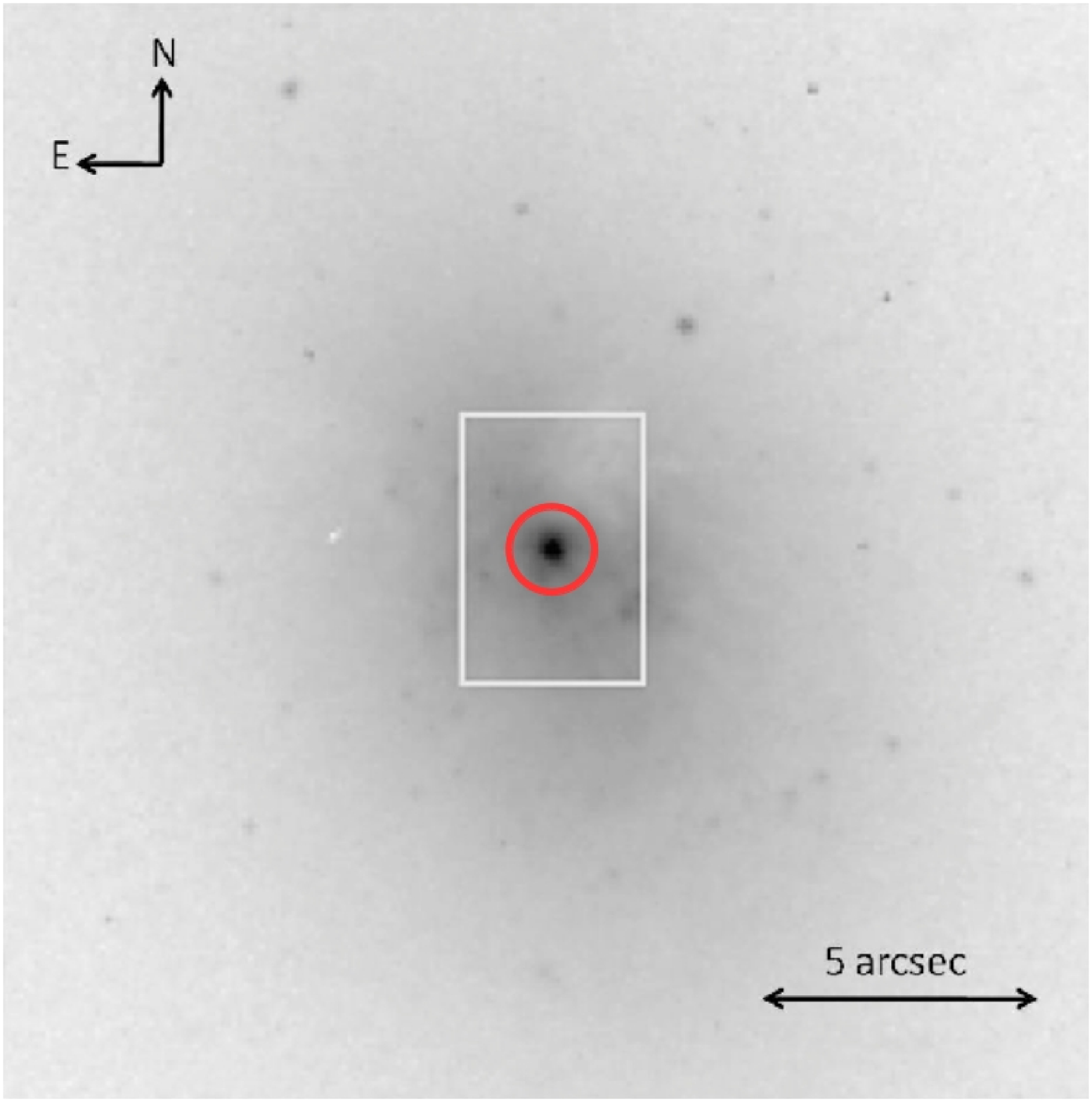}
\caption{ The GMOS 3\farcs5$\times$5\arcsec\ field of view
  superimposed on the inner 20\arcsec\ of the {\sl HST}/F569W image of
  Mrk 996 from \cite{T96} The scale and orientation are given. {\bf
    Left:} Linear contrast. {\bf Right:} Logarithmic contrast to
  emphasize the compact nucleus.  The central circle corresponds to
  the nuclear aperture of 1\farcs6 in diameter used in this work.}
\label{hst}
   \end{figure*}

   The observations were obtained with the Gemini Multi-Object
   Spectrograph (GMOS) \citep{Hook04} and the Integral Field Unit
   (IFU) \citep[][hereafter GMOS/IFU]{Allington02} at the Gemini South
   Telescope in Chile.  They were made during the nights of October
   20, 2008, using the grating B1200$+\_$G5321 (B1200) covering the
   wavelength region from 3667\AA\ to 5142\AA\ with a spectral
   resolution of 0.24\AA, and of November 6, 2008, using the grating
   R831$+\_$G5322 (R831) with a spectral resolution of 0.34\AA,
   covering the wavelength region from 5095\AA\ to 7223\AA, with an
   overlap of $\sim$50\AA\ between the red and blue spectral ranges,
   in the one-slit mode. The GMOS/IFU in this mode composes a
   pattern of 750 hexagonal elements, each with a projected diameter
   of 0\farcs2, covering a total 3\farcs5 $\times$5\arcsec\ field of
   view, where 250 of these elements are dedicated to sky
   observation. The detector is made up of three 2048$\times$4608 CCDs
   with 13.5 $\mu$m pixels, with a scale of 0\farcs073
   pixel$^{-1}$. The CCDs create a mosaic of 6144$\times$4608 pixels
   with a small gap of 37 columns between the chips.  Figure~\ref{hst}
   shows the inner 20\arcsec\ of the {\sl HST} Wide Field Camera F569W
   filter image of Mrk 996 from \cite{T96} with the location of our
   GMOS field of view superimposed on it, with a linear contrast
   (left) and with a logarithmic stretch (right) to emphasize the
   compact nuclear region.

   Table~\ref{obslog} shows the observing log which gives the
   instrumental setup, the mean airmass and exposure times, the
   dispersion, the final instrumental resolution
   ($\sigma_{inst}$=$FWHM_{inst}$/2.355), and the seeing ($FWHM$) of
   each observation.  The data were reduced using the Gemini package
   version 1.8 inside IRAF\footnote[1]{IRAF is distributed by NOAO,
     which is operated by the Association of Universities for Research
     in Astronomy, Inc., under cooperative agreement with the National
     Science Foundation.}. All science exposures, comparison lamps,
   spectroscopic twilight, and GCAL flats were overscan/bias
   subtracted and trimmed. The spectroscopic GCAL flats were processed
   by removing the calibration unit with GMOS spectral response and
   the uneven illumination of the calibration unit. Twilight flats
   were used to correct for the illumination pattern in the GCAL lamp
   flat using the task {\it gfresponse} in the GMOS package. The
   twilight spectra were divided by the response map obtained from the
   lamp flats and the resulting spectra were averaged in the
   dispersion direction, giving the ratio of sky to lamp response for
   each fiber. The final response maps were then obtained by
   multiplying the GCAL lamp flat by the derived ratio. The resulting
   extracted spectra were then wavelength calibrated, corrected by the
   relative fiber throughputs, and extracted. The residual values in
   the wavelength solution for 40 and 60 points, using a Chebyshev
   polynomial of the fourth or fifth order, typically yielded
   \textit{rms} values of $\sim$0.08\AA\ and $\sim$0.07\AA\ for the
   red and blue gratings, respectively. The final spectra cover
   wavelength intervals of $\sim$3667--5142\AA\ and
   $\sim$5095--7223\AA\ for data taken with the B1200 and R813
   gratings, respectively.


\begin{table*}
\caption{Observational setup}
\label{obslog}
\centering
\begin{tabular}{cccccccc}
\hline\hline

Observation & Grating & Central Wavelength& Airmass & Exposure Time&
          Dispersion& $\sigma_{inst}$& Seeing \\
date& & [\AA]& & [seconds]& [\AA/pixel]& [\kmsec]&
              [\arcsec]\\
(1)& (2)& (3)& (4)& (5)& (6)& (7)& (8) \\
\hline
2008 Oct 20 &B1200 & 4420 & 1.64 & 3$\times$1200 & 0.23 & 21.1 & 0.5\\
2008 Nov  6 &R831 & 6160 & 1.22 & 3$\times$900   & 0.34 & 22.8 & 0.8\\
\hline
\end{tabular}
\end{table*}

The flux calibration was performed using the sensitivity function
derived from observations of the star Feige 110 and LTT1020 for both
gratings.  The 2D data images were transformed into a 3D data cube
($x,y,\lambda$), re-sampled as square pixels with 0\farcs1 spatial
resolution and corrected for differential atmospheric refraction (DAR)
using the \textit{gfcube} routine\footnote[2]{The DAR is estimated
  using the atmospheric model from SLALIB.}. The three cubes with different
exposures were combined to produce a single data cube for each
grating. The flux maps on selected emission lines, radial velocity, and
velocity dispersion maps, as well as 1D spectra of various apertures,
were created by an extensive use of QFitsview, developed by Thomas
Ott\footnote[3]{ http://www.mpe.mpg.de/$\sim$ott/QFitsView/}. 
  Both reduced and calibrated data cubes have been made publicly available
\footnote[4]{
available in electronic form
at the CDS via anonymous ftp to cdsarc.u-strasbg.fr (130.79.128.5)
or via http://cdsweb.u-strasbg.fr/cgi-bin/qcat?J/A+A/}.

\section{Integrated spectrum}\label{sec:integrated}

\subsection{Selecting our extraction apertures}

We have simulated apertures for the extraction of the integrated
spectrum in order to compare our results with those of the similar IFU
VIMOS work of \citet{J09}, as well as those of the HST work of
\citet{T96} on the nuclear spectral properties of this peculiar
galaxy.  \citet{J09} used a $1.7 \times 2.3$\arcsec$^2$ aperture for
the core region and a $5.3 \times 6.3$\arcsec$^2$ aperture for the
outer part outside the core, and \citet{T96} obtained a nuclear
spectrum with a 0\farcs86 circular aperture with HST.  The results of
this simulated aperture analysis indicates that the VIMOS data of
\citet{J09} show similar line ratios for both the narrow and broad
components for most lines, but the absolute fluxes are a factor of 4-5
higher than our data for the nuclear aperture.  A direct comparison
with the outer region was not possible because our field of view ($3.5
\times 5.0$ \arcsec$^2$) is slightly smaller than that of VIMOS. On
the other hand, our HST simulated aperture fluxes and flux ratios give
a good match to those of the HST spectrum of \citet{T96}. Because of
the flux discrepancy with the VIMOS data, we have re-reduced the whole
data set using the more recent software that was developed by one of
us (ERC, responsible for GMOS). We have thus double-checked our
measurements by a new and independent data reduction, and confirmed
our calibration.

  Having established the accuracy of our data reduction, by both
  internal and external checks, we decided to present the results for
  the integrated spectrum using a circular aperture of 1\farcs6 in
  diameter, representative of the nuclear region of Mrk 996. This
  aperture size is consistent with the full width at zero intensity
  (FWZI) of the point spread function (PSF) of our calibration star,
  using the same instrument and setup on the same night.  The second
  aperture used in this work contains the outer region, consisting of
  the spaxels within our field of view, but not considering the inner
  1\farcs3 radius (5 pixels from the nucleus aperture).  Although both
  apertures used in the present work (the nuclear and the outer
  regions) are similar to those used by \citet{J09}, ours encompass
  smaller areas than theirs.

\subsection{Systems of emission lines with different kinematics}\label{sec:sys}

\begin{figure*}[ht]
   \centering
   \includegraphics[width=\hsize]{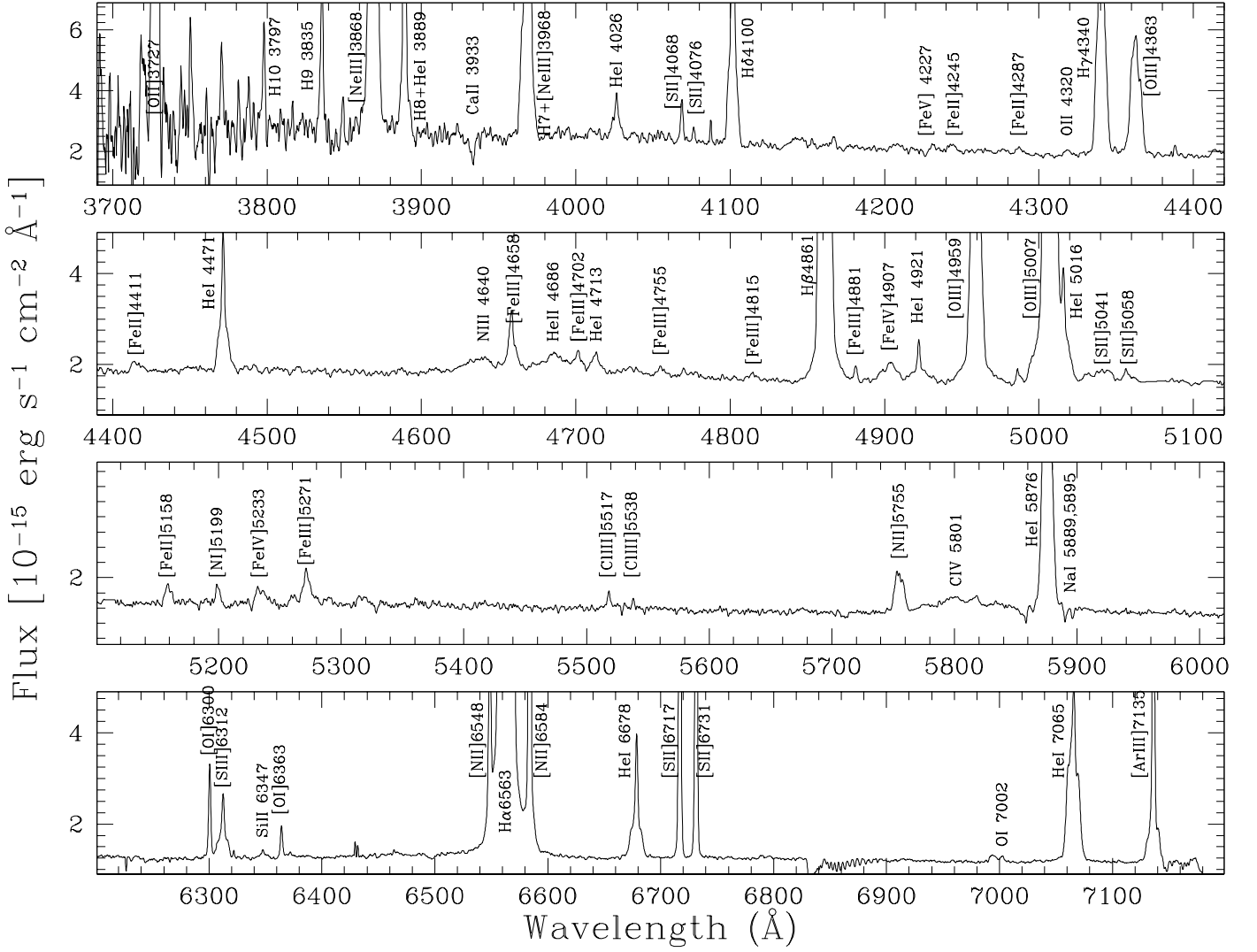}
\caption{Integrated GMOS spectrum of the nucleus of Mrk 996 within a  1\farcs6 
circular aperture for the whole observed spectral range.
\label{specfull}}
   \end{figure*}

   In Fig.~\ref{specfull}, we show the integrated nuclear spectrum of Mrk
   996 over the entire spectral range covered by the two gratings,
   within a 1\farcs6 circular aperture centered on the nucleus. While this
   aperture is very similar to the core region aperture
of \cite{J09}, as mentioned above, it is not identical.

This spectrum has a higher spectral resolution (a factor of $\sim$10 in
the blue and of $\sim$4 in the red) and a considerably higher
signal-to-noise ratio (S/N) than the spectrum of \citet{J09}. It also
covers a larger wavelength range.  The spectrum has not been
smoothed. The continuum levels of the blue and red parts of the
integrated spectrum in the overlapping region at $\sim$ 5100\AA\ match
well although the two data cubes were taken on different nights.  This,
again, confirms our  data reduction and
calibration.  We note that in the bluest part, for $\lambda$ $<$
4000\AA, the continuum is not monotonically increasing to the blue,
implying a poorer calibration due to the known low sensitivity of GMOS
in that wavelength region.  However, the remaining continuum is
monotonically increasing from the red to the blue, in agreement with
the spectrum of \citet{T96}.  By comparison, the continuum in the red
part of the \citet{J09} spectrum is nearly flat, probably indicating
 the contribution of the more spatially extended red old stellar
  population, due to the use of a considerably larger aperture
  (5\farcs3 by 6\farcs3).

The high spectral resolution of the GMOS/IFU observations allows us to
resolve the [O {\sc ii}] $\lambda$3726, 3729 doublet lines. The
hydrogen Balmer lines all show a narrow and a broad component. The
total flux in the broad component of H$\alpha$ is comparable to the
total flux in its narrow component. The He {\sc i} lines
($\lambda$4471, $\lambda$5876, $\lambda$6678, $\lambda$7065) are
clearly broadened.  The broad component is dominant in the [O {\sc
  iii}] $\lambda$4363, [N {\sc ii}] $\lambda$5755, and [S {\sc iii}]
$\lambda$6312 auroral lines,
while the low-ionization species [S {\sc ii}]$\lambda\lambda$
6717,6731, [O {\sc ii}]$\lambda\lambda$ 3726,3729, [N {\sc ii}]
$\lambda\lambda$ 6548,6584, and [O {\sc i}]$\lambda\lambda$ 6300,6363
lines are all narrow, with no broad component. These general trends
agree with those discussed by \cite{T96} and \cite{J09}.

\begin{figure}[ht]
\centering
\includegraphics[width=8cm]{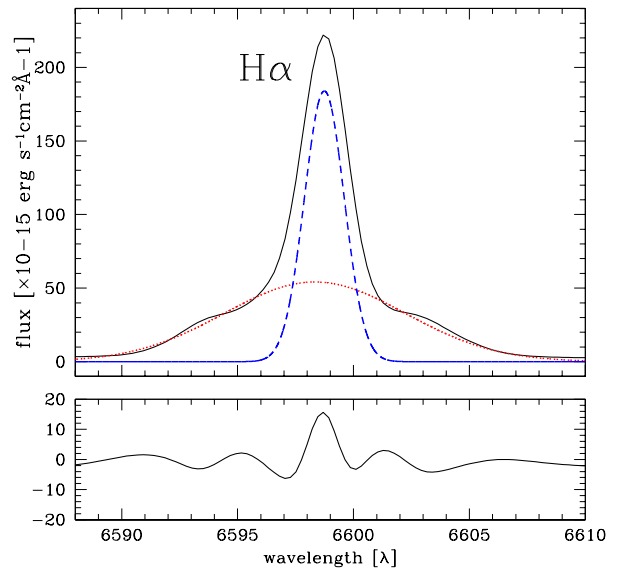}

\caption{Example of the deblend fitting procedure with the H$\alpha$
  line. Upper panel: The integrated profile of the H$\alpha$ line as
  in Figure~\ref{specfull} (black solid line). The red-dotted line shows the
  resulting fit to the broad component. The blue-dashed line shows the
  resulting fit to the narrow component. The integrated fluxes in the
  broad and narrow components in Table~\ref{tabint} are derived in
  this manner. Lower panel: residuals to the fit.}

  \label{halphafit}
\end{figure}

Thanks to the high spectral resolution of our data, the narrow and
broad components of emission lines are well separated. Therefore,
using the IRAF \textit{splot} routine we first fit the narrow
component by a single Gaussian and subtract it from the line
profile. Then we fit the broad component again by a single Gaussian.
Additionally, very broad low-intensity H$\alpha$ emission is present
with a FWZI of $\sim$100\AA, suggesting
rapid outflow with a velocity of several thousand km s$^{-1}$. This
low-intensity H$\alpha$ emission was not discussed by \citet{J09},
probably because of the lower S/N of their spectrum.  In principle, a
multi-Gaussian fitting to emission line profiles should have been
used, including more than two components for each line. However, this
approach is  subjective when more Gaussians are used
for profile-fitting,  the fit is better, without necessarily
reflecting the real physical situation.
Additionally, this procedure significantly complicates studies of the
kinematic structure.  Therefore, we have decided to fit line profiles
in the simplest way, each of their narrow and broad components being
fitted by a single Gaussian.  Fig.~\ref{halphafit} shows an example of
our deblend fits for deriving our fluxes in our line-fitting
procedure. The lower panel shows the residuals of the fit. It can be
seen that the broad component is flat on top, rather than being a
perfect Gaussian. For the purpose here the measured fluxes are little
affected by this deviation.  We show in Table \ref{tabint} the results
of the line-fitting for the integrated nucleus spectrum within the
1\farcs6 circular aperture .  In this table, $\lambda_0$ is the
rest-frame wavelength and $N_e$(crit) is the critical density of the
forbidden line.  The flux is in units of
100$\times$$F_{nar}$/$F_{nar}$(H$\beta$) for the narrow component and
of 100$\times$$F_{br}$/$F_{br}$(H$\beta$) for the broad component;
$v(rad)$ is the radial velocity in km s$^{-1}$, and $FWHM$ is the line
full width at half maximum in km s$^{-1}$. Errors in radial velocities
are negligible and are not quoted.  All measurements were performed by
hand with the task {\it splot} within IRAF, and also by running
non-interactively the profile fitting task {\it fitprofs}, providing
initial guesses for positions and widths of the lines.  The results of
these non-interactive fits are, in most cases, identical to our
measurements with {\it splot}. The advantage is that {\it fitprofs}
computes error estimates for the fitted parameters by using a Monte
Carlo technique that automatically takes into account the properties
of our data. The details of this technique is given in the {\it
  fitprofs} help pages. We have chosen a large
number of iterations for better error estimates. These are the errors
quoted in Table~\ref{tabint}.  The errors introduced by flat-fielding
the data is $<$ 1\%. A larger error of 3-4\% is introduced when
the correction by the relative fiber throughputs is performed
(response map). This will result in a 4-5\% total error, to be
added in quadrature to the listed errors in Table~\ref{tabint}.

{\footnotesize
\begin{table*}
\caption{Emission line parameters derived from the
integrated spectrum within a {\bf 1\farcs6 aperture}}\label{tabint}
\begin{tabular}{lccrrcrrcrr}
\hline\hline             

 &  & $N_e$(crit)\tablefootmark{a} &\multicolumn{2}{c}{Flux}&
&\multicolumn{2}{c}{$v(rad)$}\tablefootmark{d}&
&\multicolumn{2}{c}{$FWHM$}\tablefootmark{d} \\ \cline{4-5} \cline{7-8}
\cline{10-11}
Ion & $\lambda_0$&cm$^{-3}$&narrow\tablefootmark{b} & broad\tablefootmark{c}&&narrow & broad&
&narrow & broad\\
\hline
$[$O {\sc ii}$]$     &3726.03&4.5$\times$10$^3$&    92.39$\pm$1.34 &                 && 1640 &       &&  135$\pm$7   &                 \\
$[$O {\sc ii}$]$     &3728.82&9.8$\times$10$^2$&    93.98$\pm$1.35 &                 && 1634 &       &&  124$\pm$9   &                 \\
H10                  &3797.90&                 &     6.18$\pm$0.98 &                 && 1633 &       &&  111$\pm$8   &                 \\
H9                   &3835.39&                 &    10.83$\pm$1.07 &                 && 1633 &       &&  143$\pm$7   &                 \\
$[$Ne {\sc iii}$]$   &3868.75&8.4$\times$10$^6$&    17.14$\pm$1.09 &   80.55$\pm$1.30&& 1636 &  1606 &&   93$\pm$2   &  447.7$\pm$12   \\
H8+He {\sc i}        &3889.05&                 &    21.23$\pm$1.10 &   13.03$\pm$1.08&& 1623 &  1643 &&  112$\pm$3   &  438.2$\pm$12   \\
He {\sc i}           &4026.19&                 &     1.11$\pm$0.43 &    5.17$\pm$1.06&& 1638 &  1619 &&   82$\pm$2   &  459.5$\pm$12   \\
$[$S {\sc ii}$]$     &4068.60&1.8$\times$10$^6$&     2.67$\pm$0.46 &                 && 1631 &       &&  119$\pm$5   &                 \\
$[$S {\sc ii}$]$     &4076.35&8.6$\times$10$^5$&     0.86$\pm$0.39 &                 && 1623 &       &&  109$\pm$3   &                 \\
H$\delta$            &4101.74&                 &    24.23$\pm$0.87 &   23.39$\pm$1.13&& 1626 &  1649 &&  100$\pm$3   &  433.3$\pm$12   \\
H$\gamma$            &4340.47&                 &    41.84$\pm$1.06 &   42.50$\pm$1.58&& 1629 &  1611 &&   98$\pm$3   &  416.7$\pm$11   \\
$[$O {\sc iii}$]$    &4363.21&2.6$\times$10$^7$&                   &   27.38$\pm$1.02&&      &  1577 &&              &  471.8$\pm$13   \\
He {\sc i}           &4471.48&                 &     3.36$\pm$0.49 &    9.27$\pm$1.19&& 1624 &  1653 &&   92$\pm$2   &  481.8$\pm$13   \\
N {\sc iii} (WR)     &4640.64&                 &                   &    6.25$\pm$1.18&&      &  1385 &&              & 1344.6$\pm$36   \\
$[$Fe {\sc iii}$]$   &4658.10&                 &                   &    5.59$\pm$0.75&&      &  1598 &&              &  321.2$\pm$9    \\
He {\sc ii} (WR)     &4685.68&                 &                   &   10.39$\pm$1.98&&      &  1674 &&              & 2090.4$\pm$56   \\
H$\beta$             &4861.33&                 &   100.00$\pm$1.58 &  100.00$\pm$2.37&& 1641 &  1619 &&   91$\pm$2   &  427.7$\pm$11   \\
$[$O {\sc iii}$]$    &4958.91&6.1$\times$10$^5$&   151.05$\pm$1.97 &   47.81$\pm$1.52&& 1636 &  1662 &&   96$\pm$3   &  396.1$\pm$11   \\
$[$O {\sc iii}$]$    &5006.84&6.1$\times$10$^5$&   470.15$\pm$3.07 &  140.17$\pm$2.43&& 1635 &  1693 &&   98$\pm$3   &  519.2$\pm$14   \\
$[$Cl {\sc iii}$]$   &5517.71&7.4$\times$10$^3$&     0.76$\pm$0.53 &                 && 1649 &       &&  139$\pm$6   &                 \\
$[$Cl {\sc iii}$]$   &5537.88&2.4$\times$10$^4$&     0.39$\pm$0.41 &                 && 1635 &       &&  103$\pm$3   &                 \\
$[$N {\sc ii}$]$     &5754.64&1.2$\times$10$^7$&                   &    5.65$\pm$0.73&&      &  1638 &&              &  435.5$\pm$12   \\
C {\sc iv} (WR)      &5801.51&                 &                   &    5.90$\pm$1.22&&      &  1528 &&              & 1165.6$\pm$41   \\
He {\sc i}           &5875.67&                 &    13.89$\pm$0.90 &   35.48$\pm$1.46&& 1647 &  1596 &&   96$\pm$3   &  449.7$\pm$12   \\
$[$O {\sc i}$]$      &6300.30&1.5$\times$10$^6$&     4.83$\pm$0.52 &                 && 1641 &       &&   96$\pm$3   &                 \\
$[$S {\sc iii}$]$    &6312.10&1.4$\times$10$^7$&     1.94$\pm$0.55 &    5.93$\pm$1.20&& 1637 &  1618 &&   91$\pm$2   &  466.8$\pm$2    \\
Si {\sc ii}          &6347.09&                 &     0.23$\pm$0.39 &                 && 1653 &       &&   75$\pm$7   &                 \\
$[$O {\sc i}$]$      &6363.78&1.5$\times$10$^6$&     1.51$\pm$0.40 &                 && 1640 &       &&   92$\pm$2   &                 \\
$[$N {\sc ii}$]$     &6548.03&7.8$\times$10$^4$&    16.48$\pm$0.49 &                 && 1649 &       &&  127$\pm$3   &                 \\
H$\alpha$            &6562.82&                 &   402.06$\pm$3.01 &  505.98$\pm$4.71&& 1642 &  1623 &&   91$\pm$2   &  417.0$\pm$11   \\
$[$N {\sc ii}$]$     &6583.41&7.8$\times$10$^4$&    37.69$\pm$0.87 &                 && 1644 &       &&  102$\pm$3   &                 \\
He {\sc i}           &6678.15&                 &     3.99$\pm$0.67 &    9.40$\pm$1.38&& 1643 &  1636 &&   85$\pm$4   &  425.0$\pm$11   \\
$[$S {\sc ii}$]$     &6716.47&1.4$\times$10$^3$&    28.79$\pm$0.59 &                 && 1643 &       &&   88$\pm$4   &                 \\
$[$S {\sc ii}$]$     &6730.85&3.6$\times$10$^3$&    26.56$\pm$0.92 &                 && 1643 &       &&   90$\pm$2   &                 \\
O {\sc i}            &7002.23&                 &                   &    0.56$\pm$0.00&&      &  1621 &&              &  178.4$\pm$10   \\
He {\sc i}           &7065.28&                 &     4.83$\pm$0.81 &   29.77$\pm$1.13&& 1635 &  1609 &&  102$\pm$6   &  483.2$\pm$13   \\
$[$Ar {\sc iii}$]$   &7135.78&4.8$\times$10$^6$&    14.61$\pm$1.08 &   11.87$\pm$1.07&& 1645 &  1636 &&   86$\pm$4   &  401.1$\pm$11   \\

\hline
\end{tabular}

\tablefoottext{a}{Critical density for the upper level of the forbidden transition which is defined by the equality of all spontaneous transition rates
and all collisonal transition rates from that level.}

\tablefoottext{b}{In units of 100$\times$$F$/$F_{nar}$(H$\beta$), $F_{nar}$(H$\beta$)=(96.78$\pm$1.53)$\times$10$^{-15}$ erg s$^{-1}$ cm$^{-2}$.}

\tablefoottext{c}{In units of 100$\times$$F$/$F_{br}$(H$\beta$), $F_{br}$(H$\beta$)=(101.24$\pm$2.40)$\times$10$^{-15}$ erg s$^{-1}$ cm$^{-2}$.}

\tablefoottext{d}{Radial velocities and velocity widths are in km s$^{-1}$.}
\end{table*}
%
}

Since the critical densities for collisional deexcitation differ
according to the line (Table \ref{tabint}), different forbidden lines
trace distinct zones of the H~{\sc ii} region in Mrk
996. Additionally, ionization structure plays a role. Emission lines
of higher ionization species, [O {\sc iii}] for example, originate in
the inner part of the H~{\sc ii} region, while the emission of lower
ionization species, [O {\sc i}] for example, is produced in the outer part.
As for permitted lines of hydrogen and helium, they trace both the
inner and outer parts of the H~{\sc ii} region.  Again, broad-line
emission emerges in the inner part while narrow-line emission traces
its outer part.  Narrow-lines are also seen in the
direction of the galaxy center because all lines of sight to the
central part have to go through the outer part of the galaxy.  The
very high electron number density in the center of Mrk 996 is implied
by the extremely high broad [O {\sc iii}] $\lambda$4363/$\lambda$5007
flux ratio of $\sim$ 25\% (Table \ref{tabint}), while typical values
in high-excitation H~{\sc ii} regions are only 1 -- 3\%. Such a high
[O {\sc iii}] $\lambda$4363/$\lambda$5007 flux ratio for the broad
component in Mrk 996 occurs because the [O {\sc iii}]
$\lambda$5007 emission line is suppressed by collisional deexcitation,
while the [O {\sc iii}] $\lambda$4363 emission line is not.

Based on their radial velocities and $FWHM$s (Table \ref{tabint}), we
can identify four kinematically distinct systems of lines, in order of
increasing distance from the center and decreasing line widths.  Since
the density in the inner part of the central H~{\sc ii} region in Mrk
996 is very high, it probably cannot be resolved because of its small
linear extent. Given a constant H$\beta$ luminosity, the
radius of the emitting region scales as $\sim$
$N_e$$^{-2/3}$. Therefore, a region with an electron number density of
$\sim$ 10$^{6}$ cm$^{-3}$ will have a radius $\sim$ 500 times lower
than a region with an electron number density of $\sim$ 10$^{2}$
cm$^{-3}$ and a similar H$\beta$ luminosity.  However, spectral
information presents an advantage in that it allows the physical
conditions to be traced even in unresolved regions.  This is analogous to the
studies of broad and narrow line regions in the spectra of distant AGN
and QSOs.

The densest part of the H~{\sc ii} region appears to be located around
the central ionizing stellar cluster.  The first system of lines is
related to the Wolf-Rayet (WR) stars in this cluster. This system is
composed of the broad permitted N {\sc iii} 4640 and He {\sc ii} 4686
lines (the blue bump) and of the C {\sc iv} 5801 permitted line (the
red bump).  These WR lines have $FWHM$ $\sim$ 1300 - 2000 km s$^{-1}$
and they are blue-shifted by $\sim$ 100 - 200 km s$^{-1}$ with respect
to the narrow component of the H$\beta$ emission line.  These are
produced in the dense stellar winds of WR stars.

The second system of lines probes the innermost zone of the dense
H~{\sc ii} region. It consists of a single forbidden [O {\sc iii}]
$\lambda$4363 emission line with a critical density of 2.6 $\times$
10$^7$ cm$^{-3}$, the highest among all forbidden lines shown in Table
\ref{tabint}.  It has a $FWHM$ of $\sim$ 470 km s$^{-1}$ and is
blue-shifted by 60 km s$^{-1}$ relative to the narrow H$\beta$
emission line.  The line profile of $\lambda$4363 is not smooth and
seems to be complex.  This possibly indicates multiplicity, although
we cannot convincingly investigate this issue further 
without deciding arbitrarily on the number of line components present.
In any case, a very weak peak is seen on the top of the line profile
which coincides with the systemic velocity as given by the narrow
component of H$\beta$, and may be real. This may be the narrow
component seen in the regions outside the nucleus and is discussed
below. Its existence will be confirmed independently  by other
techniques in Sect.~\ref{sec:ON}.

The third system of lines consists of broad components of permitted
hydrogen and helium emission lines, of forbidden emission lines of
doubly ionized ions and of the auroral [N {\sc ii}] $\lambda$5755
emission line (but excluding the [O {\sc iii}] $\lambda$4363 and [Cl
{\sc iii}] $\lambda$5717, 5737 emission lines).  These lines are
blue-shifted by 20 -- 30 km s$^{-1}$ relative to the narrow
H$\beta$ emission line and have $FWHM$s of 450 -- 500 km s$^{-1}$,
similar to the $FWHM$ of the [O {\sc iii}] $\lambda$4363 emission
line.  This lower blueshift indicates that the third line system
originates in regions farther away from the center than does the
second line system.

Finally, the fourth system consists only of lines with narrow components 
($FWHM$ $\sim$ 100 km s$^{-1}$).  These are all narrow lines
and narrow components of emission lines with composite profiles.
Their radial velocities are, within the errors, the same as that 
of the narrow H$\beta$ emission line. 

A global picture consistent with the observed
properties of the above four line systems would be the following.  The
first system of lines arises in the dense circumstellar envelopes of
Wolf-Rayet stars. All other line systems originate in the H~{\sc ii}
region around the ionizing stellar cluster.  The second and third
systems of lines are formed as a result of the outflow of ionized
interstellar medium from the central part of the galaxy, and are due to
stellar winds from the WR stars. Finally, the fourth system of lines
arises in the outer less dense part of the H~{\sc ii} region that is
not perturbed by the ionized gas outflow.  We have been able to
spatially identify this fourth system in the lines of sight away from
the nucleus and extract the outer region spectrum from which a more
precise determination of the HII region abundances could be derived.
These results are presented below.

\subsection{Collisional excitation of hydrogen and helium lines}\label{sec:col}

In general, the electron number density of H~{\sc ii} regions in
star-forming galaxies is low, $\sim$ 100 cm$^{-2}$. At these
densities, the deviations of hydrogen and He {\sc i} line intensities
from their recombination values are expected to be small. Then,
deviations of the hydrogen emission line intensity ratios from their
theoretical values are attributed to extinction. However, in the case
of the dense H~{\sc ii} region in Mrk 996 the effect of collisional
excitation of hydrogen and helium lines is expected to be large,
especially in the densest part of the H~{\sc ii} region where the
broad emission lines originate, as described above.  Among the
hydrogen lines, this effect is highest for the H$\alpha$ emission
line. If collisional excitation is high, then the Balmer decrement
cannot be used for the determination of the extinction coefficient
without correction for that effect.

Table \ref{tabint} shows that the H$\alpha$/H$\beta$
flux ratios for both narrow and broad components are significantly
larger than the theoretical value of $\sim$ 2.9. However, the
H$\gamma$/H$\beta$, H$\delta$/H$\beta$ and H9/H$\beta$ flux ratios for
both narrow and broad components are close to the theoretical values.
For the narrow component, such a small
deviation can be attributed to line flux uncertainties caused by
imperfect flux calibration and differential atmospheric refraction.
For the broad component, the deviation of $\sim$50\% of the
H$\alpha$/H$\beta$ ratio from its theoretical value is too high to be
explained in this way.  We suggest that collisional excitation of
hydrogen plays an important role in the central part of Mrk 996,
enhancing the H$\alpha$/H$\beta$ ratio \citep[e.g., ][]{SI01,P07}.
Using CLOUDY photo-ionized H~{\sc ii} region models for the range of
the ionization parameter appropriate for Mrk 996 (log$U$ = --3 - --2),
a broad H$\alpha$/H$\beta$ flux ratio of $\sim$ 4.6 corresponds to an
electron number density $N_e$ $\sim$ (1 -- 5)$\times$10$^6$ cm$^{-3}$.
This range of $N_e$ is consistent with that derived by
\citet{T96,T08}, but is somewhat lower than $N_e$ $\ga$ 10$^7$
cm$^{-3}$ obtained by \citet{J09} from the analysis of the [O {\sc
    iii}] $\lambda$4363/$\lambda$1663 and $\lambda$5007/$\lambda$4363
flux ratios, although it is consistent with their lower limit of
3$\times$10$^6$ cm$^{-3}$.  Adopting an electron number density $\ga$
10$^7$ cm$^{-3}$ would lead to a broad H$\alpha$/H$\beta$ flux ratio
greater than $\sim$ 5 -- 6.  We note that our value of the broad
H$\alpha$/H$\beta$ flux ratio is not corrected for extinction. If
extinction is non-zero, then the true H$\alpha$/H$\beta$ flux ratio
would decrease, giving a smaller $N_e$. The low extinction-corrected
broad H$\alpha$/H$\beta$ flux ratio ($\sim$2.9) of \citet{J09} (their
Table 2) is inconsistent with their best estimate of high density.
Additionally, the [O {\sc iii}] $\lambda$4363/$\lambda$1663 flux ratio
is highly sensitive to the adopted extinction coefficient and the
reddening curve.  Thus, we conclude that the high central density of
$\ga$ 10$^7$ cm$^{-3}$ derived by \citet{J09} from the broad 
  [OIII] $\lambda$5007/$\lambda$4363 flux ratio is most likely
overestimated. However, the electron density derived by \citet{J09}
from the [FeIII] $\lambda$4881/$\lambda$4658 and
$\lambda$5270/$\lambda$4658 ratios, in the range
(0.5--3)$\times$10$^6$ cm$^{-3}$, is in good agreement with ours.

In addition to the hydrogen lines, the He {\sc i} emission lines are also
subject to important collisional excitation from the meta-stable 2$^3$S
level. Moreover, the He {\sc i} $\lambda$3889 line is optically
thick, as seen below.  This results in a decrease in the intensity of
this line and a fluorescent enhancement of other He {\sc i} emission
lines in the optical spectrum. If both effects are absent, then the
expected intensities of the He {\sc i} $\lambda$3889, $\lambda$4471,
$\lambda$5876, $\lambda$6678, and $\lambda$7065 emission lines
relative to the H$\beta$ line are, respectively, $\sim$0.10, 0.04, 0.11,
0.03, and 0.01 \citep[e.g., ][]{P05}. Since the He {\sc i} $\lambda$3889
emission line is blended with the H8 $\lambda$3889 emission line with
a similar intensity of $\sim$ 0.1 relative to the H$\beta$ line, the total
recombination intensity of the blend He {\sc i} + H8 $\lambda$3889 is
$\sim$ 0.2 relative to the H$\beta$ emission line.

It can be seen from Table \ref{tabint} that the narrow 
He {\sc i} emission lines 
are subject to both collisional and fluorescent 
enhancements. The importance of fluorescent enhancement is implied 
by the weakness of the He {\sc i} + H8 $\lambda$3889 line. Subtracting the
intensity of the H8 hydrogen line, we obtain an intensity of $\sim$ 0.04
for the He {\sc i} $\lambda$3889 emission line. This suggests that
the He {\sc i} $\lambda$3889 emission line is optically thick in the 
region  emitting in narrow lines. On the other hand, the intensity of the 
He {\sc i} $\lambda$7065 line is higher by a factor of $\sim$ 4. This line
is very sensitive to both collisional and fluorescent enhancements, contrary
to the other He {\sc i} $\lambda$4471, $\lambda$5876 and $\lambda$6678
emission lines.

The collisional and fluorescent enhancements of the He {\sc i}
emission lines are more pronounced in the region with broad lines. The
intensity of the He {\sc i} + H8 $\lambda$3889 blend is $\sim$ 0.13,
suggesting that He {\sc i} $\lambda$3889 emission is nearly absent
because of the high optical depth. On the other hand, the He {\sc i}
$\lambda$7065 emission line is enhanced by a factor of $\sim$ 30,
while other He {\sc i} lines are enhanced by a factor of $\sim$
3. These enhancements are much higher than those in H~{\sc ii} regions
of other blue compact dwarf galaxies \citep[see, e.g., ][]{I07} and means that
Mrk 996 is not suitable for He abundance determination.

\citet{J09} have derived the He abundance of Mrk 996, using only one
emission line, He {\sc i} $\lambda$5876, and He {\sc i} emissivities
from \citet{P05}.  They find an He abundance of 0.08 -- 0.10 in
Mrk 996, typical of dwarf emission-line galaxies, with no radial
variation. Several other He {\sc i} emission lines were also present
in the optical spectrum of \citet{J09}. However, no attempt was made
to compare He abundances derived from different lines. In addition,
\citet{P05} emissivities do not take into account fluorescent
excitation of He {\sc i} emission lines \citep{R68}, making the He
abundance determination somewhat uncertain.

\subsection{The Wolf-Rayet  population}\label{sec:wr}

Two types of WR stars are present in Mrk 996. The N {\sc iii}
$\lambda$4640 and He {\sc ii} $\lambda$4686 emission lines,
responsible for the blue bump, are due to  WNL stars,
while the C {\sc iv} $\lambda$5801 emission line, responsible for the
red bump, indicates the presence of WCE stars
\citep{G00}.  We derive the number of WR stars from the fluxes of
broad lines in the spectrum with the 1\farcs6 aperture. The maps
  in Fig.~\ref{fig:wr} (to be discussed later) also show that all of
  the Wolf-Rayet emission comes from this circular region.  We
have also checked the fluxes of these lines in larger apertures and
find that they do not change, also suggesting that all WR stars are located
in the central compact region.    No WR feature is detected
  in the integrated spectrum outside the nucleus, as presented below
  in Sect.~\ref{sec:ON}.

The observed flux of WNL stars (N {\sc iii} + He {\sc ii} emission) is
$F$(WNL) = 1.68$\times$10$^{-14}$ erg s$^{-1}$ cm$^{-2}$, and that of
WCE stars (C {\sc iv} emission) is $F$(WCE) = 5.90$\times$10$^{-15}$
erg s$^{-1}$ cm$^{-2}$ (Table \ref{tabint}).  These fluxes have not
been corrected for extinction because the collisional excitation of
the hydrogen lines makes the determination of the extinction
coefficient uncertain (see previous section).  At a distance of 21.7
Mpc, these fluxes correspond to luminosities $L$(WNL) =
9.46$\times$10$^{38}$ erg s$^{-1}$ and $L$(WCE) =
2.95$\times$10$^{38}$ erg s$^{-1}$. Adopting the luminosity of a
single WNL star to be 2.0$\times$10$^{36}$ erg s$^{-1}$, and that of a
single WCE star to be 3.0$\times$10$^{36}$ erg s$^{-1}$ \citep{SV98},
the numbers of WR stars are $N$(WNL)= 473 and $N$(WCE)= 98, their
ratio $N$(WCE)/$N$(WNL) being 0.20.  These values are typical of WR
galaxies \citep{G00}.

The total observed flux of the H$\beta$ emission line (including both
broad and narrow components, Table \ref{tabint}) is equal to
$F$(H$\beta$) = 1.98$\times$10$^{-13}$ erg s$^{-1}$ cm$^{-2}$. This
corresponds to a luminosity $L$(H$\beta$) = 1.12$\times$10$^{40}$ erg
s$^{-1}$ and a number of ionizing photons $Q$(H) =
2.34$\times$10$^{52}$ s$^{-1}$. The number of O stars can then be
derived from the equation
\begin{equation}
N({\rm O})=\frac{Q({\rm H})-N_{\rm WR}Q^{\rm WR}}{\eta_0 Q^{\rm O7V}},
\end{equation}
where $\eta_0$ is the ratio of the number of O7V stars to the number
of all OV star.  It is equal to 0.5 for a starburst age of 4 Myr
\citep[derived from the equivalent width of H$\beta$ and using the
dependence of $\eta_0$ on EW(H$\beta$) in ][]{SV98}.  The number of
ionizing photons emitted by a single WR or O7V star is $Q^{\rm WR}$ =
$Q^{\rm O7V}$ = 1$\times$10$^{49}$ s$^{-1}$ \citep{SV98}. Then, the
number of O stars in Mrk 996 is $N$(O) = 2345, giving
$N$(WR)/$N$(O+WR) = 0.19. This number of WR stars relative to that of
O stars is among the highest found for WR galaxies \citep{G00}.

Our estimates of the number of WNL and WNC stars are very similar to
those given by \citet{T96}: $N$(WNL)= 601 and $N$(WCE)= 74.  On the
other hand, our estimates of WNL, WCE, and O stars do not agree with
those derived by \citet{J09}.  Their very high values ($\sim$ 3000 WR
stars and $\sim$150 000 O stars) are partly a consequence of 
estimates  made using the flux integrated over the entire galaxy,
rather than just the core region, and partly due to their erroneously
high H$\beta$ flux (their Table 2), a factor of $\sim$ 5 higher than
ours, when duly compared with our simulated VIMOS aperture.  Such a
high H$\beta$ flux is inconsistent with our many observations of Mrk
996. Furthermore, their observed H$\alpha$ flux (narrow+broad) of
3.65$\times$10$^{-12}$ erg cm$^{-2}$s$^{-1}$ is $\sim$ 7 times higher
than the total H$\alpha$ flux of (5.4$\pm$0.7)$\times$10$^{-13}$ erg
cm$^{-2}$s$^{-1}$ obtained by \citet{gil03} from H$\alpha$ integrated
photometry over the whole extent of the line emission.  On the other
hand, our value of 9.03$\pm 0.13 \times$10$^{-13}$ erg
cm$^{-2}$s$^{-1}$ for the H$\alpha$ flux in a 1\farcs6 aperture is
more consistent with the \citet{gil03} value.

\section{Mapping the kinematics of broad and narrow lines}\label{sec:kin}

\subsection{Velocity maps}\label{velmaps}

\begin{figure*}[ht]
\centering
\includegraphics[width=6cm]{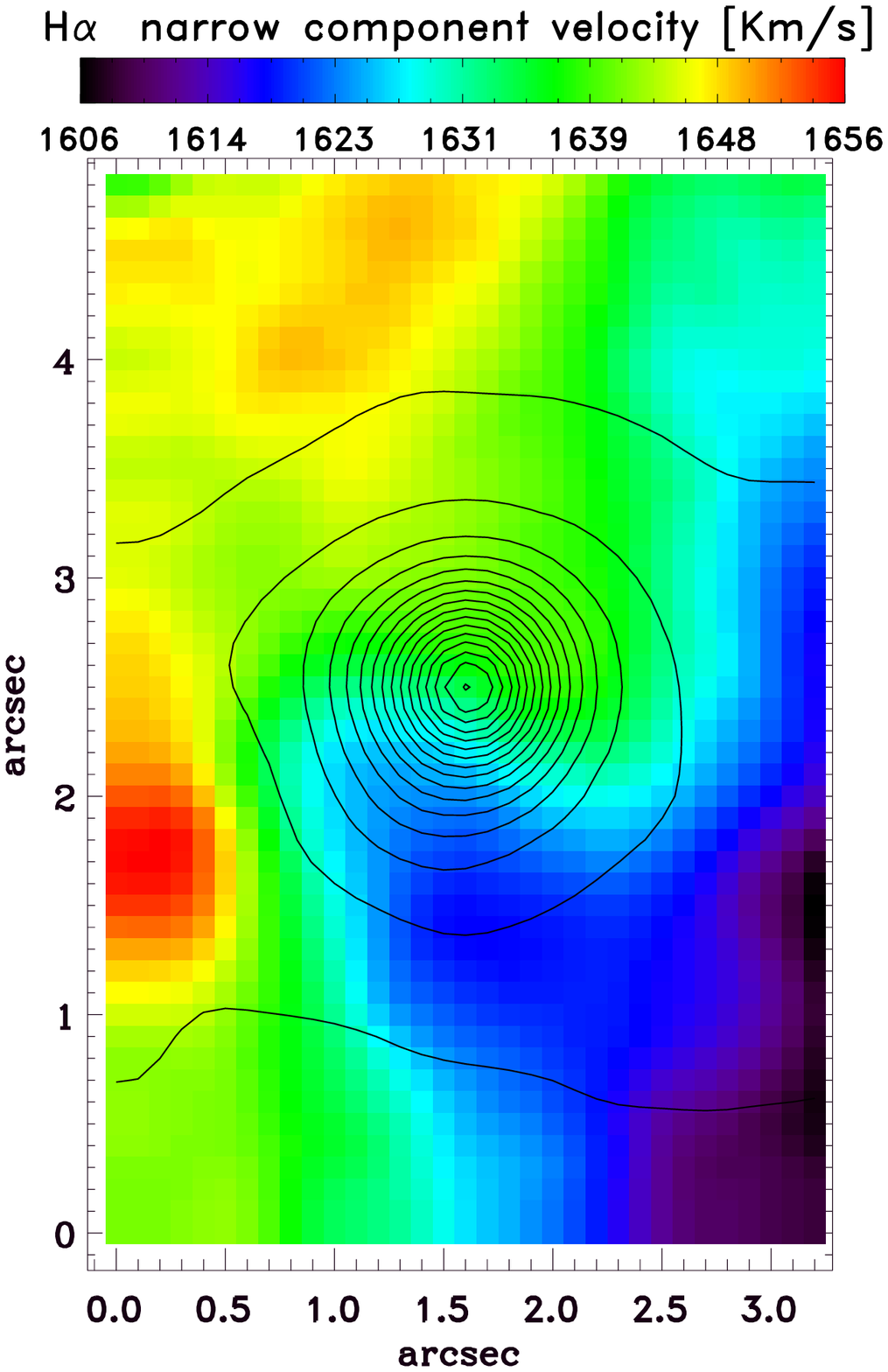}
\includegraphics[width=6cm]{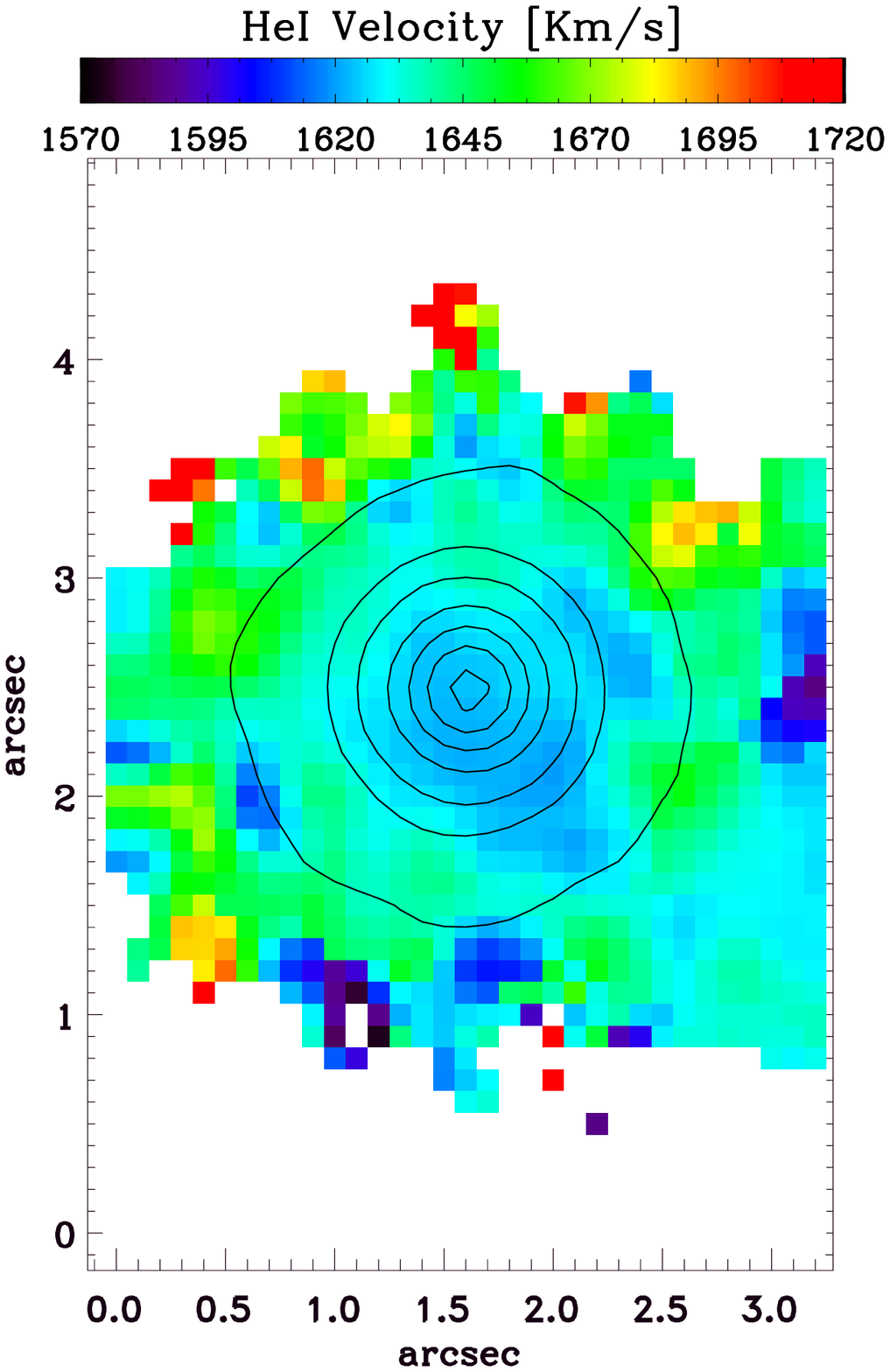}
\caption{ Radial velocity maps. {\bf (Left)} Map of the narrow
  component of the H$\alpha$ line. This map is representative of all
  other narrow-line maps (e.g., narrow [O {\sc iii}] $\lambda$5007, [O
  {\sc i}] $\lambda$6300, [S {\sc ii}] $\lambda$6717).  {\bf (Right)}
  Map of the He {\sc i} $\lambda$7065 line. This map is representative
  of other broad-line maps (e.g., [O {\sc iii}] $\lambda$4363, broad
  H$\alpha$). Contours are the monochromatic intensity of the
  corresponding emission-line from decreasing intensity intervals of
  $2\times 10^{-17}$ \ergsqcmsec in peak intensity for H$\alpha$ and
  $10^{-18}$ \ergsqcmsec for HeI. Only pixels above 3 $\sigma$ of the
  noise are shown for HeI. In both panels, north is up and east is to
  the left. \label{fig:nar_vel}}

\end{figure*}

As discussed above, there are two main regions in Mrk 996 with
distinct kinematic properties \citep[see also][]{T96,J09}:
the central high-ionization broad-line emission zone and the outer
low-ionization narrow-line emission zone. We now present maps of both
regions in the strongest emission lines and use them to discuss the
kinematics of the broad and narrow components.  All maps presented
  here made extensive use of the QFitsView astronomical
  package\footnote[5]{QFitsView is a one, two, and three dimensional FITS
    file viewer written by Thomas Ott and is used for reducing astronomical
    data.}. In particular, the function {\it velmap} goes through a
  datacube and fit a Gaussian to a line. The arguments CENTER and
  FWHM, provided by the user, are used as initial estimates for the
  {\it gauss-fit}. The task then returns the results of the best fit,
  primarily the fit line center and fit line FWHM, producing the
  corresponding radial velocity and velocity dispersion maps.

Figure~\ref{fig:nar_vel} (left panel) shows the radial velocity maps
for the narrow component of the H$\alpha$ line. Velocity maps of other
narrow lines such as [O {\sc i}] $\lambda$6300, [S {\sc ii}]
$\lambda$6717, and narrow [O {\sc iii}] $\lambda$5007 are similar to
the H$\alpha$ narrow-component map, and are not shown. All maps are
consistent with a systemic velocity of $\sim$ 1640 km s$^{-1}$, in
agreement with the velocities of the narrow lines in the integrated
spectrum (Table \ref{tabint}).  Examination of the H$\alpha$ narrow
component velocity map reveals a blueshift in the SW direction and a
redshift in the NE direction, indicative of an overall rotation
pattern.  The kinematic pattern changes in the inner 2
\arcsec\ where the symmetry axis becomes oriented in the EW
direction.  Such a twisted velocity map suggests an isotropic gas
outflow from the center, superimposed on a rotation pattern of the
underlying disc.  We also note the presence of a high-velocity
feature, perhaps produced  by an outflow blob, in the SE direction,
at the position $x$= 0\farcs2 and $y$= 1\farcs8.  This high-velocity
component is seen only in the direction diametrically opposite to the
region of high extinction in the NW discussed below.  We will discuss
this high-velocity feature further below.

Figure~\ref{fig:nar_vel} (right panel) shows the radial velocity map
for the broad He {\sc i} $\lambda$7065 line emission in the central
region.  In contrast to the narrow H$\alpha$ line emission map, it
shows no overall rotation pattern.  The velocity of the nuclear 
broad-line emission is blueshifted with respect to that of the narrow
component, in agreement with the broad-line velocities in the
integrated spectrum (Table \ref{tabint}). There is some indication of
slightly higher radial velocities in regions around the nuclear
region.  This ring-like velocity structure may indicate isotropic
motions around the central region. There is, however, one striking
exception: the radial velocity map of the [O {\sc iii}] $\lambda$4363
line does not give a systemic velocity in agreement with the one in
the nuclear region for other broad lines.  The broad [O {\sc iii}]
$\lambda$4363 line emission is concentrated in the central 2\arcsec,
and it is clearly blueshifted by $\Delta V \sim 60$ km s$^{-1}$ with
respect to the systemic velocity.  This is seen not only in the
velocity map, but also in the integrated spectrum, as discussed in
Sect.~\ref{sec:sys}. We will discuss  the kinematics of [O {\sc
    iii}] $\lambda$4363 in greater detail below.

\subsection{Velocity dispersion maps}\label{dispmaps}

\begin{figure*}[ht]
\centering
\includegraphics[width=6cm]{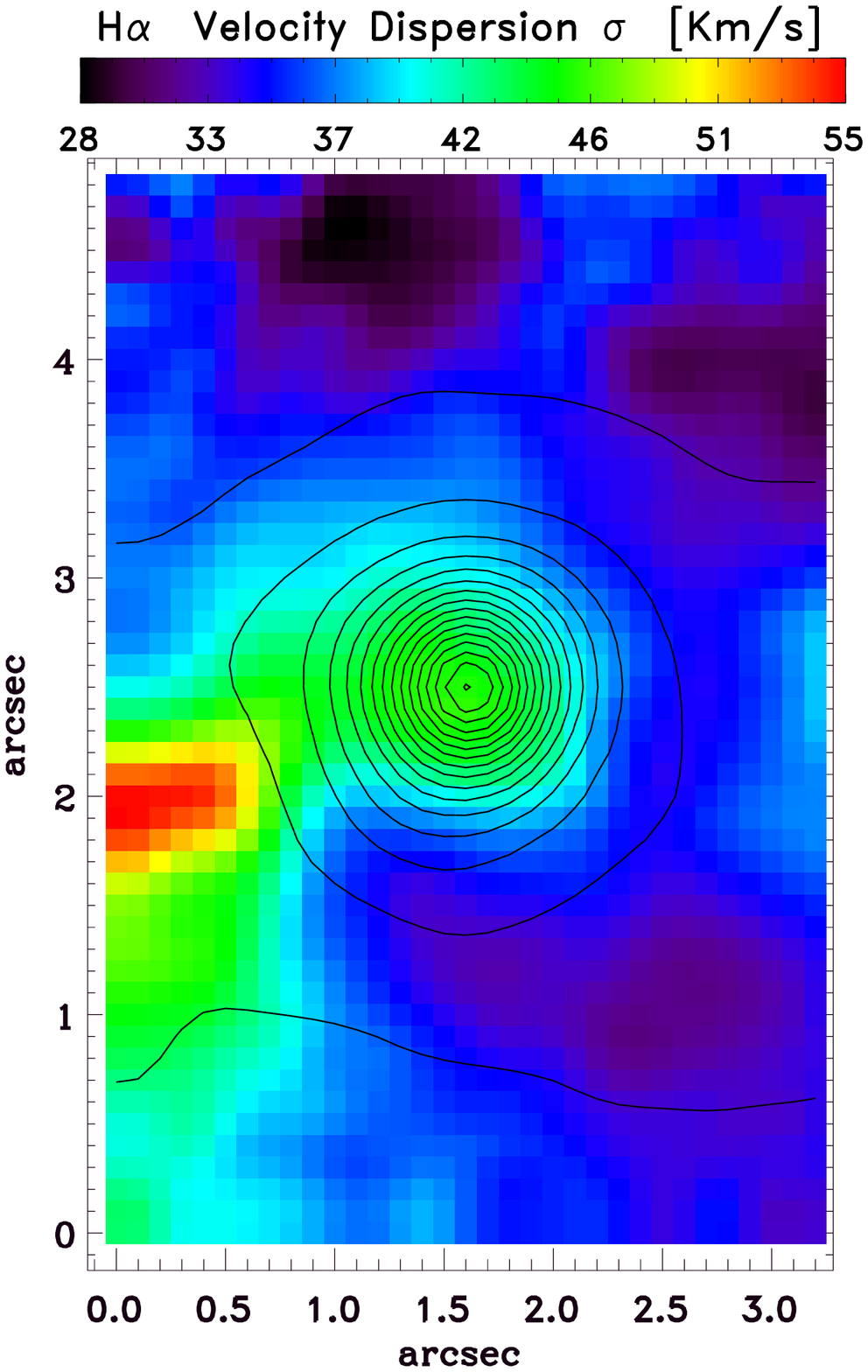}
\includegraphics[width=6cm]{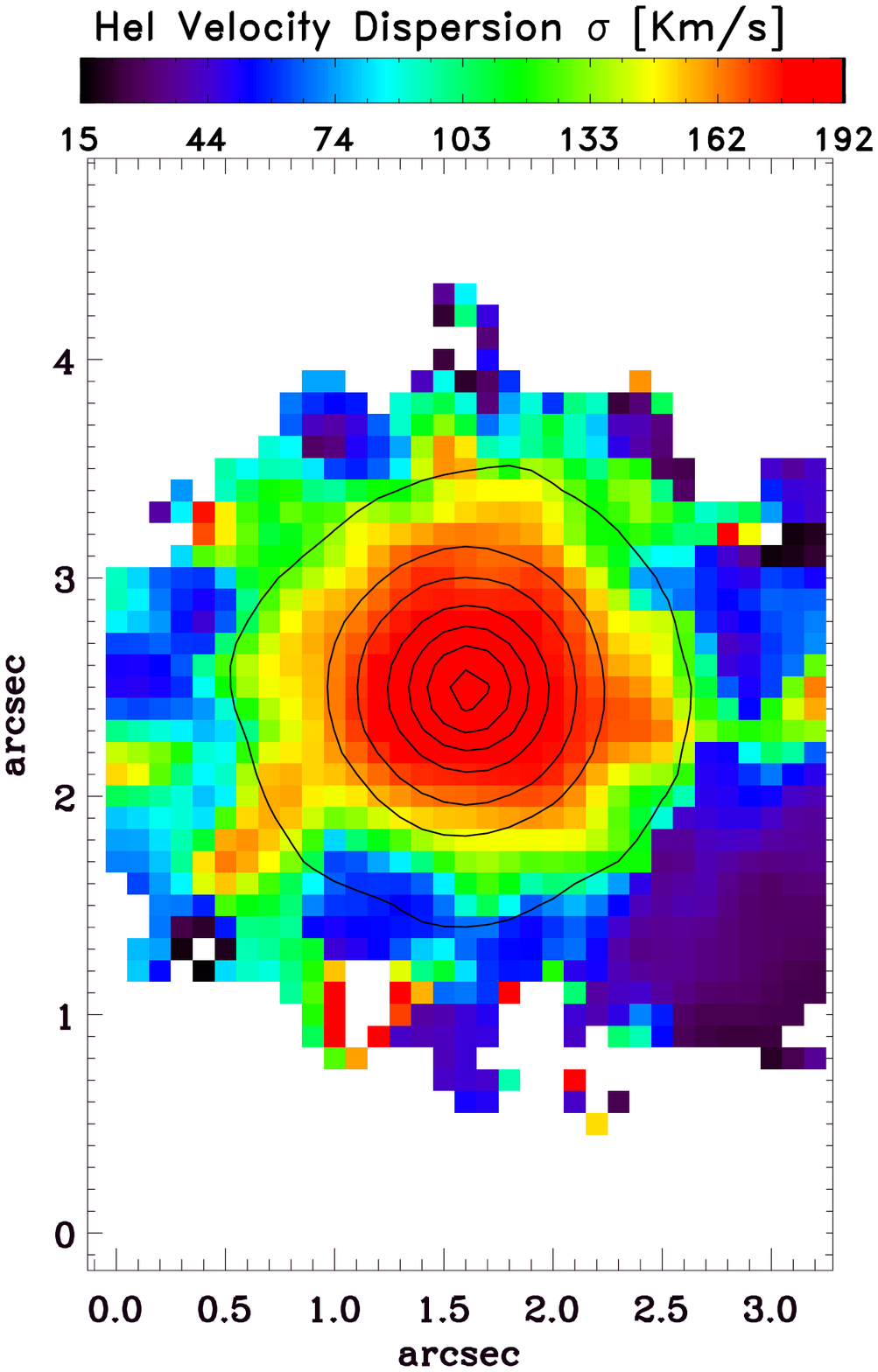}
\caption{Velocity dispersion maps. {\bf (Left)} Map of the narrow
  component of the H$\alpha$ line, representative of all other
  narrow-line maps (narrow [O {\sc iii}] $\lambda$5007, [O {\sc i}]
  $\lambda$6300, [S {\sc ii}] $\lambda$6717).  {\bf (Right)} Map of
  the He {\sc i} $\lambda$7065 line, representative of all other
  broad-line maps ([O {\sc iii}] $\lambda$4363, broad H$\alpha$).  The
  contours and orientation are the same as in Fig.~\ref{fig:nar_vel}.
\label{fig:nar_disp}}
\end{figure*}

Figure~\ref{fig:nar_disp} (left panel) shows the velocity dispersion
maps, the left panel for the narrow component of H$\alpha$, and the
right panel for the broad He {\sc i} $\lambda$7065. Again, the narrow
H$\alpha$ map is representative of all dispersion maps for the narrow
lines, such as narrow [O {\sc iii}] $\lambda,\lambda$4959, 5007, [O
  {\sc i}] $\lambda$6300 and [S {\sc ii}] $\lambda$6717.  These maps
show a high central value of $\sigma$ ($FWHM$/2.355) of $\sim$ 45 km
s$^{-1}$, in agreement with the value given by the integrated spectrum
(Table \ref{tabint}). The velocity dispersion then decreases outwards
with radius.  In the left panel, there is a clear increase in the velocity
dispersion towards the outflow blob seen in the radial velocity map,
in the SE direction. There are also two low-dispersion regions to the
NE and NW, which appear to be related to the high-extinction region
seen in the {\sl HST} color map of \cite{T96}.  An additional 
low-dispersion region is seen in the SW direction.  The NE and SW 
low-dispersion regions are aligned with the overall rotation pattern axis
seen in Figure~\ref{fig:nar_vel}, while the outflow blob shows
kinematic features about an axis perpendicular to that axis.   A
  patchy velocity dispersion map may indicate the presence of regions
  with different densities due to the presence of bubbles or shells as
  seen, for instance, in the study of the internal kinematics of the
  prototypical HII galaxy II Zw 40 \citep{bor09}.

Figure~\ref{fig:nar_disp} (right panel) shows the velocity dispersion
map for the He {\sc i} $\lambda$7065 line.  As before, this map is
similar to those of other broad lines originating from the dense
nuclear region.  The He {\sc i} velocity dispersion peaks at the
center, with $\sigma \sim 190$ km s$^{-1}$ (see also Table
\ref{tabint} for the integrated spectrum), and decreases outwards to
values typical of the narrow-line region ($\sigma$ of 50-100 km
s$^{-1}$).  The width of the He {\sc i} line is similar to that of the
broad component of the H$\alpha$ and H$\beta$ lines, and of the [O
  {\sc iii}] $\lambda$4363 and [N {\sc ii}] $\lambda$5755 auroral
lines which we discuss next.

\subsection{The peculiar kinematics of the [O {\sc iii}] $\lambda$4363
  and [N {\sc ii}] $\lambda$5755 lines}

\begin{figure*}[ht]
\centering
\includegraphics[width=6cm]{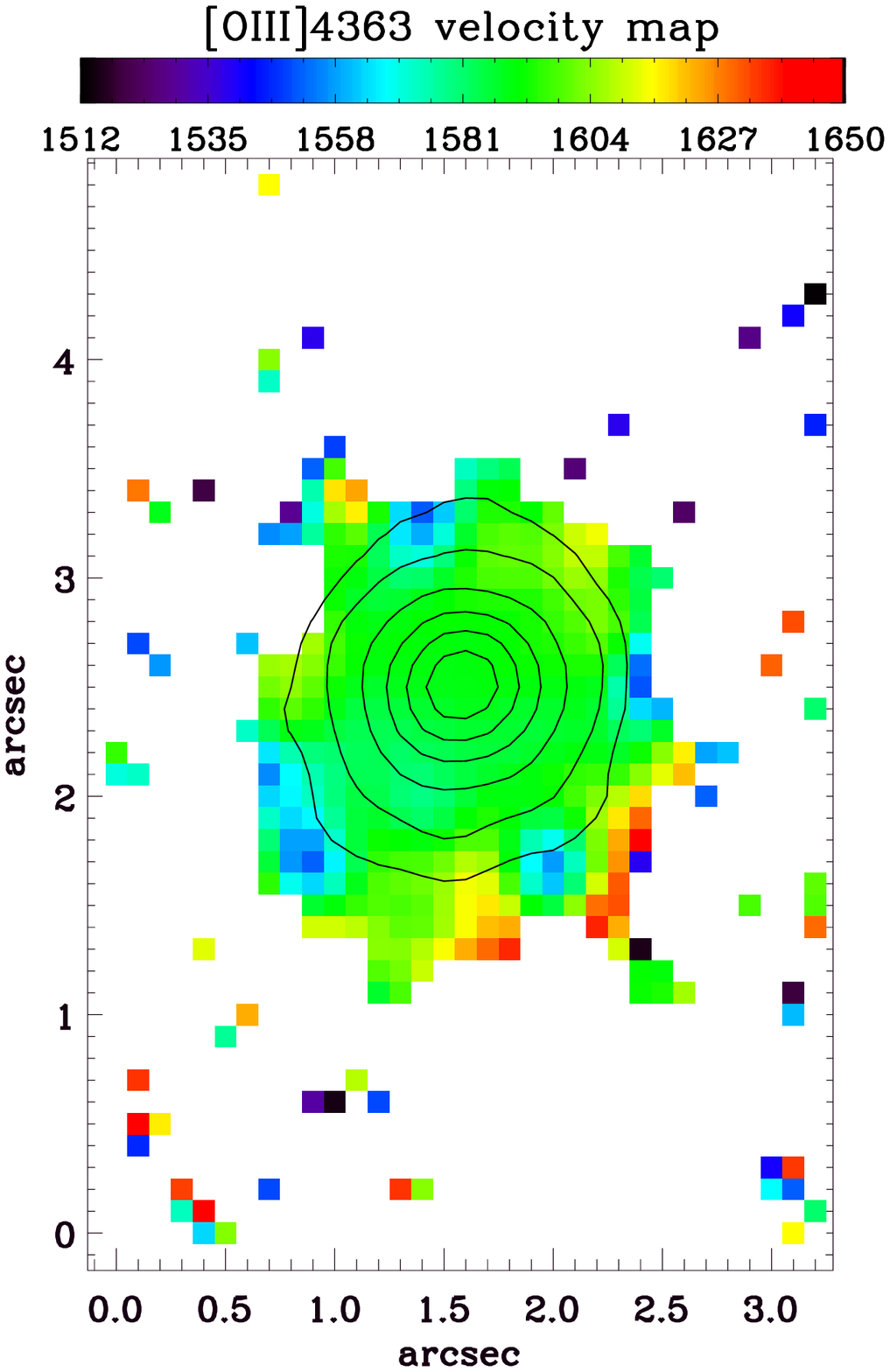}
\includegraphics[width=6cm]{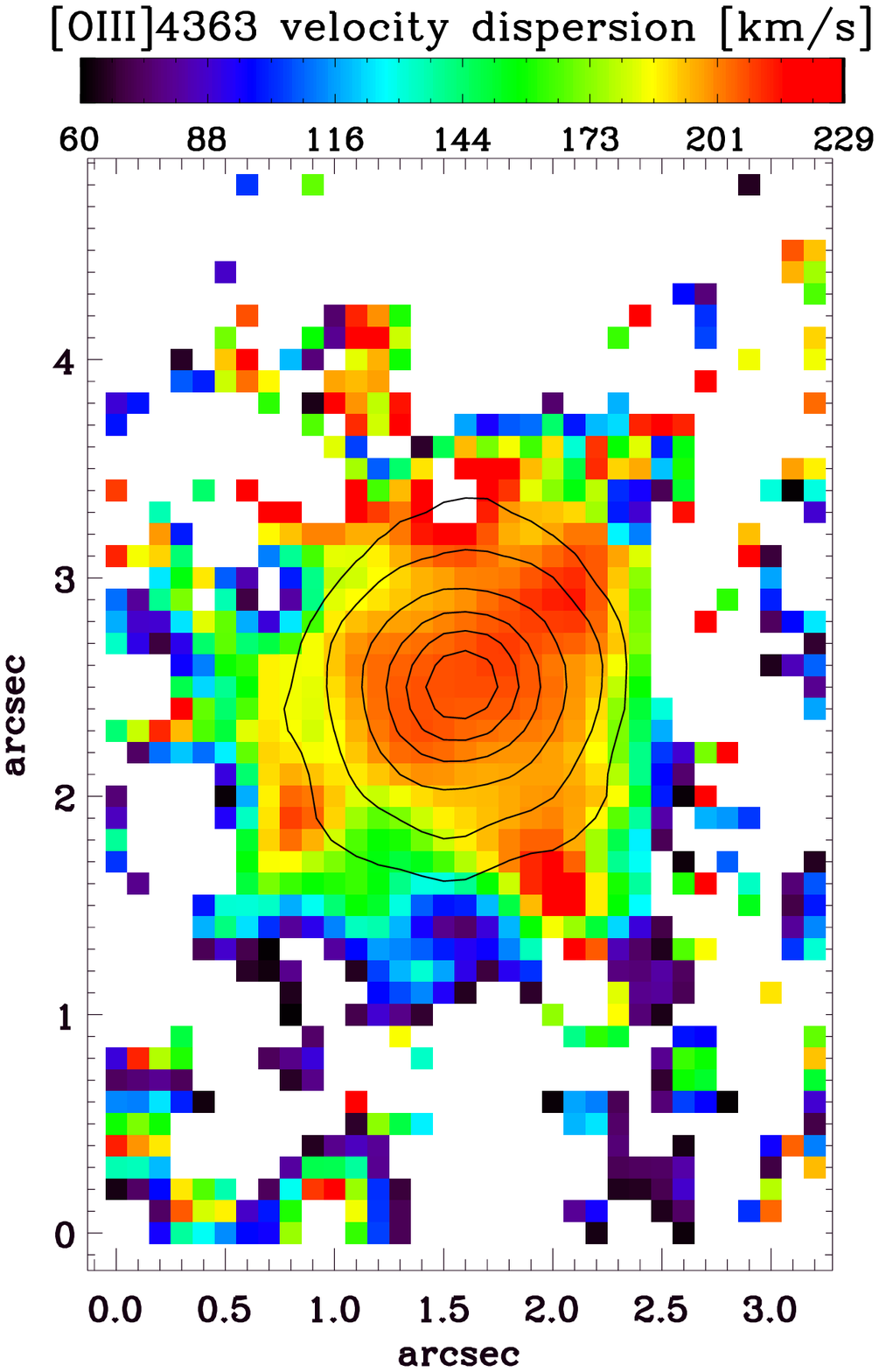}

\includegraphics[width=6cm]{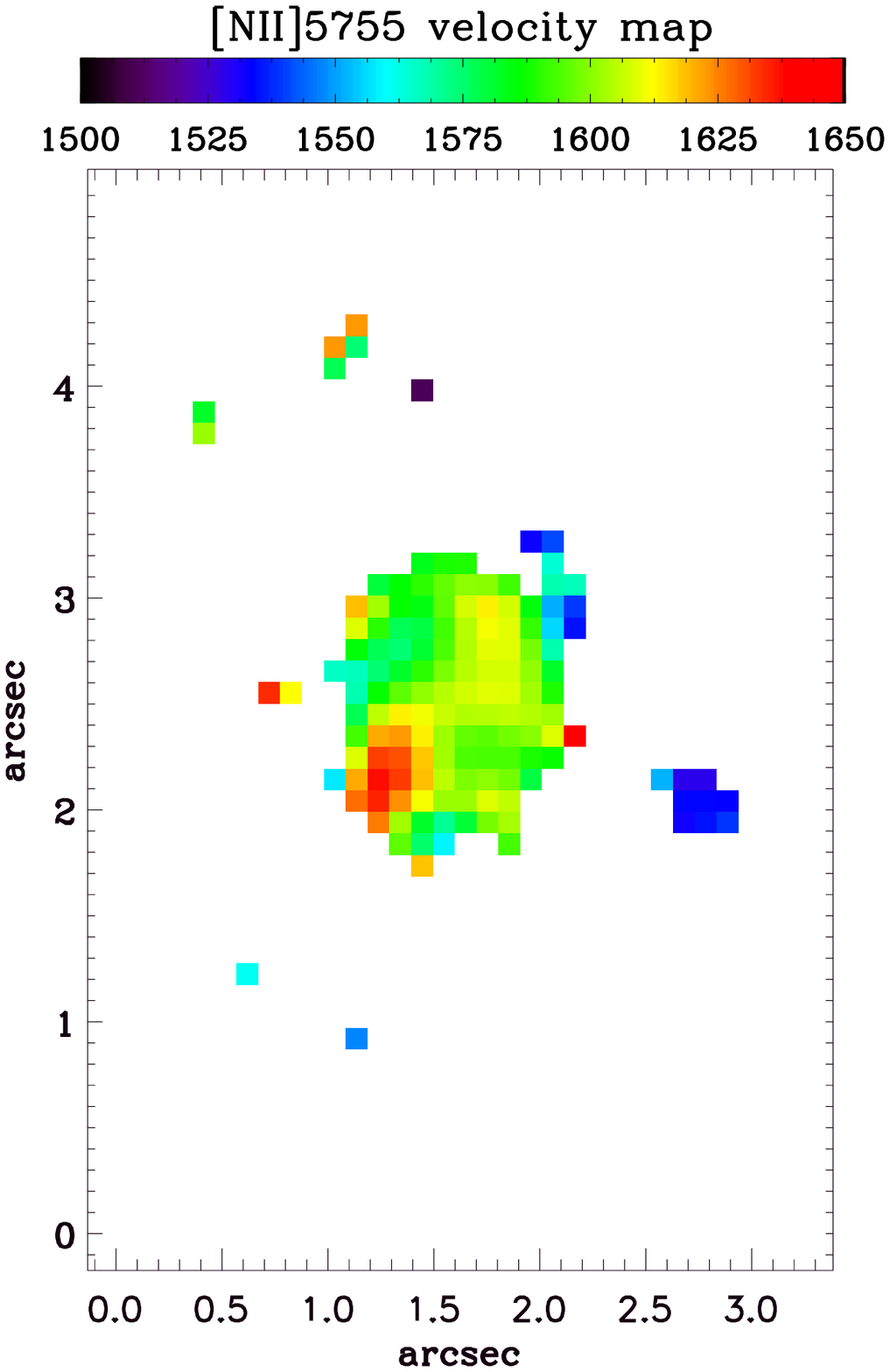}
\includegraphics[width=6cm]{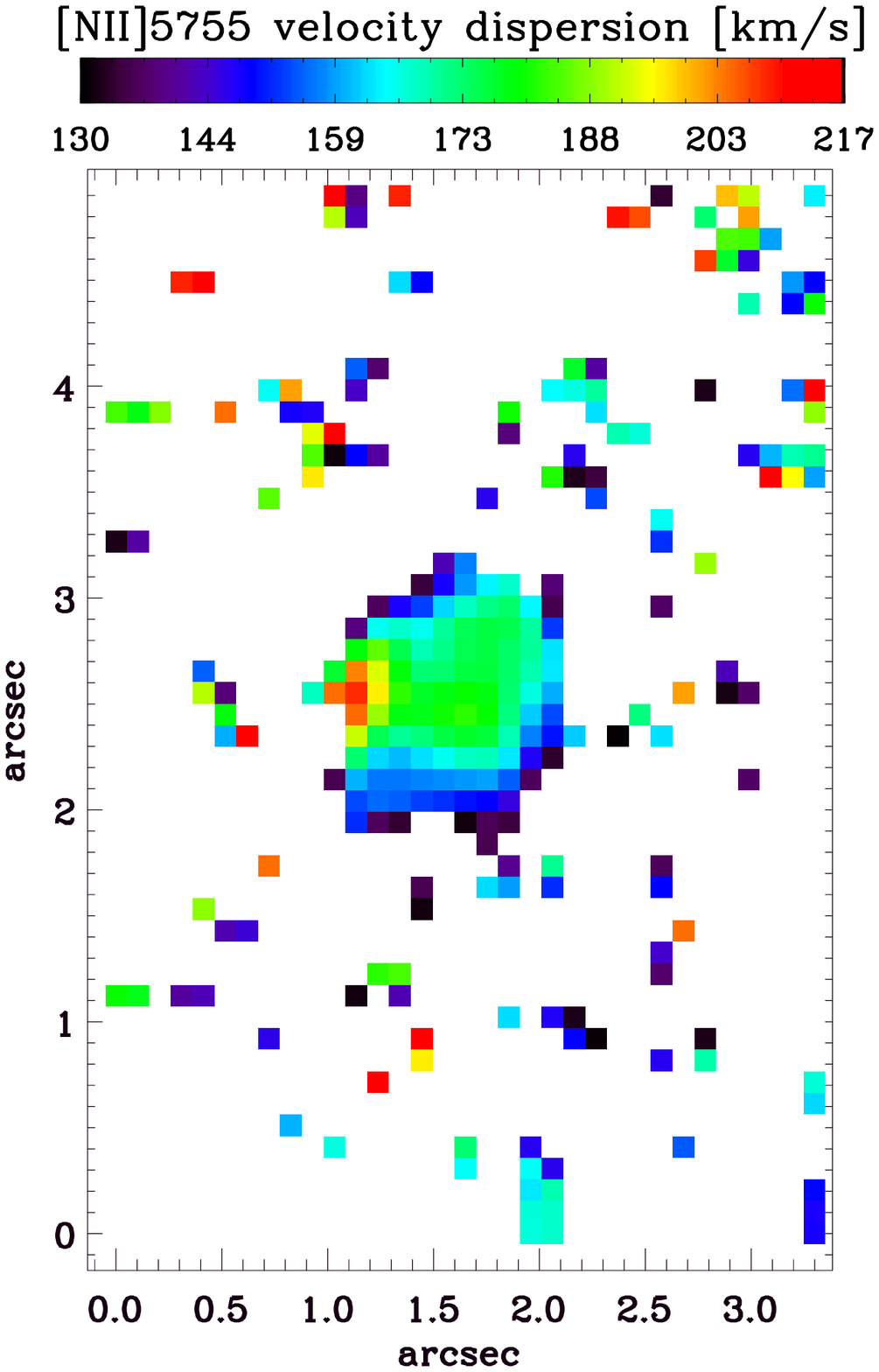}
\caption{ The peculiar kinematics of the [O {\sc iii}] $\lambda$4363
  line (upper panels) and of [N {\sc ii}]$\lambda$5755 (lower
  panels). Contours are the monochromatic intensity of the
  corresponding emission-line from decreasing intensity intervals of
  $10^{-18}$ \ergsqcmsec in peak intensity. Pixels within the contours
  are  3 $\sigma$ above noise.  North is up and east is
  left.}. \label{fig:oiiinii}
\end{figure*}

Figure~\ref{fig:oiiinii} shows the velocity fields of [O {\sc iii}]
$\lambda$4363 and [N {\sc ii}] $\lambda$5755.  Both lines are totally
dominated by the broad emission: the narrow lines, if present, are
undetected in the spatially resolved maps.  The velocity dispersion of
[O {\sc iii}] $\lambda$4363 is very similar to that of He {\sc i}
$\lambda$7065 (Figure~\ref{fig:nar_disp}), with some structures in the
SE-NW direction and an integrated $\sigma \sim 200$ km s$^{-1}$.  A
peak is also seen in the SW direction, but in a region of lower S/N
per pixel, so we will not consider it real.  The [N {\sc ii}]
$\lambda$5755 line is also broad, though with a somewhat smaller integrated
value.  The radial velocity maps of these lines have one
intriguing peculiarity: both lines are blueshifted with respect to the
systemic velocity of the galaxy, by some 60 km s$^{-1}$ in the case of
[O {\sc iii}] $\lambda$4363, and by $\sim$20 km s$^{-1}$ in the case
of [N {\sc ii}] $\lambda$5755.  In addition, [N {\sc ii}]
$\lambda$5755 also shows some structures in the SE-NW in the radial
velocity map, suggestive of bipolar outflow motions from the nucleus.
The kinematics of the narrow-line region outside the nucleus are,
however, not correlated with those of the inner regions, suggesting
that the strong motions associated with the broad lines observed are
decoupled from motions in the narrow-line region.  This interpretation
is consistent with the assumption that the outflow motions originate
from WR stars in the nuclear region, as is also implied from the flux and
velocity maps of the WR blue and red bumps observed in the original
data cube.  This is also in agreement with the interpretation drawn in
Sect.~\ref{sec:sys} for the different systems of emission lines seen in
the integrated spectrum.

\section{Matching the spatially resolved kinematics with the
  two-density model}\label{sec:den}

\begin{figure*}[ht]
\centering
\includegraphics[width=6cm]{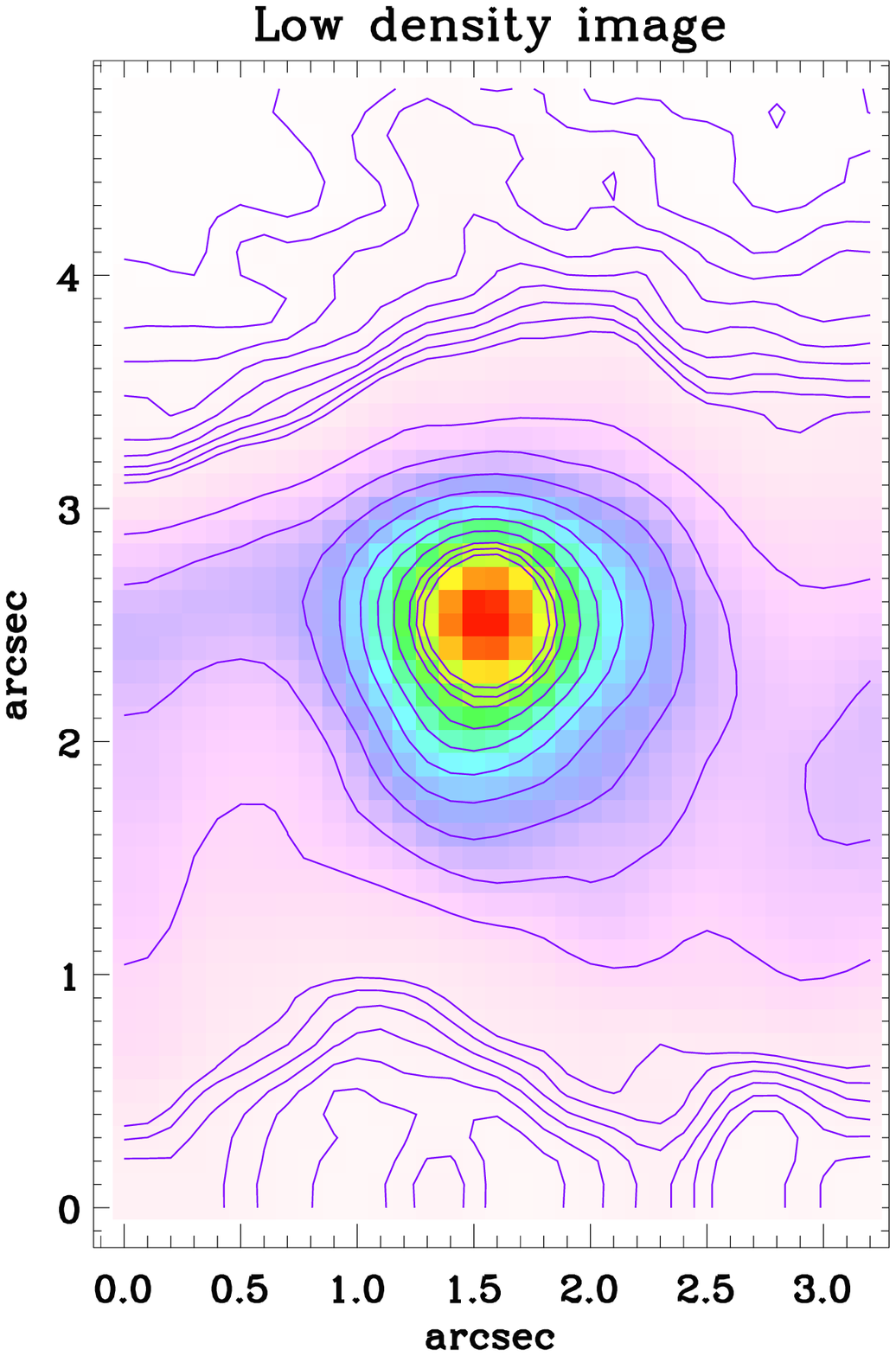}
\includegraphics[width=6cm]{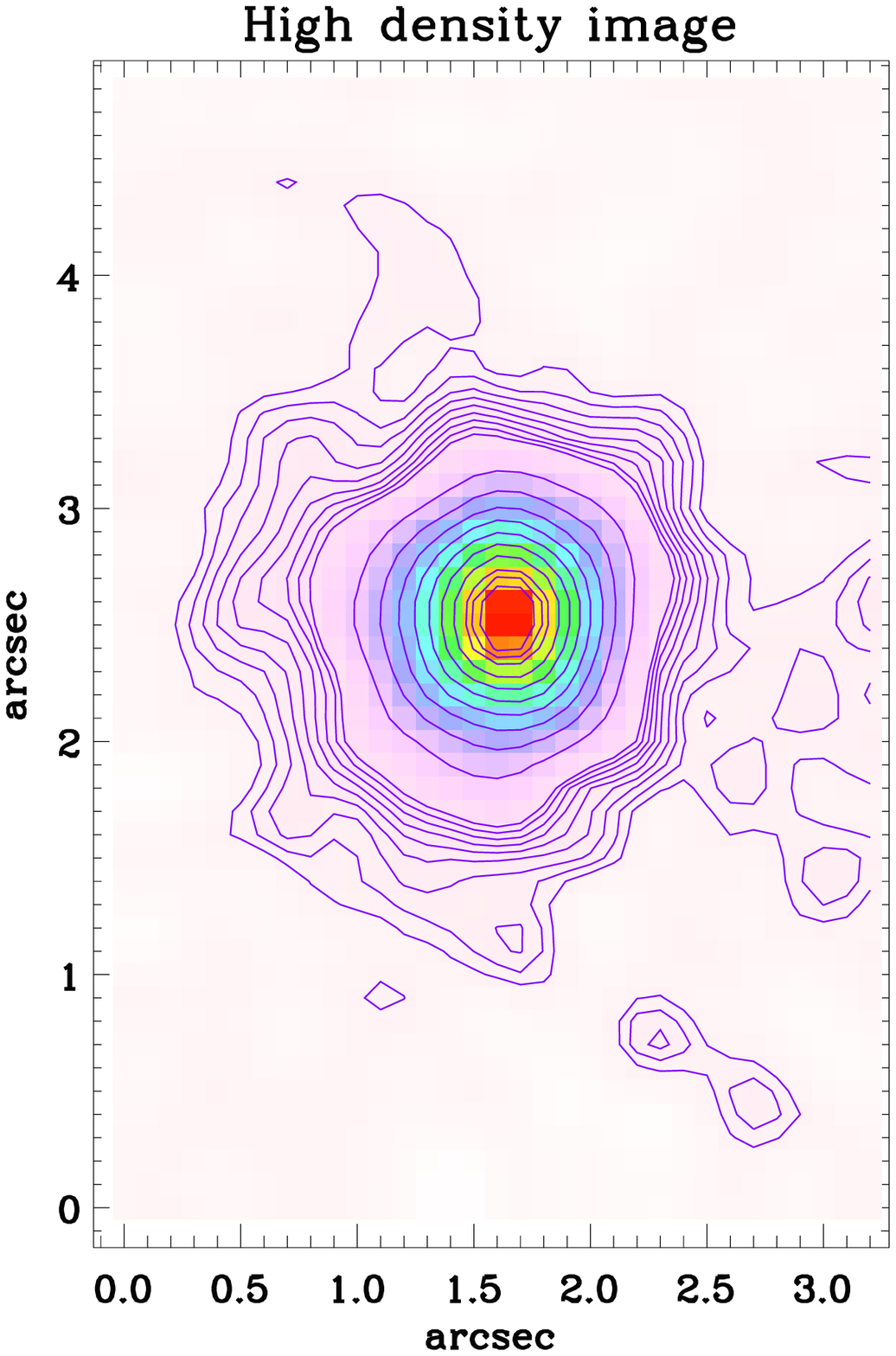}

\caption{Maps of low {\bf (left)} and high {\bf (right)} electron density emission.
 The orientation is the same as in Fig. \ref{fig:nar_vel}.
\label{fig:steiner}}
\end{figure*}

The [S {\sc ii}] $\lambda$6717/$\lambda$6731 ratio has been widely
used in aperture spectroscopy to derive the electron density in H~{\sc
  ii} regions.  The disadvantage is that it only gives an average
value of the physical conditions in the region, masking any electron
density variation or gradient or any nonuniformity and inhomogeneity
in the ionization structure.
That the density structure in Mrk 996 is not uniform has been
discussed by \cite{T96}, \cite{T08}, and \cite{J09}. \citet{T96,T08} had
to invoke a CLOUDY model with two zones of different electron
densities to account for the integrated optical, near-, and
mid-infrared spectra of Mrk 996.

Here, we use a method for mapping low- and high-density clouds in
astrophysical nebulae recently devised by \cite{smro09}.  This method
aims to distinguish regions of low electron densities from those of
high electron densities by using individual forbidden line emission
images, as opposed to their ratio which may often have low S/N in the
outer regions and give untrustworthy results. We will apply this
method to the [S {\sc ii}] $\lambda$6717 and [S {\sc ii}]
$\lambda$6731 emission line images of Mrk 996 to test the hypothesis
of a two-density model for the galaxy.  These two images are
transformed into new images of low- ($I_{ij}(ld)$) and high-
($I_{ij}(hd)$) density emission by applying the \cite{smro09} formula

\[\left |\begin{array}{c} 
          I_{ij}(ld)\\
	I_{ij}(hd)
         \end{array}\right | = 
\begin{array}{c}
 \underline{~~~~~1~~~~~} \\
R_{ld} - R_{hd} 
\end{array} \times
\left | \begin{array}{cc}
1 & -R_{hd} \\
-1 & R_{ld}
\end{array}\right | \times
 \left |\begin{array}{c} 
          I_{ij}(\lambda_{ld})\\
	I_{ij}(\lambda_{hd})
         \end{array}\right | ,\]

\noindent 
where  $I_{ij}(\lambda_{ld})$ is the [S {\sc ii}]$\lambda$6717 image,
$I_{ij}(\lambda_{hd})$ is the [S {\sc ii}]$\lambda$6731 image,
$R_{ld}$ is the low-density limit ratio of 
[S {\sc ii}] $\lambda$6717/$\lambda$6731, and
$R_{hd}$ is its high-density limit ratio. From Table 1 in \cite{smro09}, 
we find $R_{ld} = 1.44$ and $R_{hd} = 0.44$,
 corresponding to number densities of 81 cm $^{-3}$ and 5900 cm $^{-3}$,
respectively.  The latter value should be considered  an upper limit
to the number density of the gas to which the [S {\sc ii}] diagnostics
can be applied  
since the critical densities for collisional deexcitation for [S {\sc ii}]
$\lambda$6717 and $\lambda$6731 are 1400 and 3600 cm$^{-3}$, respectively 
(Table \ref{tabint}).

We can apply this method to our integral field observations to assess
density variations along the line of sight to the central region of
Mrk 996. Figure~\ref{fig:steiner} (left panel) shows the low-density
image ($I_{ij}(ld)$) and the right panel shows the high-density image
$I_{ij}(hd)$. If the [S {\sc ii}] emission came only from low-density
clouds, all emission would be seen in the left panel only, and none
would be seen in the right one. We see clearly that, in Mrk 996, we do
have emission along the line of sight from a low-density cloud (left
panel) which covers the whole field and emits more in the EW
direction.  The high-density image 
 (right panel) shows emission concentrated only in the nuclear region.
This technique shows clearly that
a single low-density regime is ruled out and that  an additional 
regime of high density must be present.
These clouds in the nuclear region are probably associated with the
broad-line emission shown by some ionic species
(e.g., Fig.~\ref{fig:nar_vel}b) and with the WR stars discussed below
(see Fig.~\ref{fig:wr}).
By performing surface photometry on the high-density image
$I_{ij}(hd)$, we derive the diameter of the high-density region to be
$\sim$ 1\farcs6 or $\sim$ 160 pc, about the size of the inner spiral
structure  discussed by \cite{T96} and \cite{J09}. This size coincides
  with our chosen aperture for the integrated nuclear spectrum shown in
  Fig.~\ref{specfull}.

\section{PCA tomography}\label{PCA results}

\begin{figure*}[ht]  
\centering           
\includegraphics[width=6cm]{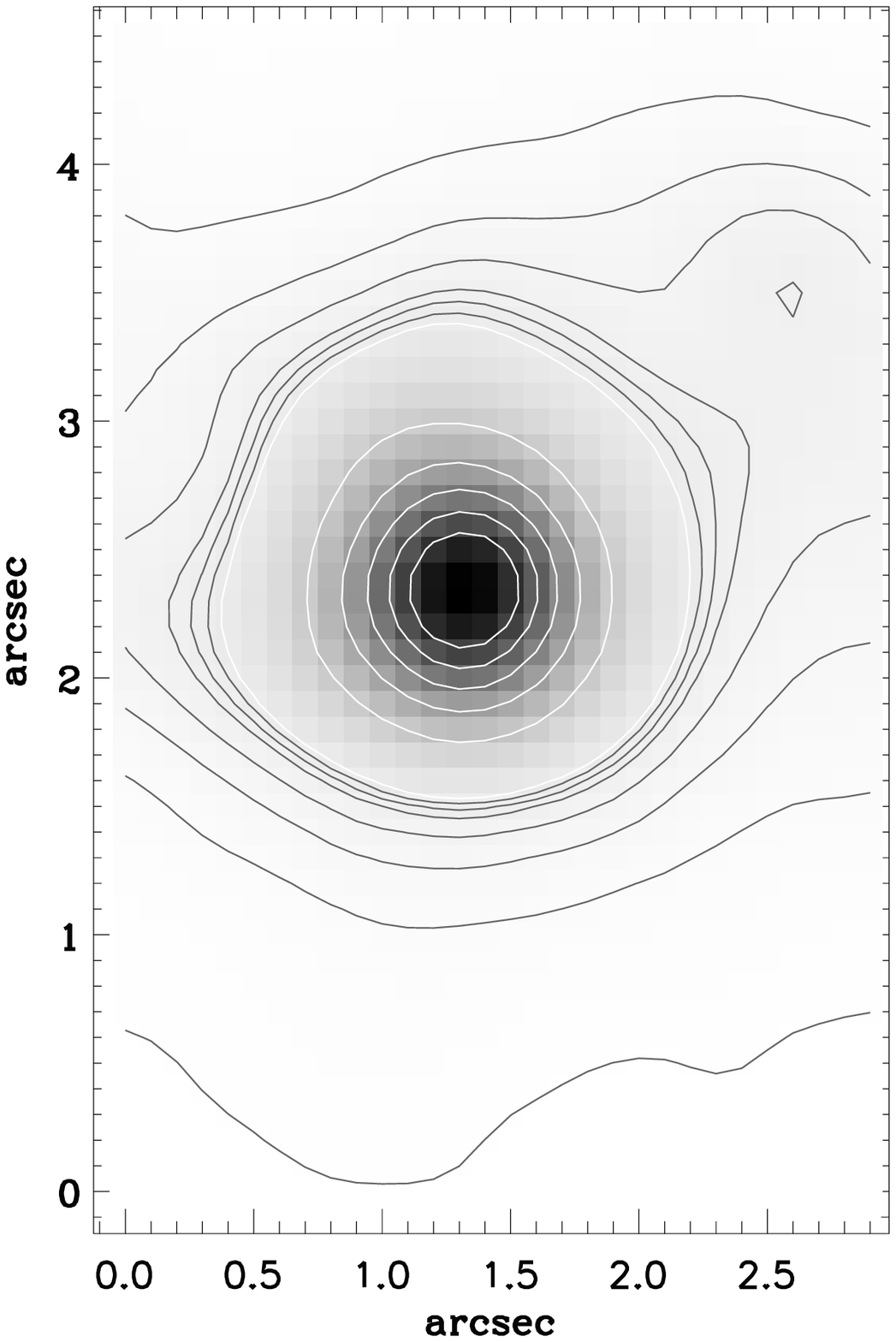}
\includegraphics[width=10cm]{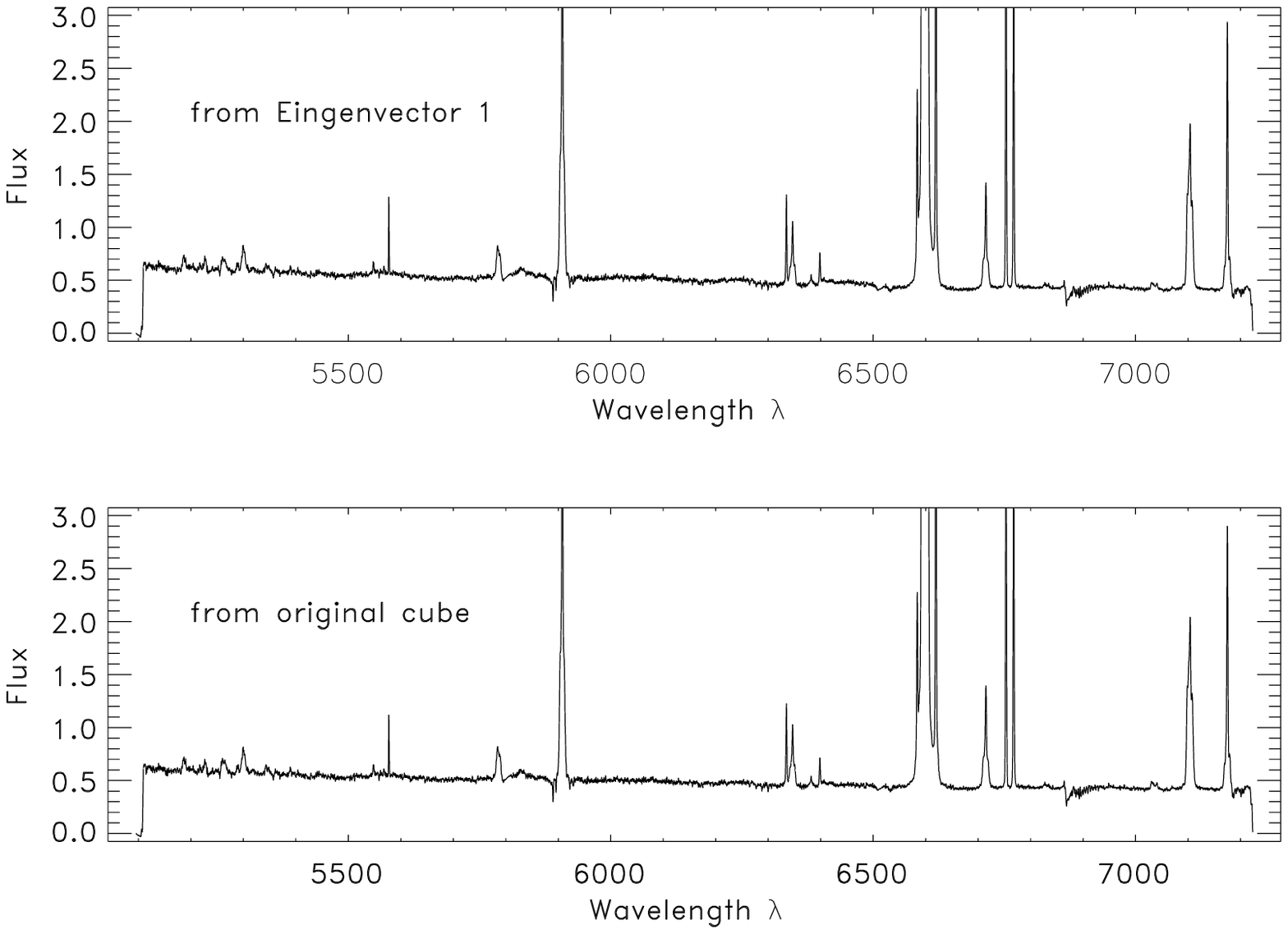}
\caption{Tomogram (left) and eingeinvector (E$_1$) in red cube
  (5700-7200\AA).\label{eig1} {\bf (bottom right)} the spectrum from
  the original data cube; {\bf (upper right)} the extracted spectrum
  from the reconstructed data cube, using the PCA tomography results
  (eigenvector 1 and tomogram 1) \citep[see][]{smro09b}.}
\end{figure*}

\begin{figure*}[ht]  
\centering           
\includegraphics[width=6cm]{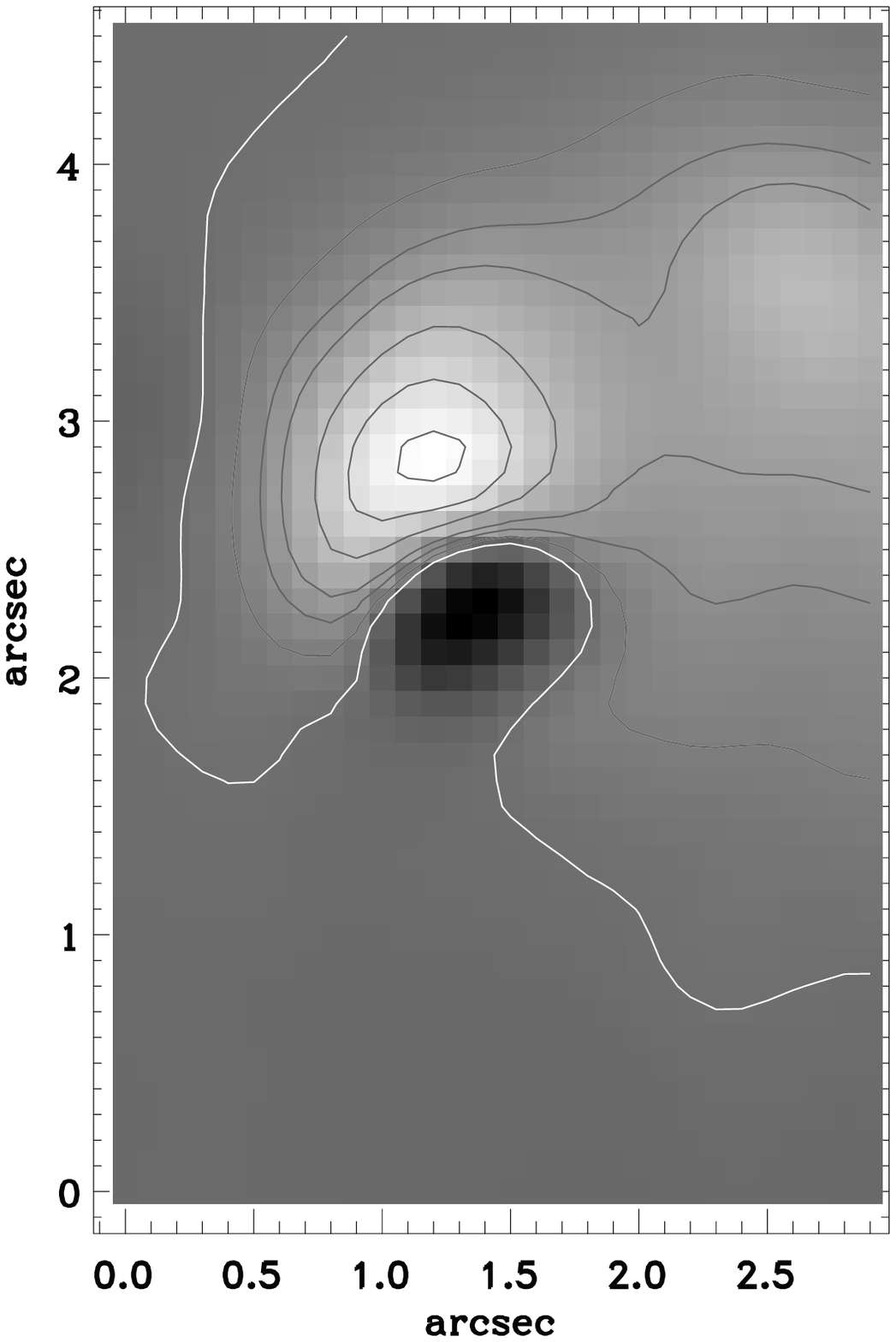}
\includegraphics[width=8cm]{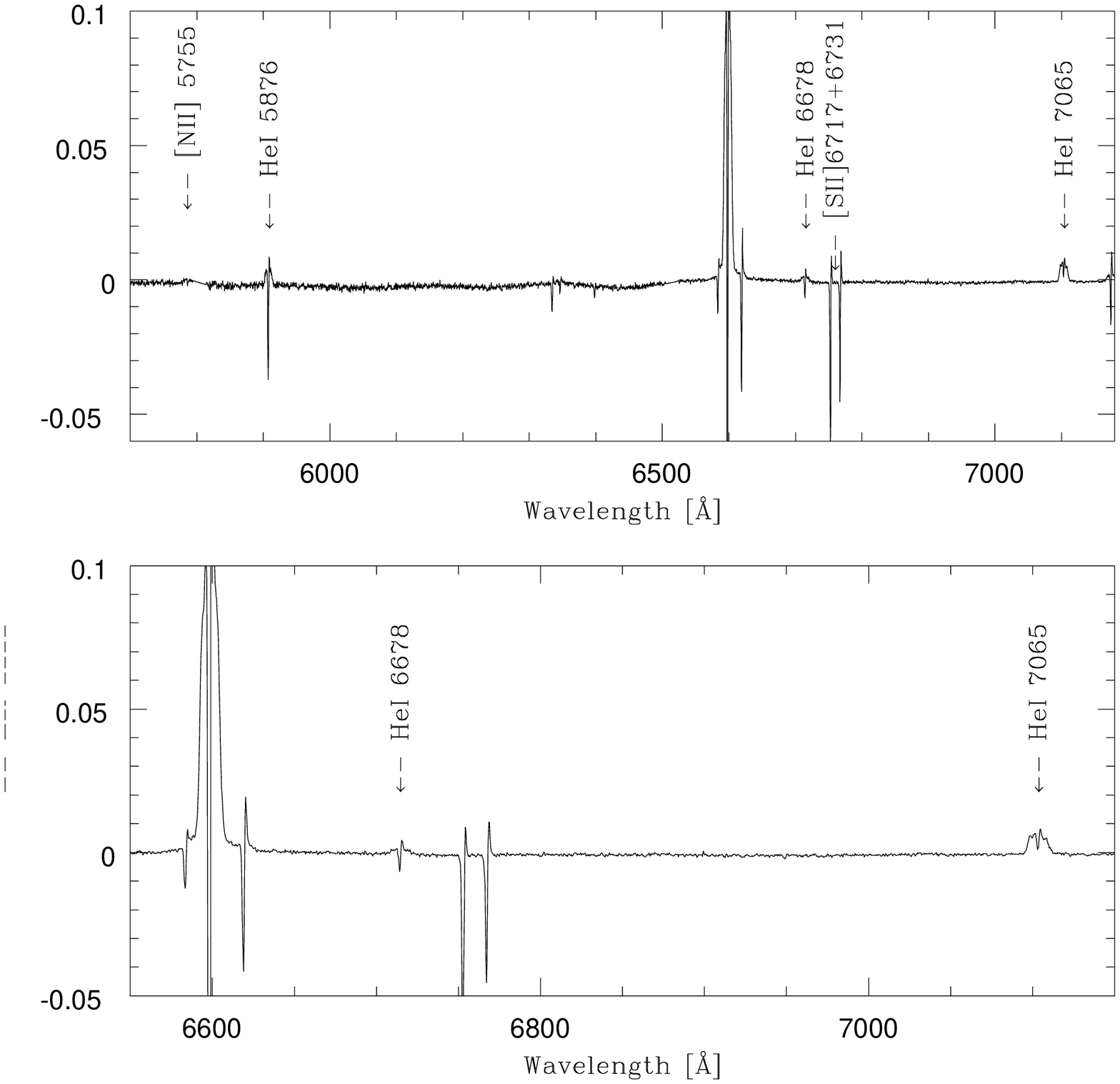}
\caption{Tomogram (left) and eingeinvector (E$_2$) in red cube
  5700-7200\AA~ (right). The eigenvector shown in the lower-right panel
  is  a zoom in a shorter wavelength range for a better
  visualization of the features of interest.\label{eig2}}
\end{figure*}

\begin{figure*}[ht]
\centering
\includegraphics[width=6cm]{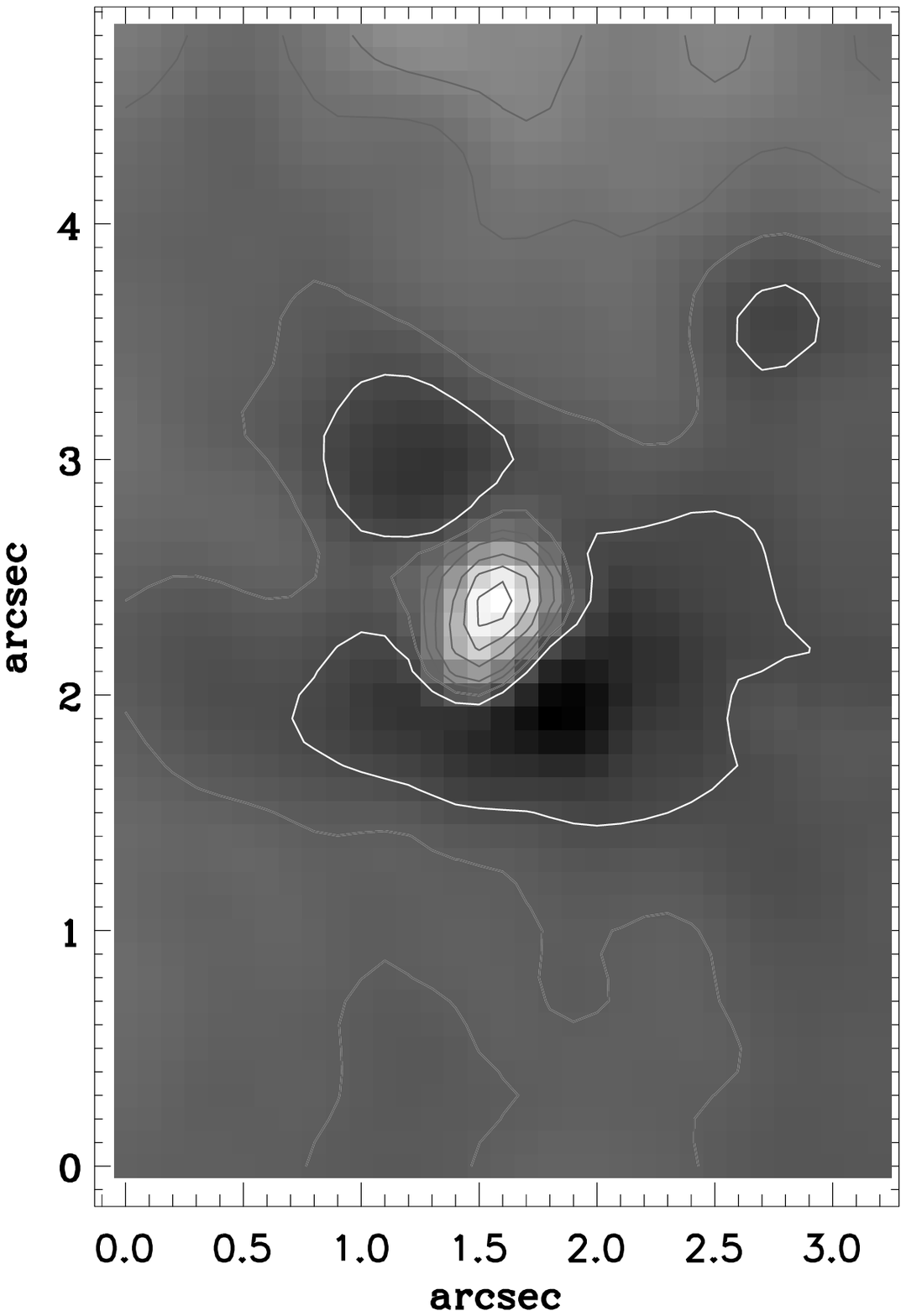}
\includegraphics[width=8cm]{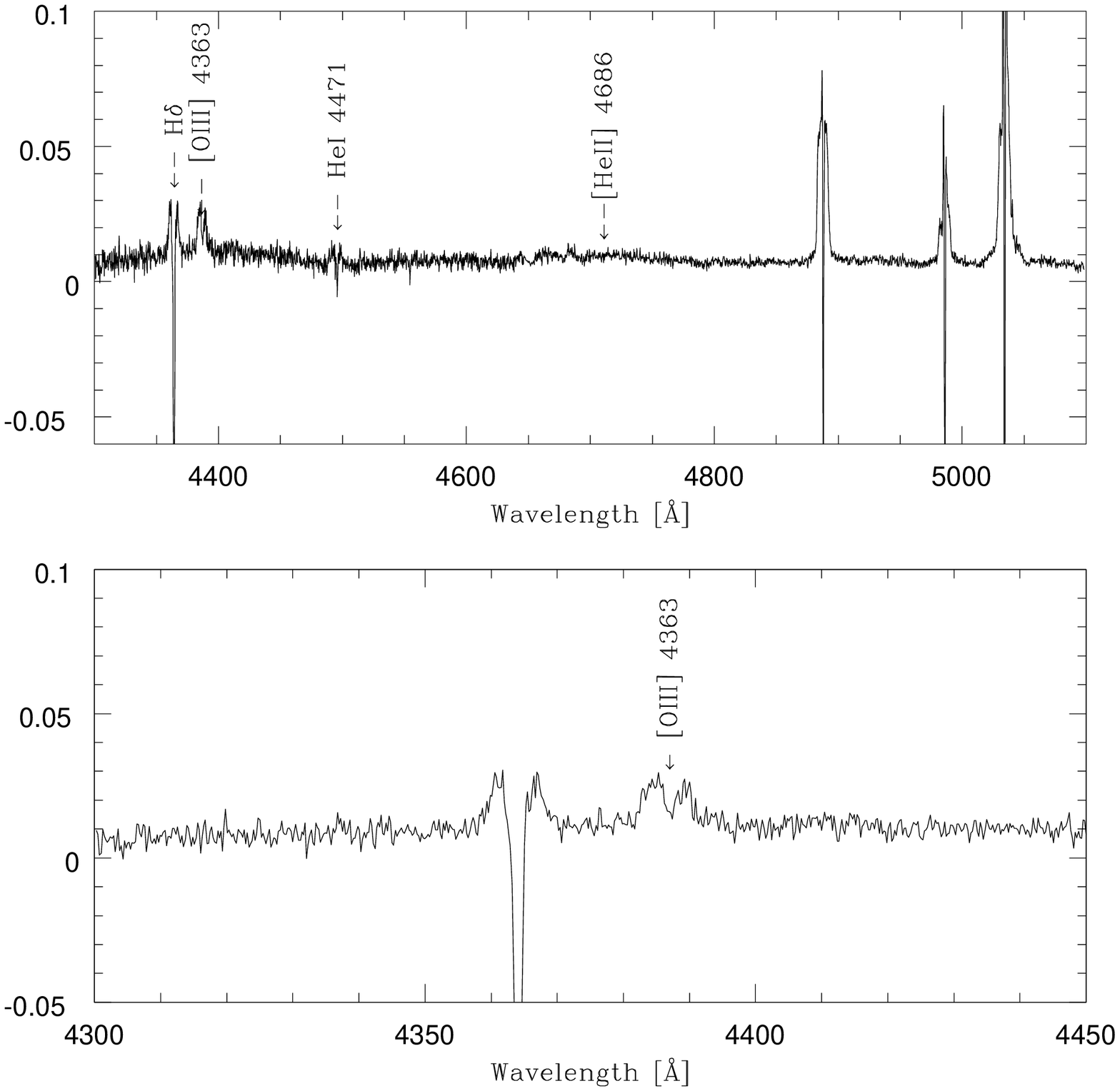}
\caption{Tomogram (left) and eingeinvector (E$_3$) in blue cube
4300-5100\AA~ (right). The eigenvector shown in the lower-right panel
  is  a zoom in a shorter wavelength range for a better
  visualization of the features of interest.\label{eig3}}
\end{figure*}

To fully exploit the wealth and complexity of the information that
integral field spectroscopic (IFS) data provide, we need  analysis techniques
that are more
sophisticated than those commonly used in
one-dimensional long-slit spectroscopy.  However, this new methodology
is still scarce, and most IFS studies simply reduce the data cube to
one-dimensional spectra and two-dimensional maps, so that the usual
well-known spectroscopy and imaging techniques can be applied to
analyze the data.  Here, we use a technique that has been introduced
recently to extract spatial and spectral information from data cubes
in a statistical manner, so that it can be used to derive physical
information.  This technique is called principal component analysis
 tomography.  It combines the statistical PCA analysis, widely
used in astronomy, with tomography which is also used in astronomy and
other sciences to represent certain types of information derived from
imaging techniques.   A short presentation of the technique
can be found in \cite{smro10}.

What PCA tomography basically does is to extract hidden information by
transforming a large set of correlated data, in our case the
wavelength pixels, into a new set of uncorrelated variables, ordered
by their eigenvalues.  Each new component, or eigenvector, carries the
combined information of the original data, ordered by their
significance as measured by their relative variance.  The new
coordinates can then be represented by their eigenvector and their
respective projection called a tomogram.  The combined analysis of the
eigenvectors and tomograms allows for interpretations of physical
phenomena that may not be directly seen in an usual spectrum or image.

\cite{hair98} devised a so-called scree test, which was used by
\cite{smro09b}, and it allows the assessment of the most interesting
eigenvectors and tomograms, those that contain most of the relevant
information from the data.  This test applied to our PCA results shows
that the first five eigenvectors and tomograms are sufficient to
reconstruct the data cube with the most significant information on
the uncorrelated physical properties of the original data.

\begin{table}
 \caption{PCA Eigenvalues (blue and red cubes)}\label{eigenvalues2}  
\begin{tabular}{lcc} \hline \hline
Eigenvector & Eigenvalues & Eigenvalues\\
E$_k$ & Variance (\%)& Variance (\%)\\
&5700-7200\AA & 4300-5100\AA       \\
\hline
E$_1$&        98.15 &      97.51\\
E$_2$&        1.027 &      1.552 \\
E$_3$&       0.5372 &     0.2891\\
E$_4$&      0.06388 &      0.172\\
E$_5$&      0.04478 &    0.06334\\
E$_6$&       0.0148 &    0.05601\\
E$_7$&      0.01219 &     0.0308\\
E$_8$&      0.01171 &    0.01628\\
E$_9$&     0.008297 &    0.01367\\
E$_{10}$&  0.005329 &    0.01098\\
\hline
   \end{tabular} 
\end{table}

Table~\ref{eigenvalues2} presents the PCA results from the analysis of the
GMOS/IFU data cubes for the red and blue gratings.
Column 1 shows the eigenvalue numbers and Cols. 2 and 3 
show the resulting variances
for the first 10 principal components in the cases of  
the red and blue gratings, respectively. From these, one can
see that the first four principal components carry 99.5\% of all uncorrelated
information contained in the data.

The first eigenvector in the red cube, which accounts for 98.15\% of
the data cube variance, is shown in Figure~\ref{eig1}. We note that
the reconstructed data cube, using only the first principal component
(Eigenvector 1 and Tomogram 1), is able to reproduce the original
integrated spectrum.   The upper-right panel in Fig.~\ref{eig1}
  shows the representative integrated spectrum obtained by using a
  reconstructed data cube with only the first eigenvector.  This spectrum is
  identical to the one extracted from the original data
  cube.   The added contribution of eigenvectors
2-5 accounts for only $\le$ 2\% of the variance in the data cube.
With the simultaneous analysis of tomogram 1 and the corresponding
eigenvector 1, we can reproduce most of the information that can be
obtained from a direct broadband image and from an integrated
spectrum of the corresponding field of view. This shows the great
redundancy of this type of data, allowing  for
discriminating non-redundant information. The power of this technique
lies in the fact that by removing the effects of the strongest
correlations, one can look for the less significant ones.

Eigenvector 2 still contributes significantly ($>$1\%) and its
tomogram reveals distinct bipolar motions originating from the
nucleus (Fig.~\ref{eig2}).  In this case, the y-axis
does not represent flux, and the eigenvector is not a spectrum. 
  The point to note is that the eigenvectors are not spectra but
  vectors of correlations.  Here, the anti-correlations (up and down
  spikes) are seen in the narrow lines only. This means that this
  particular phenomenon is affecting only the narrow lines.  In
  addition, the feature shows an anti-behavior of the
  blue side vs. the red side of the lines which leads us to interpret that
  we are seeing motions in the narrow lines only (e.g., [S {\sc
    ii}] 6717,6730).  The respective tomogram in Figure~\ref{eig2}
  (left) shows higher order kinematics of the narrow line.
  This feature cannot be seen in a classical way as if it were in a
  direct spectrum, but rather as a residual hidden phenomenon carrying
  only 1\% of the variance. 
One may interpret this anti-correlation as representing the 
second-order rotation of a low-density cloud system in the circumnuclear
region.   Therefore, the broad lines observed in Mrk 996 are
  probably not due to a turbulent mixing layer, as postulated by
  \cite{J09}. The present interpretation is more consistent with the
hypothesis that the broad lines originate from stellar wind outflows
from WR stars in the nuclear region, as implied from the flux and
velocity maps of the WR blue and red bumps observed in the original
data cube, convoluted with some rotation of the low-density gas within
this unresolved inner region. Our analysis is based on two
independent data sets, the blue and the red cubes.  The features
observed in the eigenvectors and tomograms of both data sets are all
very similar, which lends credibility to our interpretations.

\begin{figure*}[ht]
\centering
\includegraphics[width=6cm]{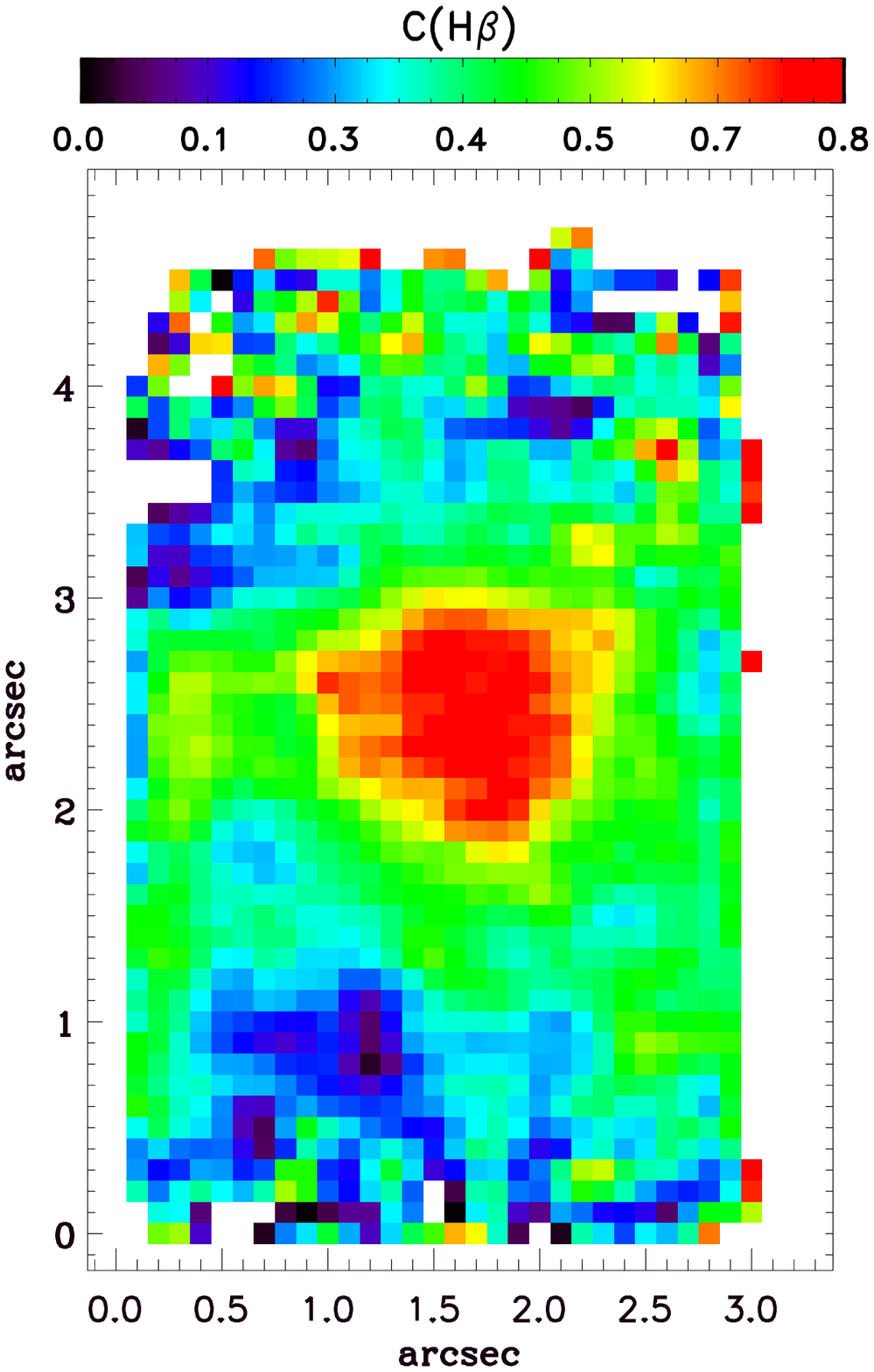}
\includegraphics[width=6cm]{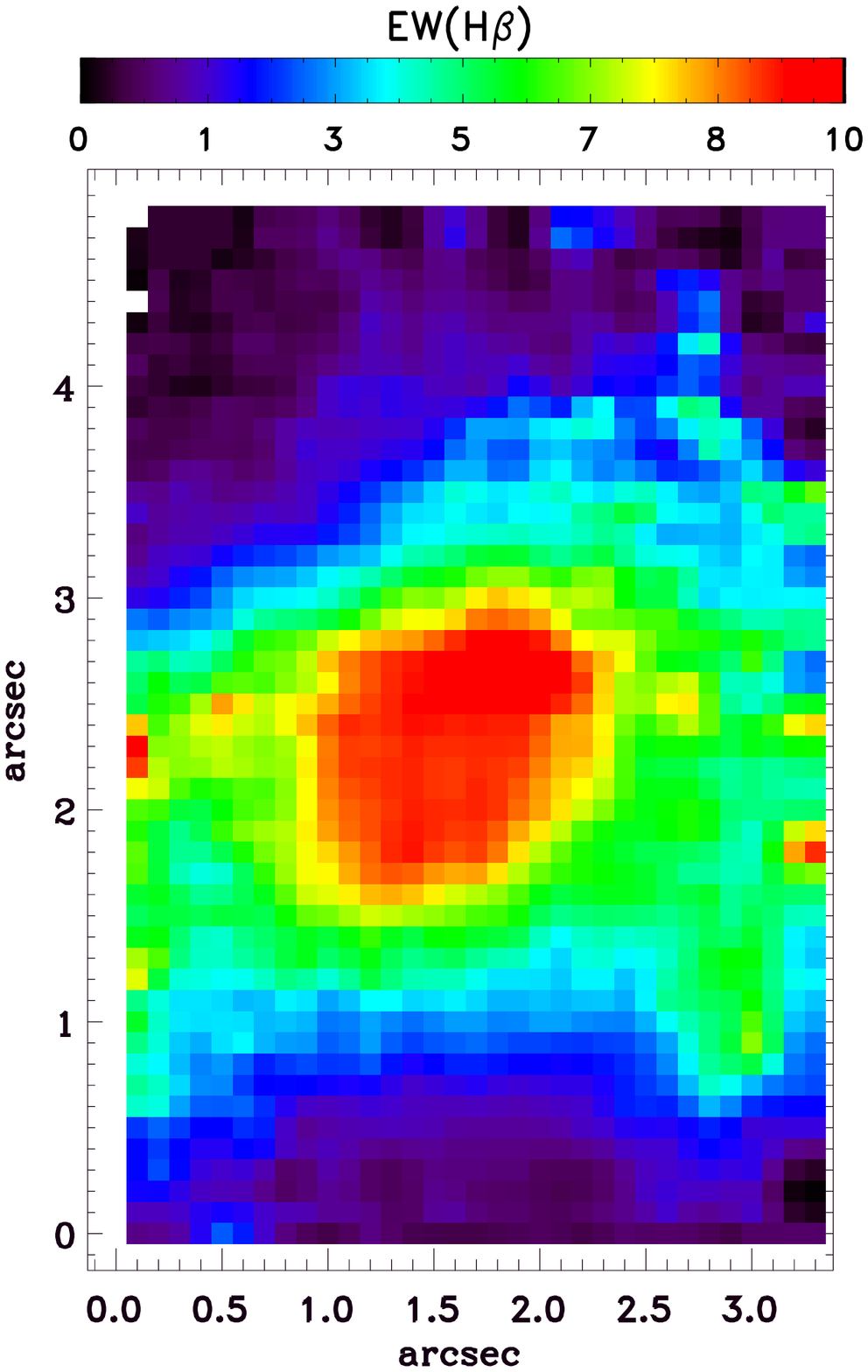}
\caption{{\bf (Left)} Logarithmic extinction coefficient map derived
  from the Balmer decrement using integrated line profiles (broad +
  narrow components). Pixels with high noise at the border were masked
  out.
  {\bf (Right)} H$\beta$ equivalent width map (per pixel in \AA).  The
  orientation is the same as in
  Fig. \ref{fig:nar_vel}.} \label{fig:chb}
\end{figure*}

Eigenvector 3 contributes about 0.5\% of the variance of the blue
cube.  It also reveals a strong feature from its tomogram
(Figure~\ref{eig3}, left) which seems to indicate a distinct intensity
contribution from the narrow lines originating in the region
surrounding the nucleus, which is different than that of the broad
lines.  It is noteworthy that the highest intensity features
correspond to the positive correlation shown in the broad lines.
These features are mapped as white and black pixels in the tomogram and
coincide with the broad-line and high-density emission region as
extensively discussed above.  The lower intensity negative correlation
corresponds to the narrow-line and low-density emission region as
mapped with gray pixels in the tomogram,  but the most important
feature can be seen in the zoomed eigenvector (Figure~\ref{eig3},
bottom right) where a {\it narrow} contribution to the [O {\sc
    iii}]$\lambda$ 4363 line is unequivocally detected.  This narrow
component is not directly seen in the integrated spectrum, but it
shows up in this third eigenvector as a dip feature.  It is a
fundamental finding for our purpose and validates our effort to
extract a spatially resolved integrated spectrum which excludes the
nuclear region.  This will allow for the direct and precise
determination of chemical abundances in Mrk 996 (see below) without
having to resort to modeling as  \cite{T96} did.  We note that this
direct and precise mapping of the narrow component of the [O {\sc
    iii}]$\lambda$ 4363 line could not be achieved with the data of
\cite{J09} because the narrow component was not detected in their
observations with a lower signal-to-noise ratio.

Eigenvector 4 (not shown) contributes only $\sim$0.1\% of the
variance, but it contains a visible feature in the NW direction.
This feature is not related to the emission lines but to the
continuum.  It may represent, in a statistical way, the effect of dust
obscuration.  Extinction in that direction has been noted above from
analysis of the original cube, through mapping of the Balmer
decrement.
It can also be seen from the {\sl HST} broadband images \citep{T96}.
Eigenvector 5 (not shown) contributes less than 0.1\% of the variance,
and it carries virtually no additional information on uncorrelated
physical properties.  The higher order eigenvectors and tomograms
become more difficult to interpret and/or reach noise features or
fingerprints that may not be real, but are instead associated with detector
defects and other artificial features.   We note that the PCA
  tomography technique can be used to eliminate higher order noise of
  the original data by the suppression of detector defects
  (fingerprints), noise, and by the reconstruction of the data cube
  accounting only for the first meaningful eigenvectors.  A lengthier
  discussion of this topic is beyond the scope of the present
  paper. Admittedly, the interpretation of the PCA tomography is not
  straightforward, but it becomes robust when combined with all other
  information available. The use of PCA tomography for 
  analyses similar to the one presented here can be found in more
  recent works, such as  \citet{stein13}, \citet{san13}, \citet{riffel11},
  \citet{ricci11a}, and \citet{schnorr11}.

\section{Mapping the physical conditions}\label{results}

\subsection{Extinction and H$\beta$ equivalent width maps} \label{extinction}

The extinction map in Figure~\ref{fig:chb} (left panel) was derived by
the ratio of the broad+narrow components of the H$\alpha$ and H$\beta$
emission lines.  It shows that the nuclear region of Mrk 996 has a
higher extinction ($C$(H$\beta$) $\sim$ 0.7). It is surrounded by a
region of lower extinction, with $C$(H$\beta$) decreasing to zero.
Some higher extinction extension towards the NW and E directions are
seen in this map. 
The central part of the high $C$(H$\beta$) region
coincides with the broad-line high-density region.  As emphasized in
Sect.~\ref{sec:col}, the H$\alpha$/H$\beta$ ratio in this region cannot be
used for extinction measurements, as
collisional excitation of hydrogen in the nuclear region may make the
Balmer decrement and the derived extinction value there artificially
high (Fig. \ref{fig:chb}, left).  The regions outside the nucleus show
$C$(H$\beta$) $\sim$ 0.4, in good agreement with the extinction derived
from the integrated
spectrum outside the nucleus (see below Table~\ref{tab:outer}).
 Figure~\ref{fig:chb} (right panel) shows the map of H$\beta$
  equivalent widths [EW(H$\beta$)] per pixel.  It appears that the highest
  values are not centered on the nucleus but rather in the NW
  direction, or in a circular ring just outside the nuclear region.
  Since the broad component in H$\beta$ is primarily seen in the
  central part of the galaxy, the integrated EW(H$\beta$) per pixel
  drops steeply outwards.

\subsection{Maps of line ratios sensitive to electron temperature and density}\label{densmaps}

\begin{figure}[ht]
\centering
\includegraphics[width=6cm]{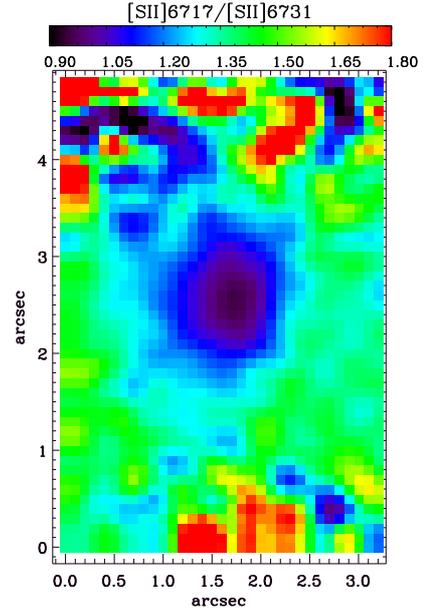}
\caption{Map of 
the [S {\sc ii}] $\lambda$6717/[S {\sc ii}] $\lambda$6731 ratio. 
 The orientation is the same as in Fig. \ref{fig:nar_vel}.
\label{fig:ne_te}}
\end{figure}

Figure~\ref{fig:ne_te} shows the spatial distribution of
the [S {\sc ii}] $\lambda$6717/[S {\sc ii}] $\lambda$6731 ratio.
Low values (corresponding to the high-density regime, $N_e \ge 10^3$
cm$^{-3}$) of this ratio are seen in the central region, with a
gradient towards higher values (corresponding to the low-density
regime, $N_e \le 10^2$ cm$^{-3}$) outside the nucleus, in good
agreement with Figure~\ref{fig:steiner}.

 The ratio of the [O {\sc iii}] emission from an upper level (the
  auroral line, $\lambda$4363) relative to that from lower levels
  (the nebular lines, $\lambda\lambda4959,5007$) is known to be
  highly temperature sensitive. For the low-density region, the
  electron temperatures with the low-density approximation, typical of
  the warm ionized gas in H~{\sc ii} regions, range from 10000-20000
  K. However, as mentioned above, the very high flux ratios we find in
  the central regions is due to the very high density and thus the
  low-density approximation for temperature determination is no longer
  applicable.  Mapping this ratio over the whole extent of our FOV is
  more difficult in our case, since in the nuclear region the
  narrow line is unresolved and in the outer region the broad line is
  not present.  The integrated (broad+narrow) line ratio would yield
  an unphysical result since, as mentioned in \S~\ref{sec:sys}, the
  different line systems originate from regions with completely
  different physical conditions.  We have only used this ratio in the
  integrated spectrum of the region surrounding the nucleus (outer
  spectrum) in order to derive the chemical abundances directly, as
  described below.

\subsection{Diagnostic diagrams}

\begin{figure*}[ht]
\centering
\includegraphics[width=5.4cm]{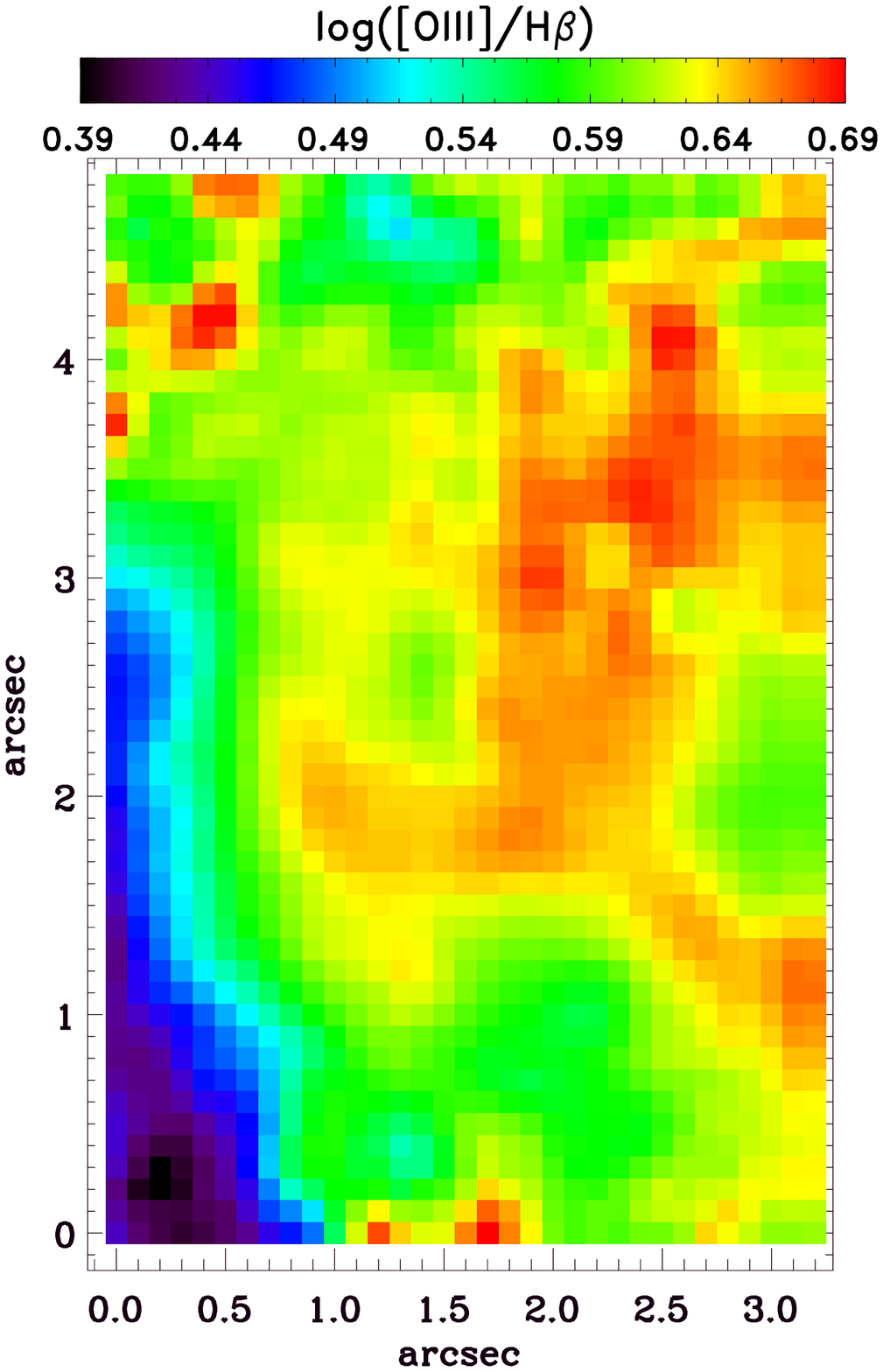}
\includegraphics[width=5.4cm]{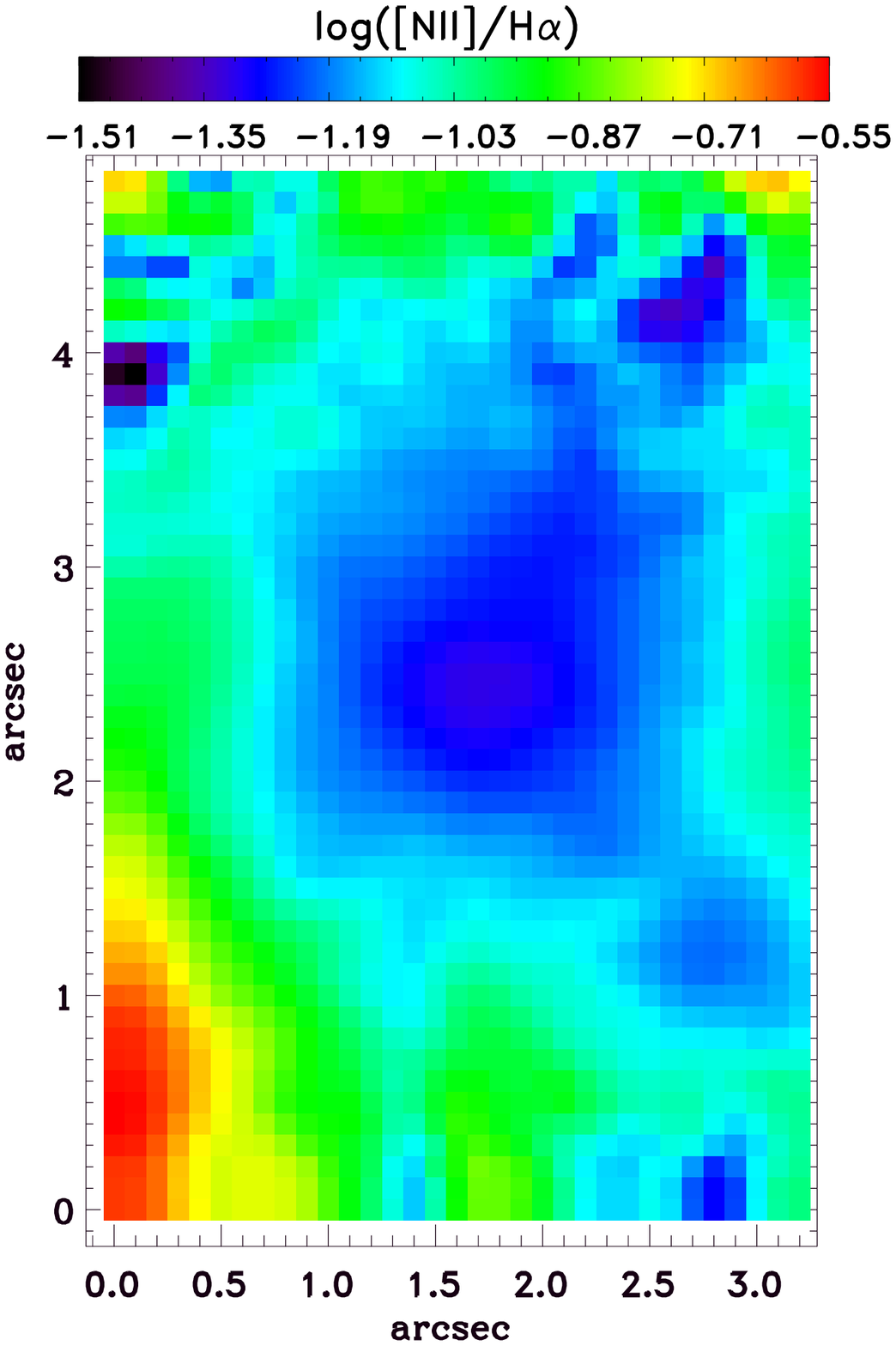}
\includegraphics[width=5.4cm]{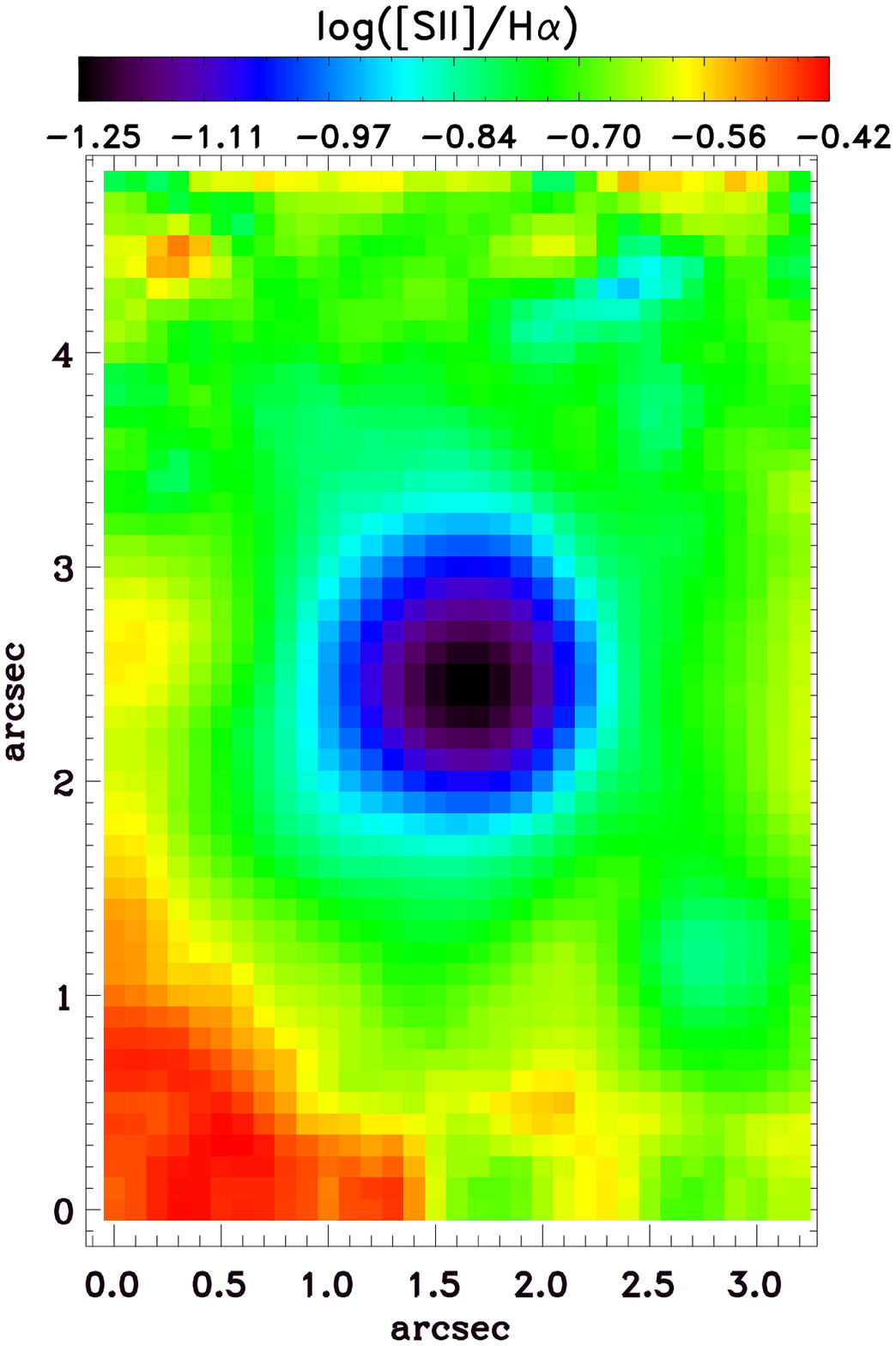}

\caption{Excitation maps: BPT diagram of narrow emission line
  ratios. {\bf (Left)} log([O {\sc iii}] $\lambda$5007/H$\beta$).
  {\bf (Middle)} log([N {\sc ii}] $\lambda$6584/H$\alpha$. {\bf
    (Right)} log([S {\sc ii}] $\lambda$6717,6731/H$\alpha$.  The
  orientation is the same as in
  Fig. \ref{fig:nar_vel}.} \label{fig:bpt}
\end{figure*}

Figure~\ref{fig:bpt} shows maps of some narrow emission-line ratios of
interest such as log([O {\sc iii}] $\lambda$5007/H$\beta$) (left
panel), log([N {\sc ii}] $\lambda$6584/H$\alpha$) (middle panel), and
log([S {\sc ii}] $\lambda$6717,6731/H$\alpha$) (right panel).  These
ratios are often used in a diagnostic diagram, known as the
\citet{bpt81} (BPT) diagram, to identify the source of excitation in
narrow emission-line galaxies.  We have used the mapping of these
ratios
to assess the source of ionization in individual patches of the
interstellar medium (ISM) in Mrk 996 \citep[see
  also][]{lag09,lag12}.  Examination of the spatial distribution of
these diagnostic ratios shows that regions of intense emission of
high-ionization species ([O {\sc iii}]) are coincident with those of
weak emission of low-ionization species ([N {\sc ii}], [S {\sc ii}]),
implying a single source of ionization, namely the UV radiation from massive
stars.  The conclusion is the same if we plot the BPT diagram pixel by
pixel (not shown here) instead of using maps of individual ratios.
Similar results and conclusions were obtained by \cite{J09}. All
points fall in the locus predicted by models of photo-ionization by
massive stars \citep[e.g.,][]{ostfer06}.  However, it has been shown by
\citet{S06} and \citet{Gr06} that an AGN hosted by a low-metallicity
galaxy would not be easily distinguished in such a diagram, even if
the active nucleus contributed significantly to the emission lines. In
fact, the presence of the [O {\sc iv}] $\lambda$25.89 $\mu$m in the
MIR spectrum of Mrk 996 \citep{T08} implies the presence of harder
ionizing radiation than the stellar one. \citet{T08}
analyzed several possible sources of this radiation, including the
presence of an AGN, and concluded that the most probable source is
photo-ionization by fast shocks plowing through a dense ISM.  In a
Chandra X-ray study of Mrk 996, \cite{G11} did not find evidence for
an AGN and also attributed the [O {\sc iv}] $\lambda$25.89 $\mu$m
emission to shocks associated with supernova explosions and stellar
winds.

\subsection{Wolf-Rayet star emission}\label{wolfrayet}

\begin{figure*}[ht]
\centering
\includegraphics[width=6cm]{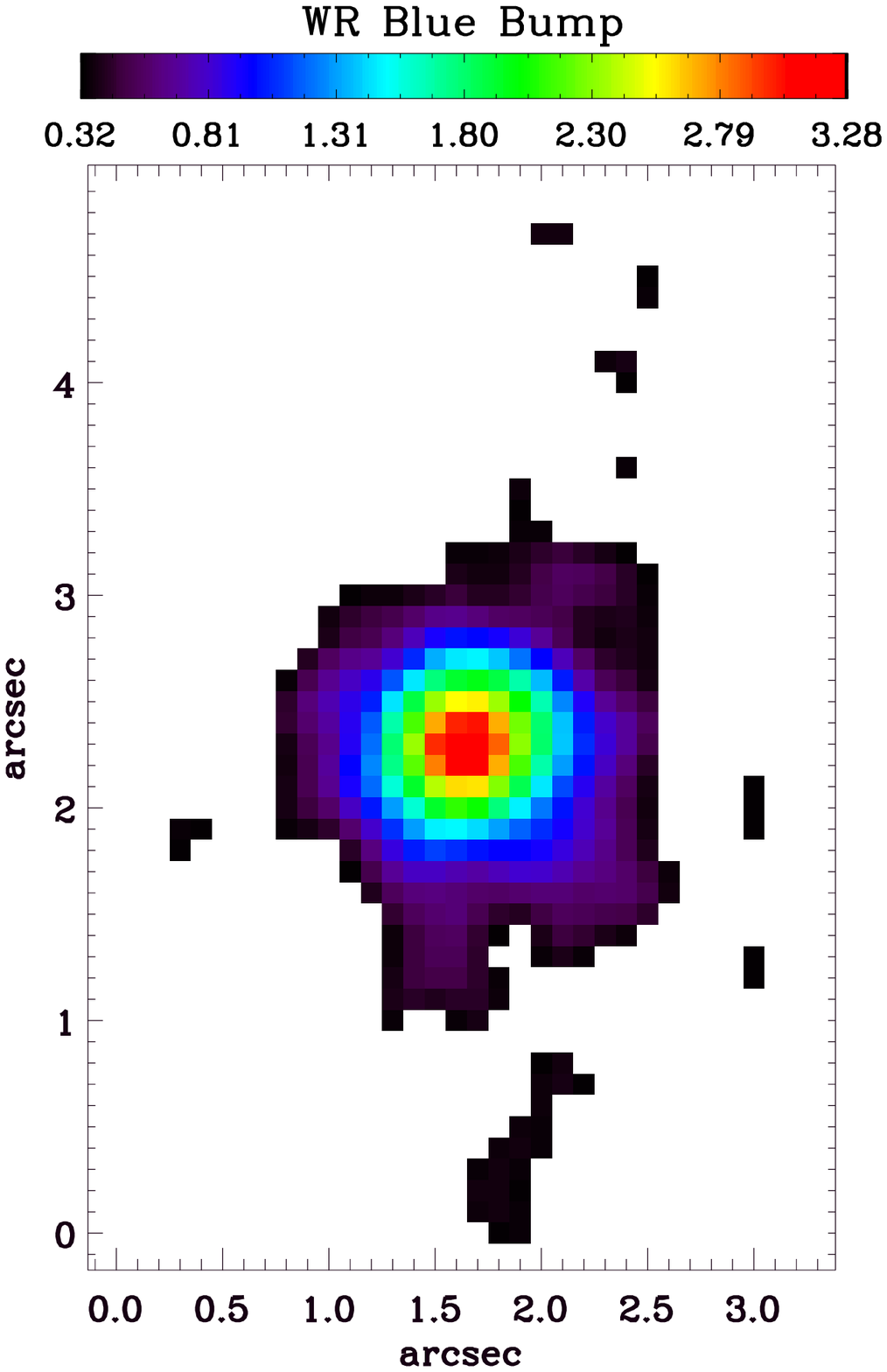}
\includegraphics[width=6cm]{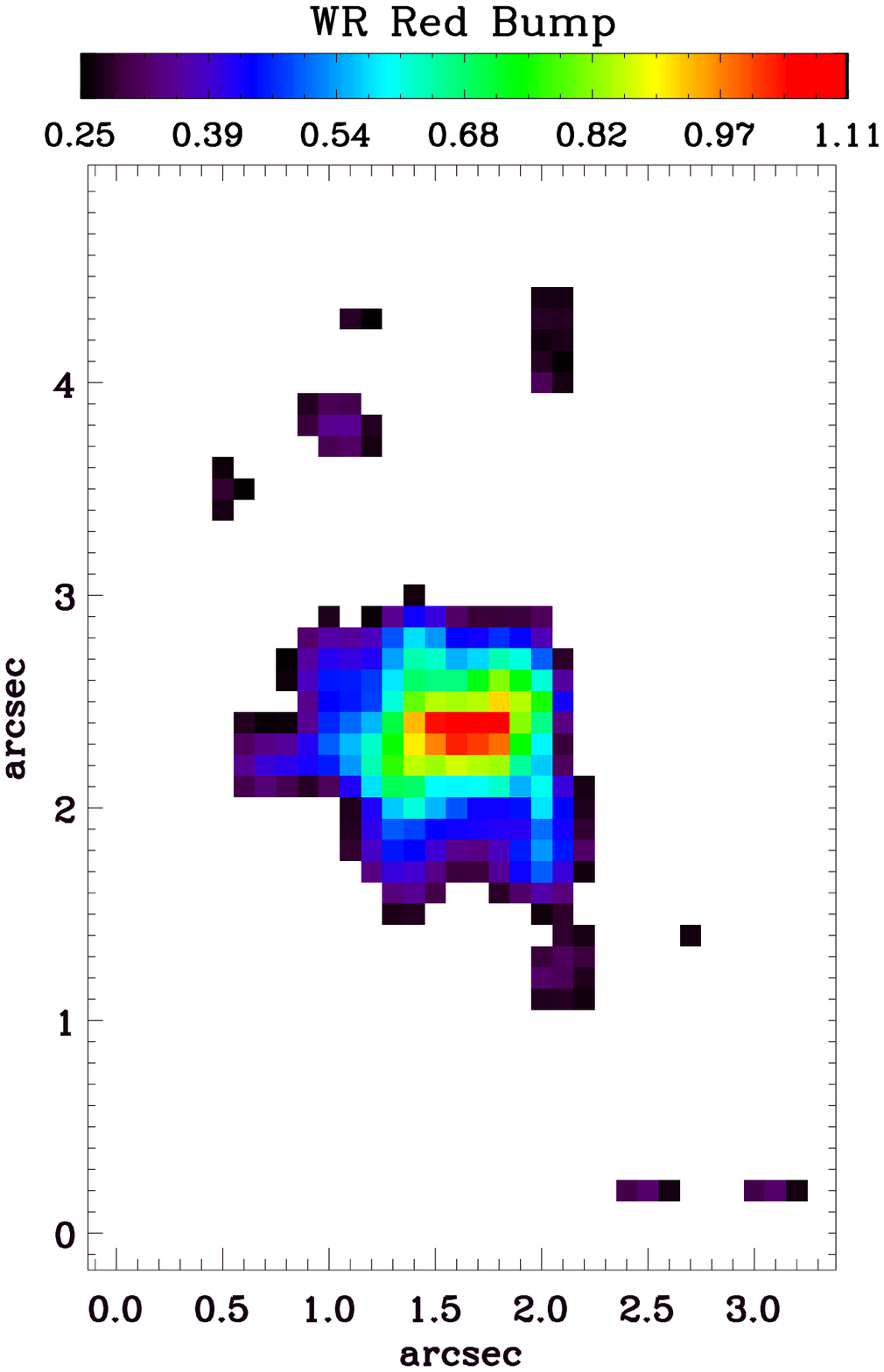}
\caption{Maps of Wolf-Rayet spectral features. {\bf (Left)} 
Blue bump (without the 
[Fe {\sc iii}] emission).
{\bf (Right)} Red bump. The fluxes in the 
bumps are given in units of  
10$^{-16}$ erg s$^{-1}$ cm$^{-2}$.
The orientation is the same as in Fig. \ref{fig:nar_vel}.
\label{fig:wr}}
\end{figure*}

\citet{T96} and \cite{J09} found a large population of
Wolf-Rayet  stars in the central part of Mrk
996. Figure~\ref{fig:wr} shows maps of the spectral features
associated with this Wolf-Rayet stellar population. 
These maps were created by summing the fluxes within the wavelength
range of interest, and subtracting the adjacent continuum.   We
have excluded from it the narrow [Fe {\sc iii}] $\lambda$4658 line.

The left panel shows the map of the blue bump which includes the [N
{\sc iii}] $\lambda$4640 and He {\sc ii} $\lambda$4686 emission
features.  The right panel shows the map of the red bump due to the
weaker C {\sc iv} $\lambda$5808 emission.  The blue bump emission
appears to be more extended spatially than the red bump emission,
though this is just a consequence of the lower S/N in the red bump
feature.  The maps show that the blue and red bumps are coincident
spatially and that their spatial distributions are identical to those
of the broad He {\sc i} and H$\alpha$ emissions.  This implies that
the WR stars are located and concentrated solely in the nuclear region
of Mrk 996.

The total observed (uncorrected for extinction) flux of the blue bump
is $3.87 \times 10^{-14}$ erg s$^{-1}$ cm$^{-2}$, while that of the
red bump is $1.14 \times 10^{-14}$ erg s$^{-1}$ cm$^{-2}$.  These
fluxes derived directly from the maps are a factor of $\sim$ 2 larger
than those derived from the integrated spectrum as described in
\S~\ref{sec:wr}.  The origin of this discrepancy is probably due to
our not performing here a proper deblending of the nebular lines
(i.e., [Fe {\sc iii}] $\lambda$4658, [Fe {\sc iii}] $\lambda$4702, and
He {\sc i} 4713). Thus, a more detailed quantitative comparison with
the integrated spectrum is not warranted. With the maps, we wish only
to show the region where the WR emission originates.

\subsection{Physical conditions and oxygen and nitrogen 
abundances of the outer narrow-line region}\label{sec:ON}

\begin{figure*}[ht]
\centering
\includegraphics[width=6cm]{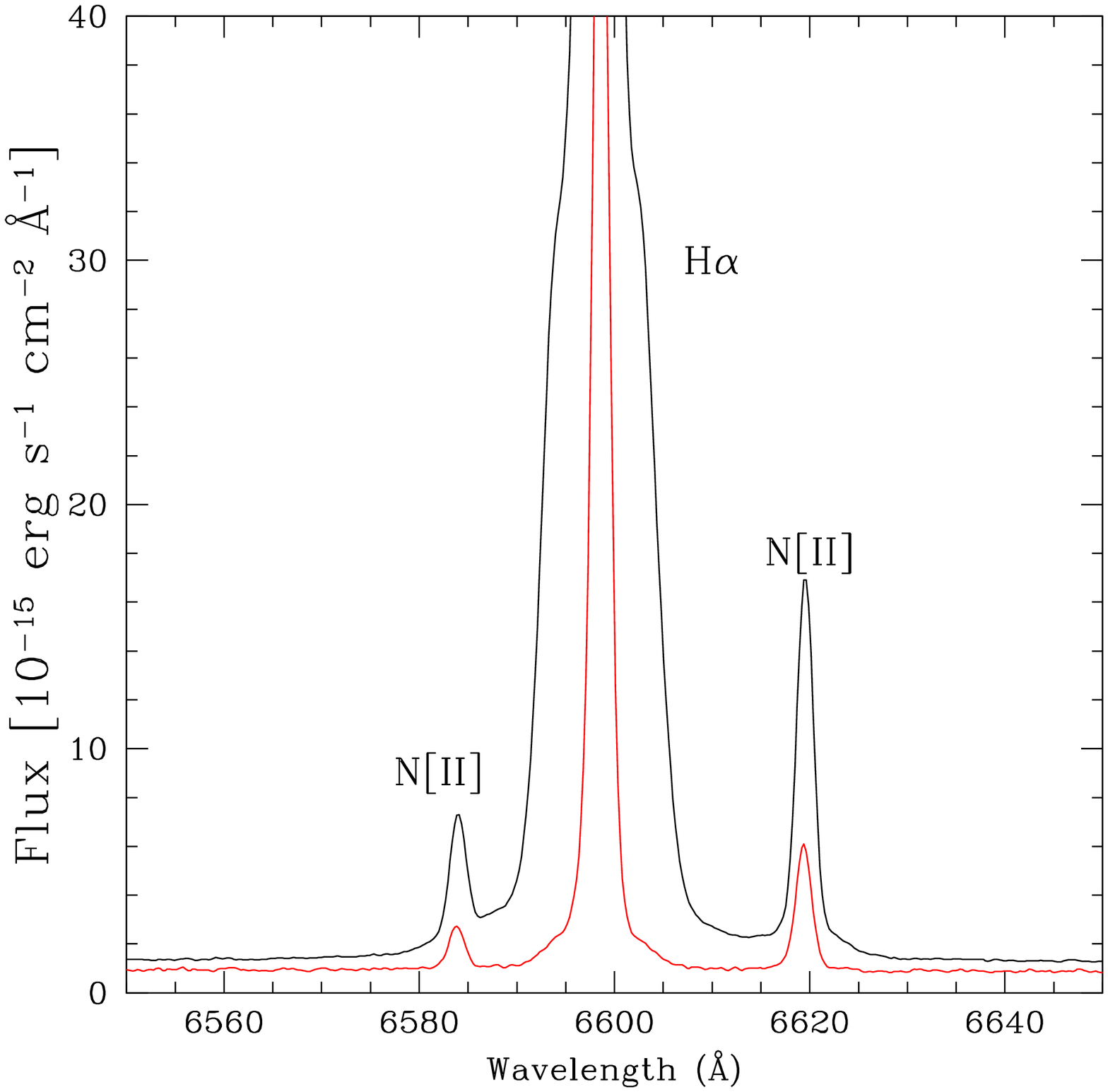}
\includegraphics[width=6cm]{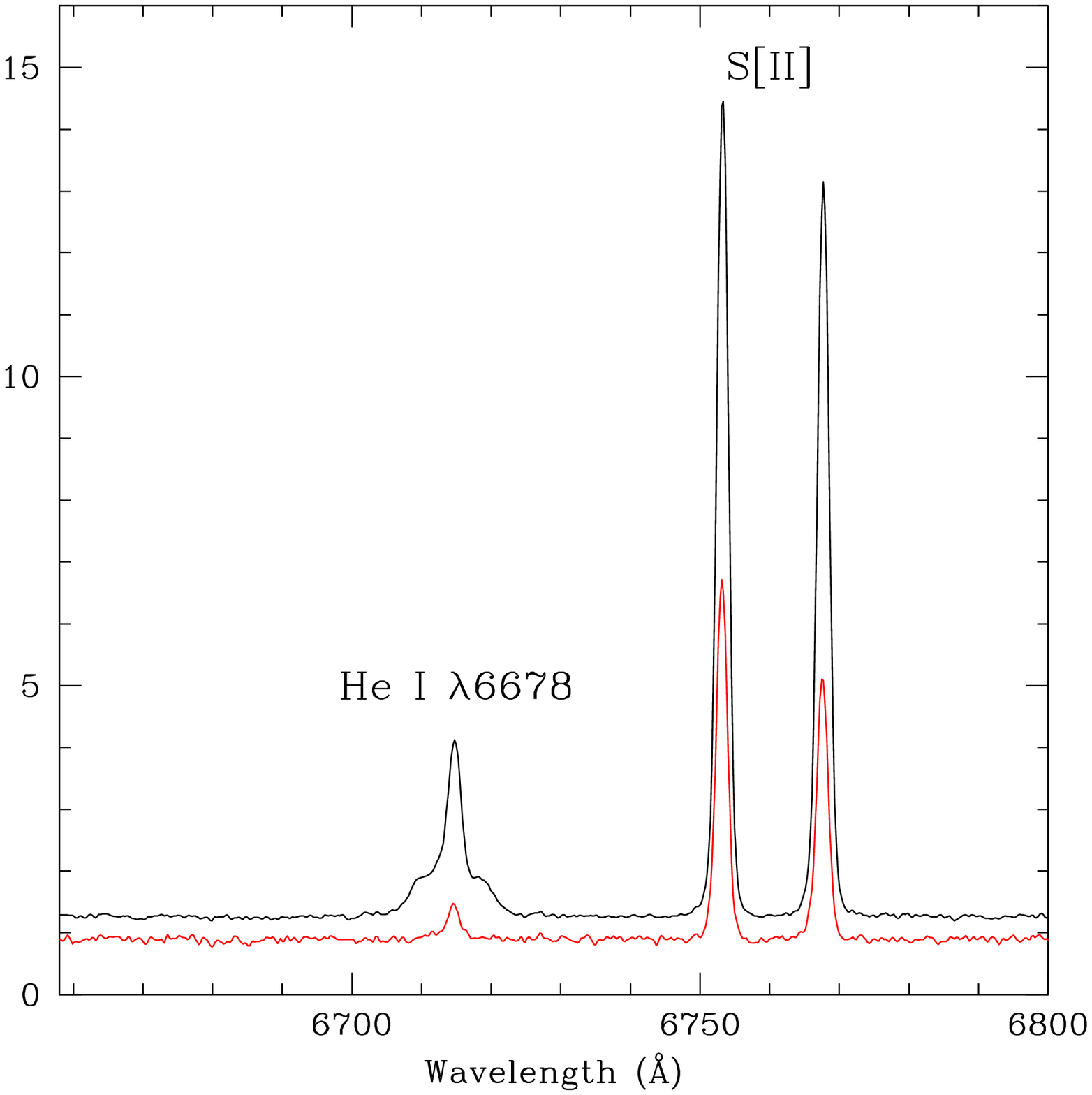}
\includegraphics[width=6cm]{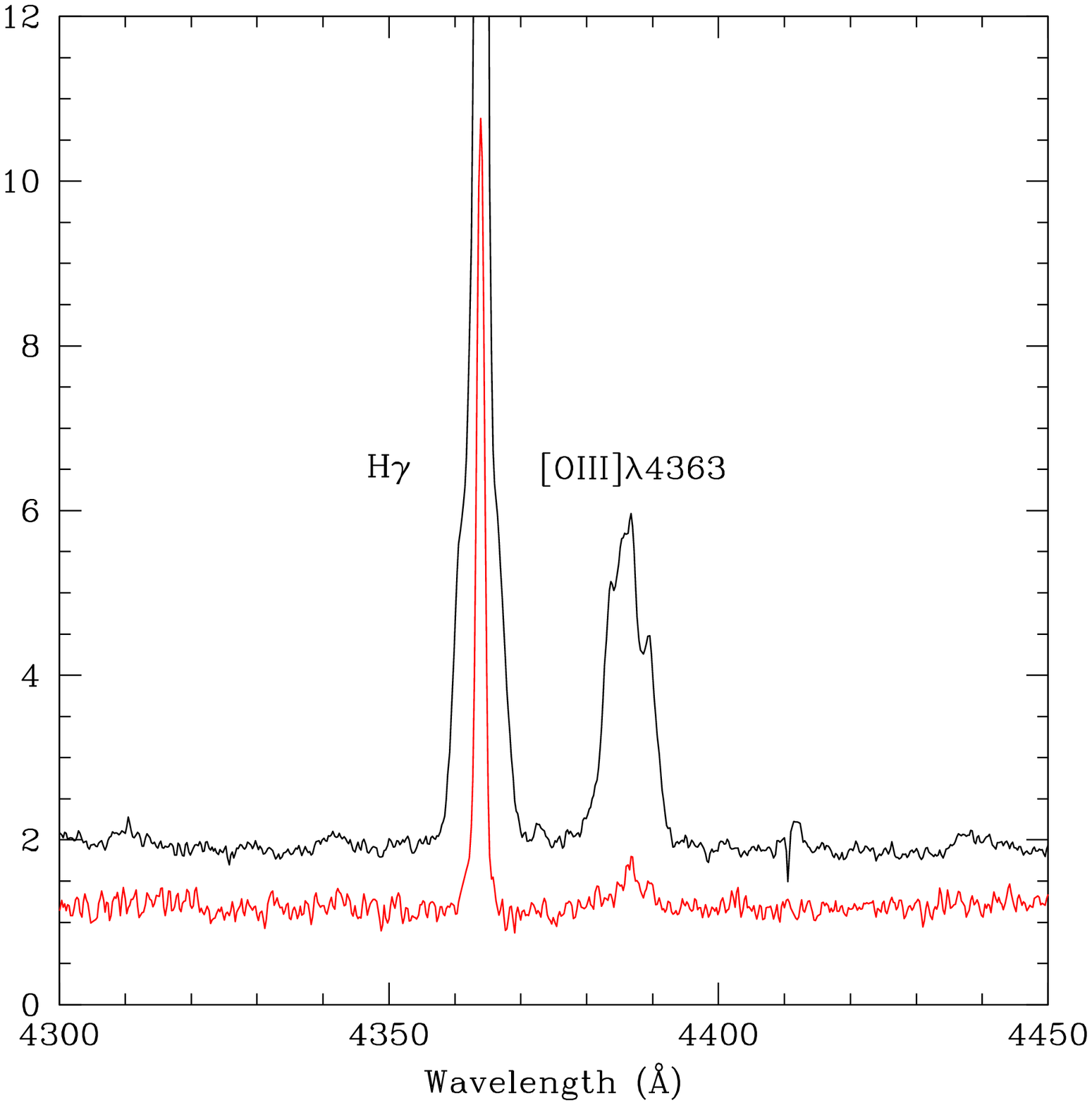}
\caption{Spatially narrow-line region spectrum (lower red spectrum) 
displayed with the integrated spectrum of the nuclear region
 (upper black spectrum).\label{fig:outer}}
\end{figure*}

The integrated nucleus spectrum of Figure~\ref{specfull} clearly shows
the presence of blended broad and narrow lines.  Table~\ref{tabint}
presents the emission line fluxes of both broad and narrow components
derived from this integrated nucleus spectrum (obtained through a
$\sim$1\farcs6 aperture) after line deblending.
As discussed in Sect.~\ref{sec:den}, we have been able to separate
spatially the nuclear broad-line region from the surrounding 
narrow-line region by using the electron density diagnostic map
(Figure~\ref{fig:steiner}). The broad-line region, where the He {\sc
  i}, [O {\sc iii}] $\lambda$4363, and [N {\sc ii}] $\lambda$5755
lines, and the Wolf-Rayet blue and red bumps originate,
coincides with the high-density region,  with a diameter of $\sim$
  1\farcs6, in the electron density diagnostic maps.  The diameter
of $\sim$160pc of the nuclear broad-line and high electron
density region is consistent with the size derived by \cite{T96} by
CLOUDY photo-ionization modeling of the nuclear emission of Mrk 996.

Having spatially resolved the broad-line and high-density region of
Mrk 996 by the use of the various diagnostic maps, we can now go one
step further: we can exclude the nuclear region and extract an
integrated spectrum of the light that comes exclusively from the
narrow-line region surrounding the nucleus.  Figure~\ref{fig:outer}
shows the outer narrow-line region spectrum in red lines (lower
spectrum). For comparison, the integrated spectrum of the nuclear
region is shown by a black line (upper spectrum).  We emphasize that
the narrow lines are {\bf not} derived from a decomposition of the
blended lines, but they are measured from the actual integrated
spectrum in the region outside the nucleus.  The three panels in
Fig.~\ref{fig:outer} show several lines of interest, and are labeled
[O {\sc iii}] $\lambda$4363, H$\alpha$, and He {\sc i} $\lambda$6678.
It can be seen that the narrow component of [O {\sc iii}]$\lambda$4363
in the outer region is very weak, but clearly detected at a 7 $\sigma$
level.  This weak line is swamped by the broad component in the
nuclear region and not detectable at lower S/N observations.  One can
also see that the narrow line of [O {\sc iii}] $\lambda$4363 in the
outer region, as opposed to the broad line of [O {\sc iii}]
$\lambda$4363 in the integrated nucleus region, is not blueshifted.
It falls in the same systemic recession velocity derived from the narrow
components of the lines in the integrated nucleus region.  The
existence of this narrow component originating from the low-density
region has also been demonstrated by a completely independent
technique, that of PCA tomography, as discussed in Sect.~\ref{PCA
  results}.  The [N {\sc ii}] and [S {\sc ii}] lines
show a narrow component everywhere, independent of the density. We
note the high value of the [S {\sc ii}] $\lambda$6717 / $\lambda$6731
ratio in the narrow-line region, indicative of a low electron density.

{\footnotesize
\begin{table}
\caption{Emission line parameters derived from the
outer spectrum within a  2.7 x 4.4\arcsec excluding the inner circle with 1.3\arcsec radius}\label{tab:outer}
\begin{tabular}{lcrcr}
\hline\hline             

Ion & $\lambda_0$&  F($\lambda$)/F(H$\beta$)\tablefootmark{a} &$v(rad)$&$FWHM$ \\
\hline
$[$O {\sc ii}$]$    &  3726.03 &  88.84$\pm$ 20.8 & 1628 &     99$\pm$3    \\   
$[$O {\sc ii}$]$    &  3728.82 & 146.52$\pm$ 26.1 & 1625 &    131$\pm$3    \\
$[$Ne {\sc iii}$]$  &  3868.75 &  40.53$\pm$ 16.6 & 1627 &    156$\pm$4    \\
H8+He {\sc i}       &  3889.05 &  25.09$\pm$ 15.7 & 1614 &    114$\pm$3    \\
$[$S {\sc ii}$]$    &  4068.60 &  46.87$\pm$ 7.05 & 1603 &     78$\pm$2    \\
H$\delta$           &  4101.74 &  23.50$\pm$ 5.20 & 1623 &     98$\pm$3    \\
H$\gamma$           &  4340.47 &  42.01$\pm$ 3.71 & 1620 &    101$\pm$3    \\
$[$O {\sc iii}$]$   &  4363.21 &   3.34$\pm$ 2.01 & 1627 &    143$\pm$4    \\
He {\sc i}          &  4471.48 &   3.87$\pm$ 2.04 & 1615 &    114$\pm$3    \\
H$\beta$            &  4861.33 & 100.00$\pm$ 2.93 & 1626 &    123$\pm$3    \\
$[$O {\sc iii}$]$   &  4958.91 &  96.60$\pm$ 2.72 & 1627 &    121$\pm$3    \\
$[$O {\sc iii}$]$   &  5006.85 & 276.46$\pm$ 3.84 & 1626 &    120$\pm$3    \\
He {\sc i}          &  5875.67 &  13.24$\pm$ 1.52 & 1633 &    109$\pm$3    \\
$[$O {\sc i}$]$     &  6300.30 &   4.85$\pm$ 1.17 & 1631 &     99$\pm$3    \\
$[$S {\sc iii}$]$   &  6312.10 &   2.01$\pm$ 1.15 & 1633 &    110$\pm$3    \\
$[$O {\sc i}$]$     &  6363.78 &   1.64$\pm$ 1.09 & 1635 &    100$\pm$3    \\
$[$N {\sc ii}$]$    &  6548.03 &  10.80$\pm$ 0.98 & 1641 &     93$\pm$2    \\
H$\alpha$           &  6562.82 & 366.07$\pm$ 3.71 & 1634 &     91$\pm$2    \\
$[$N {\sc ii}$]$    &  6583.41 &  29.83$\pm$ 1.75 & 1639 &     90$\pm$2    \\
He {\sc i}          &  6678.15 &   3.95$\pm$ 1.07 & 1637 &    109$\pm$3    \\
$[$S {\sc ii}$]$    &  6716.47 &  33.34$\pm$ 1.47 & 1638 &     85$\pm$2    \\
$[$S {\sc ii}$]$    &  6730.85 &  24.84$\pm$ 1.82 & 1638 &     88$\pm$2    \\
He {\sc i}          &  7065.28 &   1.91$\pm$ 1.36 & 1635 &     83$\pm$2    \\
$[$Ar {\sc iii}$]$  &  7135.78 &   9.90$\pm$ 1.42 & 1637 &     84$\pm$2    \\
 \hline
\end{tabular}

$F_{nar}$(H$\beta$)=(35.60$\pm$2.10)$\times$10$^{-15}$ erg s$^{-1}$ cm$^{-2}$\\
$c($H$\beta_{narrow}) = 0.31$

\tablefoottext{a}{In units 100$\times$$F$/$F_{narrow}$(H$\beta$)}

\end{table}
%
}

\begin{figure}[ht]
\centering
\includegraphics[width=3.0in]{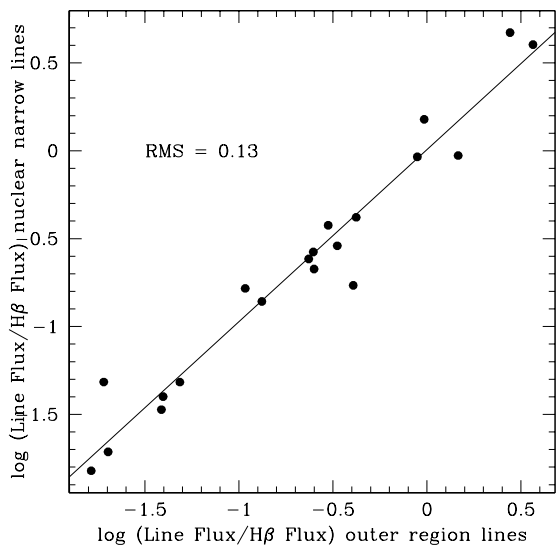}
\caption{Comparison of the measured line fluxes in the integrated
  spectrum of the outer region outside the nucleus
  (Table~\ref{tab:outer}) with the fluxes measured in the narrow
  components of the integrated spectrum of the nucleus of Mrk 996
  (Table~\ref{tabint}). \label{fig:out_nuc}}
\end{figure}

 We can now use the measured line fluxes, shown in
 Table~\ref{tab:outer} from the spatially separated narrow-line region
 spectrum (red-line lower spectra in
 Figure~\ref{fig:outer}) to derive the physical conditions and element
 abundances in the low-density outer region.  As expected, the line
 fluxes measured outside the nucleus are in good agreement with the
 measured fluxes of the narrow components of the integrated spectrum
 in the nuclear region, as shown in Figure~\ref{fig:out_nuc},
 corroborating our statement that this zone of low density is on the
 line of sight of the inner broad-line dense nucleus. By scaling the
 intensity of the [O {\sc iii}] $\lambda$4363 in the outer region with
 the narrow H$\beta$ flux ratio of the outer region to the nuclear
 region, we estimate that the narrow component of the [O {\sc iii}]
 $\lambda$4363 in the nuclear region must be $\sim$ 4 fainter than the
 integrated line, which explains the difficulty in detecting this line
 directly in the integrated nuclear spectrum.

\subsubsection{Oxygen abundances} 

We make use of the {\it nebular} package available in the {\it stsdas}
external package under IRAF to derive abundances. The tasks in this
package are based on a five-level atom model developed by
\cite{derobertis87}.  The detection of [O {\sc iii}] $\lambda$4363
allows a direct determination of the electron temperature
$T_e$(O$^{++}$)=$1.29\times10^4$K, while the ratio of the [S {\sc ii}]
lines permits us to determine a low electron density in this narrow-line
region of $N_e \sim$71 cm$^{-3}$.
We obtain 12+log(O/H)=7.94$\pm0.30$ using $C$(H$\beta$)=0.31, that is,
Z$_\odot$/6 by adopting the solar calibration of \cite{A09}.  The
  same result 12+log(O/H)=7.88$\pm0.36$ is obtained using the T$_e$
  direct method with the prescriptions of \cite{pag92},
  \cite{IZ94}, and \cite{TH95}.

This oxygen abundance is in agreement with the value of \cite{T96} who
derived 12+log(O/H)=8.0 with a CLOUDY two-zone model of Mrk 996.  
  It is, however, considerably less (by a factor of at least 3) than the
  lower limit of 12+log(O/H) $\sim$ 8.37 obtained by \cite{J09} for
  the broad-line region, based on an assumed electron temperature of
  10,000 K. Our value of the oxygen abundance is more reliable since
  the [O {\sc iii}] $\lambda$4363 line intensity, and hence the
  electron temperature, was determined directly in the low-density
  region outside the nucleus.

\subsubsection{Nitrogen abundances}

The derived nitrogen abundance for the low-density region is 
log(N/O)=$-$1.53$\pm$0.15, in good agreement with the value of $-$1.43 
obtained by \cite{J09}. It is typical of values derived for BCDs \citep{I06}.   

However, the nuclear region is nitrogen enhanced by a factor of $\sim$
20, with log(N/O)=$-$0.15$\pm 0.1$.  In this case, we assumed a
two-density model in {\em nebular}, with $N_e$(low)=450 cm$^{-3}$,
$N_e$(high)=$10^6$ cm$^{-3}$, $T_e$(low)=$10^4$K and
$T_e$(high)=$4\times10^4$K.  $T_e$(high) is derived from the
  extinction corrected ratio of [OIII] broad-line fluxes assuming
  $N_e$(high).
Our high N/O in the broad-line region
is in agreement with that obtained by \cite{T96} from CLOUDY modeling
(model 2), and with the value of $-$0.13 obtained by \cite{J09}.  This
nitrogen enhancement is probably due to local pollution from WR stars.
 Similar cases for a nitrogen enhancement have been observed, for
  example, in the central region of the nearby dwarf starburst galaxy NGC
  5253 by \citet{wes13}, and for the very metal-deficient (12+ log
  (O/H)=7.64) luminous (MB= -18.1m) blue compact galaxy (BCG) HS
  0837+4717 by \citet{pul04}. \citet{brin08}, in turn, have noted an
  elevated N/O for galaxies with Wolf-Rayet features in their optical
  spectrum from SDSS survey. \citet{lo10} have also detected a high N/O
  ratio in objects showing strong WR features (HCG 31 AC, UM 420, IRAS
  0828+2816, III Zw 107, ESO 566-8, and NGC 5253). They claim that the
  ejecta of the WR stars may be the origin of the N enrichment in
  these galaxies.  However, as pointed out by \citet{I06}, this
high N/O cannot be caused by WR nitrogen-enriched ejecta with a number
density $N_e$ similar to that of the ambient gas, but only by
considerably denser ejecta since the emissivity of the [N {\sc ii}]
lines is proportional to $N_e^2$.

\section{Summary and conclusions}\label{conclusions}

The galaxy Mrk 996 is an extraordinary blue compact dwarf galaxy with a dense
unresolved nucleus ($<160$ pc) and a central density of $\sim$10$^6$
cm$^{-3}$. The dense nucleus is surrounded by a lower density star-forming  
region, with a density of
$\sim$10$^2$ cm$^{-3}$.  We present here integral field spectroscopy obtained 
with GEMINI-SOUTH/GMOS/IFU which allows us to study in 2D the physical
conditions, ionization structures, and 
kinematic properties of these two regions, as well as their relationship.

 We have made an extensive comparison with the results of the previous
 similar IFU work by \citet{J09} on Mrk 996. Our results fully agree
 with theirs on the spatial variation of the physical properties and
 detected internal structures of Mrk 996.  However, our quantitative
 results differ somewhat in the number of massive stars present in the
 star forming nucleus, and in the determination of the chemical
 abundances.  The first disagreement may be partly due to absolute
 calibration differences.  Their fluxes are a factor of $\sim$ 5
 brighter than our simulated VIMOS aperture.  We believe our reduction
 and calibration procedures to be correct as they have been
 double-checked independently by us using different standard stars.
 The red cube and blue data cubes were derived from observations of
 different nights and the calibrations agreed, without the need for
 averaging the data.  Other external checks have also shown
 consistency. More importantly, we have made a more precise
 determination of oxygen abundance in the low-density gas by directly
 detecting the narrow line of [OIII]$\lambda$4363 outside the nucleus.
 This detection was confirmed, independently, by the use of a new
 innovative method of data cube analysis.

We have thus obtained the following results:

1) The integrated spectrum shows four kinematically distinct systems
of emission lines, with line-widths decreasing outwards from the
center.  The first system shows both broad and narrow lines,
originating from the nuclear region. The broad component is probably
associated with the circumnuclear envelopes around WR stars, while the
narrow component is probably related to the circumnuclear envelopes
around O stars.  The second system comes from the innermost and
densest part of the star-forming region.  It consists of the peculiar
emission of the [O {\sc iii}]$\lambda$4363 line which is broad
($\sigma \sim 200$ \kmsec) and blueshifted by 60 \kmsec\ relative to
the narrow H$\beta$ line.  The third system consists of the permitted
hydrogen and helium lines, and of the auroral [N {\sc ii}] line.
These also show large widths, similar to those in the second system,
but are less blueshifted ($\sim 20$ \kmsec).  They are likely to
originate in regions farther away from the center than the second line
system. The fourth system consists only of narrow lines of hydrogen,
helium, doubly ionized ions, and of forbidden lines of neutral and
singly ionized species.  These all have the same radial velocities as
the H$\beta$ emission line.

2) Most of the observed physical conditions and kinematics of the nucleus
of Mrk 996 appear to be related to the presence of a Wolf-Rayet 
stellar population, as 
inferred by the presence of the blue and red
bump spectral features. We estimate, from the integrated spectrum as
well as from the spatially resolved maps of these features, that 
the nuclear region of Mrk 996 contains 
$\sim$ 473 WNL and $\sim$ 98 WCE stars, with 
$N$(WR)/$N$/(O+WR)=0.19, at the high end for WR galaxies.

3) The monochromatic emission maps and the line ratio, velocity, and
dispersion maps of the nuclear region suggest an isotropic ionized gas
outflow from the center ($<$160 pc), superposed on an underlying
rotation pattern. This outflow is probably associated with the ejecta
from the WR stellar population.  The narrow ($\sigma \sim 45$\kmsec)
and broad ($\sigma \sim 200$\kmsec) lines from the nucleus are
supersonic.   The [O {\sc iii}]$\lambda$ 4363 and [N {\sc ii}]$\lambda$
5755 lines show peculiar kinematics, suggestive of outflow motions
from the nucleus, or multiplicity due to a patchy ISM.

4) We have also performed a PCA tomography analysis that
corroborates, in a completely independent manner, the kinematic
picture outlined above for Mrk 996: an outflow from the inner region,
associated with winds from WR stars, superposed on an underlying
rotation pattern, affecting the motions of low-density clouds.  This
recently developed statistical method for handling data cubes has also
resulted in the independent detection of the narrow component of the
[O {\sc iii}]$\lambda$ 4363 line, not seen previously in integrated
spectra.  This detection allows a reliable and direct measurement of
the chemical abundances in the region outside the nucleus.

5) We obtain an oxygen abundance 12+log(O/H)=7.90$\pm0.30$ ($\sim$ 0.2
$Z_\odot$) for the low-density  narrow-line region around the
nucleus, in agreement with the abundance derived by \cite{T96}, but
considerably less than the lower limit of 12+logO/H=8.37 derived
  by \cite{J09} for the broad-line region, based on an assumed
  electron temperature. The N/O ratio in the low-density region is
typical of BCDs.  However, as discussed by previous investigators,
there is a nitrogen enhancement by a factor of $\sim$ 20 in the
nuclear region, probably due to nitrogen-enriched WR ejecta, but also
to enhanced nitrogen line emission in a high-density environment.
 The presence of a large number of WR stars in the nucleus of this
  dwarf starburst galaxy greatly affecting its local kinematics is
  somewhat unexpected because of its sub-solar oxygen abundances. If we
  take the results of stellar population models at face value \citep[see,
  e.g.,][]{cer94}, this implies that we are witnessing a very
  short phase of the evolution of the starburst in Mrk 996.

6) Finally, we have used a method recently proposed by \cite{smro09}
to map low- and high-density regions in Mrk 996.  By performing
surface photometry on the resulting high-density image, we have
obtained an upper limit of $\sim$ 1\farcs6 ($\sim$ 160 pc) for the
diameter of the nuclear region. This region is where the broad lines
originate.  The emission line ratios in the BPT diagnostic diagram
appear to indicate ionization from UV radiation of massive stars
alone.  The presence of harder ionizing radiation, as implied by the
presence of the [O {\sc iv}] $\lambda$25.9 $\mu$m emission-line in its
MIR spectrum \citep{T08}, is most likely due to fast radiative shocks
propagating in a dense interstellar medium.  This is in agreement
  with the Chandra X-ray study of Mrk 996 by \cite{G11} who rule out
  an AGN, but also favor shocks as the ionization source for the [O
  {\sc iv}] $\lambda$25.9 $\mu$m emission.  We intend to 
  investigate this issue further with new integral field observations in the
  near-IR with adaptive optics.  This will allow us to set
  more stringent upper limits to the size of the broad-line emitting
  region, and also to study additional physical properties of the nebular
  emission in this wavelength range including possible H$_2$ emission
  that probe the warm molecular gas in the shells around the
  photo-dissociation regions.

The advent of Integral Field Units has allowed spatially 
resolved spectroscopy, with   
the simultaneous acquisition of a complete set of spectra covering 
a spatially resolved field of view. Despite this great instrumental 
advance, the superior and abundant data that are obtained 
with IFUs are still mostly analyzed with the traditional techniques 
of 1D slit spectroscopy.  Although 
these traditional methods are useful for a first look at the data, 
they do not fully exploit the wealth of information
given by integral field spectroscopy. In some cases, they may
not even be the proper tools to use, particularly when assumptions on
boundaries of regions need to be made, or other structure-dependent
physical equilibria need to be assumed. We have attempted to apply 
a few new techniques to the data in this paper, and they have 
yielded useful results. More new methodologies for analysis are required,
and we will apply them as they are devised.

\acknowledgements

E.T. acknowledges the US Gemini Fellowship by AURA that supported his
visit to the Astronomy Department of the University of Virginia where
most of this work was carried out. T.X.T is grateful for the hospitality of the
Institut d'Astrophysique de Paris.  We thank Roberto Cid Fernandes,
Linda J. Smith, Mark Whittle, and Fran\c cois Cuisinier (in memorium)
for reading the manuscript and providing fruitful comments. We are
also indebted to Joao Steiner and Tiago Ricci for their comprehensive
help with the application and interpretation of our PCA tomography
results.  We are specially thankful to the referee for his/her careful
analysis of this manuscript and criticisms and suggestions that
greatly improved the presentation of our results.

\end{document}